\documentclass{svjour3}                     
\smartqed  
\usepackage{graphicx}
\begin{document}

\title{Control of a nonlocal entanglement in the micromaser via two quanta non-linear
processes induced by dynamic Stark shift
%
}

\titlerunning{}        

\author{M. S. Ateto 
}


\institute{M. S. Ateto \at
              Mathematics Department, Faculty of Science, South Valley
University, 83523 Qena, Egypt\\
             \email{omersog@yahoo.com}             \\
}

\date{Received: date / Accepted: date}

\maketitle

\begin{abstract}
We show that, under certain conditions, the micromaser can act as an effective source of
highly correlated atoms. It is possible to create an extended robust entanglement between two successive, initially unentangled atoms passing through a cavity filled with with a nonlinear medium taking into consideration a slight level shift. Information is transfered from the cavity to the atoms in order to build up entanglement. The scheme has an advantage over conventional creation of entanglement if the two atoms (qubits) are so far apart that a direct interaction is difficult to achieve. The interaction of the atoms with the micromaser occurs under the influence of a two-quantum transition process. Interesting phenomena are observed, and an extended robust entangled state is obtained for different values of the system parameters. Illustrative variational calculations are performed to demonstrate the effect within an analytically tractable two-qubit model.
\end{abstract}

\section{Introduction}
\label{intro}
Entanglement usually arises from quantum correlations between
separated subsystems that cannot be created by local actions on
each subsystem. Preparation of quantum entanglement between distant
parties is an important task required for quantum communication and
quantum information processing~\cite{Nielsen00}. In such processing one usually
needs to find the entanglement properties and a way to control it,
therefore the study of the dynamic properties of entanglement is useful
for processing quantum information.\\
In the last years there has been
an intensive research in the field of quantum communication that has 
yielded a variety of methods to distribute bipartite entanglement
~\cite{ZZHE93,DBCZ99,D99,A04,VPC04,GS04,OA07}. Nevertheless, due to the lack of a complete understanding of
mixed state entanglement and multi-partite entanglement, it is not
always clear what is the optimal way to distribute entanglement
among distant parties.\\
A great effort has been devoted to the  generation of 
atomic entanglement and entanglement between cavity modes through atom-photon interactions~\cite{PWK04,BBPS96,BDSW96,VP96,V02,VT99,VPRK95,W98,GMN06,GMN07,JCPZ94,SPPK91,JPLK91,KLMCK93} and some
notable experimental demonstrations have also been performed,
for instance Refs.~\cite{RBH01,MCJLKK06,DGMN04,TDD03,M02,TDFD03,HMNWBRH97,KLAK02}. Of particular interest is the generation of entangled states in two-atom systems, since
they can represent two qubits, the building blocks of the quantum gates that are essential to implement quantum protocols in quantum information processing.\\
A number of studies have shown that entanglement can be created between two objects that do not interact directly with each other, but interact with a common field, heat bath or thermal cavity field~\cite{GMN07,KLAK02,Braun02}. The formation of
atom-photon entanglement and the subsequent generation of correlations between spatially separated atoms have been shown using the micromaser~ \cite{M02,HMNWBRH97,LEW96,AMNN01,RSKW90}.\\
The micromaser~\cite{PB93,FIJAME86,PK88} is appreciated as a practical device
for processing information. It stores radiation for
times significantly longer than the duration of the interaction
with any single atom~\cite{PB93}. The interaction of
an atom with the intracavity field of a micromaser will leave the atom-field system in an entangled
state. The long cavity lifetime implies that the memory of
this entanglement can influence the interaction with subsequent
atoms and nonlocal correlations between these
successive atoms can be induced leading to a violation
of Bell’s inequality~\cite{PB93}. In other words, the successive
atoms can interact with the field left by earlier atoms,
in this manner gain can be seen with the most diffuse of
gain media containing, on average, less than one atom
at any given moment~\cite{DMHW85} and the dynamics of an atom
inside the cavity will modify the evolution of the later
atoms~\cite{RSKW90,FJM86}.\\
For a field that is interacting with more than one atom at a time, the atom-field entanglement was investigated~\cite{KLMCK93}. In this scheme~\cite{KLMCK93}, an atom of fixed position situated at a peak of the cavity field becomes entangled with a second atom, at a variable distance away from the first atom, via their mutual cavity field with which they interact. Furthermore, the entanglement between a pair of
atoms pumped at the same time through a micromaser
has been analyzed in Ref.~\cite{M02}. In practice it is rather difficult to realize such a setup though. The genuine one-atom micromaser, on the other hand, can be operated over a
reasonably large region of parameter space, and is thus a feasible device~\cite{RSKW90} for generating entanglement between two or more atoms.\\
It has become well known that the degree of quantum entanglement depends crucially on the physical nature
of the interacting objects and the character of their
mutual coupling. It has been noticed that the above investigations
involved mostly the absorption or emission
of a single photon in an atomic transition. However,
involvement of more than one photon, in particular,
two photons in the transition between two atomic
levels has been known for a long time~\cite{LNT65}. The output
radiation from such interactions exhibits nonclassical
properties such as strong sub-Poissonian photon statistics as well as fields with increased photon number~\cite{MCKRD94}.
Needless to say, the idea of squeezed light has originated
from two-photon process~\cite{Yuen76}. Thus, it would be
interesting to study the properties of two-atom entanglement
in the framework of a two-photon process. Moreover,
the two-photon process introduces a dynamic Stark shift
in the atomic transition which is related to the magnitude
of the electric field of the radiation inside the cavity.\\
However, it should be noticed that all these results have been obtained
for the case where the Kerr medium and the
Stark shift are ignored.
The Kerr medium~\cite{FALI95,AFAT02,WLG06} can
be modelled as an anharmonic oscillator with frequency
$\omega$. Physically this model may be realized as if the cavity
contains two different species of atoms, one of which
behaves like a two-level atom and the other behaves like
an anharmonic oscillator in the single-mode field of frequency $\omega$~\cite{JOPU92}. Such a model is interesting by itself. The cavity mode is coupled to the Kerr medium as well as
to the two-level atoms.\\
A Kerr-like medium can be useful
in many respects, such as detection of nonclassical
states~\cite{Hillaer91}, quantum nondemolition measurement~\cite{CHCOWO92},
investigation of quantum fluctuations~\cite{ZHGCLM00}, generation
of entangled macroscopic quantum states\cite{Gerry99,PKH03}, and
quantum information processing~\cite{PACH00,VFT00}.\\
In a previous study~\cite{Braun02} it has been shown that the entanglement
between two qubits that do not interact directly with
each other can be created for a very short time after the
interaction is switched on.\\
Our purpose here is to demonstrate
analytically the increase of the entanglement time
created for two atoms (qubits) if the cavity
field is filled with a nonlinear medium and a slight level shift
is taken into consideration. In section
II, we introduce our model and the obtained wave function that
controls the model in a general form at time $t>0$. A brief discussion
of the technique we are going to use to compute the entanglement (concurrence) of
mixed quantum states is introduced in section III. The reduced
density matrices of some special cases of our final state vector at
any time $t>0$
depending on various initial states of the system are computed in section IV, supported by discussions of the study. Our conclusions are summarized in section V.
\section{The full system and its solution}
\label{sec:1}
We consider a scheme of the micromaser-type where
two two-level atoms traverse a high-$Q$ ($Q\approx10^9$)
single-mode cavity one after the other in a
manner that their flights through the cavity do not overlap~\cite{GMN07,DGMN04,LEW96,AMNN01,PB93,Ateto07,EGSWH96,MHNWBRH97,ELSW002}. There is no direct
interaction between the two atoms, although secondary
correlations develop between them. The entanglement
of their wave functions with the cavity photons can
be used to formulate local-realist bounds on the detection
probabilities for the two atoms~\cite{LEW96,PB93}. The
generation of nonlocal correlations between the two
atomic states emerging from the cavity can in general
be understood using the Horodecki theorem~\cite{HHH95}. The
cavity mode is assumed to be filled with a Kerr-like
medium~\cite{FALI95,AFAT02,WLG06}. Each atom has energy levels $|1_{i}\rangle$ and $|0_{i}\rangle$
(i=1,2) such that $E_{1_{i}}-E_{0_{i}}=\hbar \omega_{0}$. We assume that the
two atoms make individually two-photon transitions of frequency $2\omega$
between the nondegenerate states $|0_{i}\rangle$ (the ground state, energy $%
E_{0_{i}}$, $i=1,2$) and $|1_{i}\rangle$ (the excited state, energy $%
E_{1_{i}}$). The transitions are mediated by a single intermediate level $%
|k\rangle$ (energy $E_{k}$, with $E_{1_{i}}>E_{k}>E_{0_{i}}$); the
frequencies for $|0_{i}\rangle\rightarrow|k\rangle$ and $|1_{i}\rangle%
\rightarrow|k\rangle$ are $\omega-\Delta$ and $\omega+\Delta$, respectively.
The frequency $\omega_{0}$ includes a spontaneous contribution to Stark shift due to a
direct dipole transition from the intermediate level $|k\rangle$ to $%
|0_{i}\rangle$ and $|1_{i}\rangle$. The coupling constants $\kappa_{1}$ (for $|0_{i}\rangle\rightarrow|k\rangle$), $\kappa_{2}$ (for $%
|1_{i}\rangle\rightarrow|k\rangle$), and $\Delta$ determine the Stark shift
parameters $\beta_{1}$ and $\beta_{2}$ of the two levels and also the
coupling $\kappa$ between the effective two level atoms, states $%
|0_{i}\rangle$ and $|1_{i}\rangle$, and the field mode
$$
\beta_{1}=\kappa_{1}^{2}\Delta^{-1},~ \beta_{2}=\kappa_{2}^{2}\Delta^{-1},~
\kappa=\kappa_{1}\kappa_{2}\Delta^{-1}.\eqno(1)
$$
The atom-field interaction is governed by the Jaynes-Cummings
(JC) model via a two-quantum process~\cite{FALI95,AFAT02,FNLU96}. It is assumed that the atom-field interaction time is shorter than the lifetime of the cavity, so that the cavity relaxation will not be considered. The cavity field is assumed to be filled with a nonlinear medium, namely, Kerr medium~\cite{FALI95,AFAT02,WLG06}, while the atoms are assumed to have a shift in their levels due to the interaction with the radiation field. Assuming for simplicity the photon mode to be in resonance with the atoms, the model Hamiltonian under the rotating wave approximation (RWA) reads
$$
\hat{\emph{H}}=\omega_{0}\hat{S}_{3}+\omega \hat{A}^{\dagger}\hat{A}+\chi \hat{A}^{\dagger 2}\hat{A}^{2}+%
\frac{1}{2}\hat{A}^{\dagger}\hat{A}\bigl[\beta_{1}(1-\hat{S}_{3})+\beta_{2}(1+\hat{S}_{3})\bigr]
$$
$$
+
\kappa\bigl(\hat{A}^{\dagger 2}~\hat{S}_{-}+\hat{A}^{2}~\hat{S}_{+}\bigr);~~~~~~~~(\hbar=1),\eqno(2)
$$
where $\omega$ is the cavity frequency, $\omega_{0}=2\omega$ is the
frequency of two atomic energy level difference and $\kappa=\kappa_{1}%
\kappa_{2}\Delta^{-1}$ is the coupling parameter that connects the field with
the atomic system. We denote by $\chi$ the dispersive part of the third-order susceptibility of the Kerr-like medium~\cite{FALI95,AFAT02,WLG06}. The operator $\hat{A}^{\dagger} (\hat{A})$ is the field creation
(annihilation) operator, which satisfies the commutation relation $[\hat{A},
\hat{A}^{\dagger}]=1$. The operators $\hat{S}_{+}$, $\hat{S}_{-}$ and $\hat{S}_{3}$ are the usual raising, lowering and inversion operators for the two-level atomic system,
which satisfy the commutation relations $[\hat{S}_{3}, \hat{S}_{\pm}]=\pm 2\hat{S}_{\pm}$ and $[\hat{S}_{+}, \hat{S}_{-}]=\hat{S}_{3}$.\newline
The Hamiltonian given by Eq. (2) can be written in the form
$$
\hat{\emph{H}}=\hat{\emph{H}}_{0}+\hat{\emph{H}}_{int},\eqno(3)
$$
where $H_{0}$ represents the unperturbed Hamiltonian that is given by
$$
\hat{\emph{H}}_{0}=\omega(\hat{A}^{\dagger}\hat{A}+\hat{S}_{3}),\eqno(4)
$$
while the perturbed Hamiltonian is given by:
$$
\hat{\emph{H}}=\chi \hat{A}^{\dagger 2}\hat{A}^{2}+\frac{1}{2}\hat{A}^{\dagger}\hat{A}\bigl[%
\beta_{1}(1-\hat{S}_{3})+\beta_{2}(1+\hat{S}_{3})\bigr]
+ \kappa\bigl(\hat{A}^{\dagger 2}~\hat{S}_{-}+\hat{A}^{2}~\hat{S}_{+}\bigr),\eqno(5)
$$
The state vector $|\psi_{f}(t=0)\rangle$ of the field is represented by a linear
superposition of the number state $|n\rangle$, i.e.,
$$
|\psi_{f}(t=0)\rangle=\sum_{n=0}^{\infty} F_{n}|n\rangle, \eqno(6)
$$
where $|n\rangle$ is an eigenstate of the number operator $\hat{A}^{\dagger}\hat{A}=n$%
; $\hat{A}^{\dagger}\hat{A} |n\rangle=n|n\rangle$, and $F_{n}$ is, in general, complex
and gives the probability of the field to have $n$ photons by the relation:
$$
P(n)=\langle n|\psi_{f}(t=0)\rangle \langle
\psi_{f}(t=0)|n\rangle=|F_{n}|^{2}.\eqno(7)
$$
As already indicated above, we consider a pair of two-level atoms going
through the cavity mode one after another. Then the initial state vector of
the interacting first atom-field system is given by
$$
|\psi_{a-f}(t=0)\rangle=|\psi_{a}(t=0)\rangle \otimes
|\psi_{f}(t=0)\rangle=\sum_{n=0}^{\infty} F_{n}|n, 1_{1}\rangle, \eqno(8)
$$
where $|1_{1}\rangle$ represents the state vector of the first atom being
in excited state. At any instant of time $t$ the joint state vector of the
field and the first atom can be obtained from the solution of the time-dependent  Schr\"{o}dinger equation
$$
i \frac {d}{dt}|\psi_{a-f}(t)\rangle=\hat{\emph{H}}~|\psi_{a-f}(t)\rangle,%
\eqno(9)
$$

The time of flight through the cavity $t$ is the same for every
atom~\cite{GMN07,DGMN04,LEW96,AMNN01,PB93,Ateto07,EGSWH96,MHNWBRH97,ELSW002}, and the joint state vector of both the two atoms and the field may be denoted by $|\psi_{a-a-f}(t)\rangle$,  and the corresponding atom-atom-field pure-state density operator is:
$$
\rho(t)=|\psi(t)\rangle \langle\psi(t)|;~~~~~~
|\psi(t)\rangle=|\psi_{a-a-f}(t)\rangle.\eqno(10)
$$
In order to quantify the degree of entanglement between the two atoms, the
field variables must be traced out. One may write the reduced mixed-state density
matrix of the two atoms after taking the trace over the field variables as:
$$
\rho_{a-a}(t)=Tr_{field}~\rho(t).\eqno(11)
$$
Under the initial condition (6), by solving the Schr\"{o}dinger equation (9),
we obtain directly the time-dependent wave function of the atom-field system that
evolves according to the form
$$
|\psi_{a-f}(t)\rangle=\sum_{n}^{\infty} F_{n}e^{-it\Lambda_{n}}[K_{n}(t)~|n,1_{1}
\rangle
+~ R_{n+2}(t)~|n+2,0_{1} \rangle],~~~\eqno(12)
$$
with the amplitudes $K_{n}(t)$ and $R_{n}(t)$ given by:
$$
K_{n}(t)=\cos(\Upsilon_{n}t)+i\kappa\bigl[\frac{\chi}{\kappa}~(2n+1)
+\frac{n(r^{2}-1)+2r^{2}}{2r%
}\bigr]\frac{\sin(\Upsilon_{n}t)}{\Upsilon_{n}},~~~\eqno(13)
$$
and,
$$
R_{n-2}(t)=-i\kappa\sqrt{n(n-1)}~\frac{\sin ~(\Upsilon_{n-2}t)}{\Upsilon_{n-2}}%
,~~~~~\eqno(14)
$$
where $\Upsilon_{n}$ is given by
$$
\Upsilon_{n}=\kappa\sqrt{\bigl[\frac{\chi}{\kappa} (2n+1)+\frac{n(r^{2}-1)+2r^{2}}{2r}\bigr]^{2}+(n+1)(n+2)}~.~~~~~~~~~~~\eqno(15)
$$
and $\Lambda_{n}$ reads:\newline
$$
\Lambda_{n}=\kappa\bigl[\frac{\chi}{\kappa} n(n+1)+\frac{n(r^{2}+1)+2r^{2}}{2r}\bigr],\eqno(16)
$$
with $r=\kappa_{1}/\kappa_{2}$.
\newline
Note that within the delay time between the two atoms the field evolves towards a thermal steady state, moreover, repetition of the instant in which the later atoms enter the cavity means the same field repeats at this instants precisely when successive atoms exit the cavity~\cite{FIJAME86}.\\
If an additional atom is prepared in a superposition as
$$
|\psi_{a}(t>0)\rangle_{0}=a|1_{2}\rangle+ (a-1)^{2}|0_{2}\rangle,\eqno(17)
$$
this atom will interact with the field that has been modified by the passage of the first atom. Assuming the flight time $t$ of the two atoms through the cavity to be
the same, the joint time-evolved wave vector of the tripartite system of the two atoms and the cavity system after the second atom leaves the cavity is obtained by solving the Schr\"{o}dinger equation,
$$
i\frac {d}{dt}|\psi_{a-a-f}(t)\rangle=H_{int}|\psi_{a-a-f}(t)\rangle,\eqno%
(18)
$$
which is expressed as:
\[
|\psi_{a-a-f}(t)\rangle=\sum_{n} F_{n}\biggl\{a~ \biggl(
e^{-2it\Lambda_{n}}[H_{n}(t)|n,1_{1},1_{2} \rangle
+T_{n+2}(t)|n+2,1_{1}, 0_{2} \rangle]~~~~~~~~~~
\]
\[
~~~~~~~~~~~~~~~~~~~+e^{-it\Lambda_{n+2}}e^{-it%
\Lambda_{n}}[J_{n+2}(t)|n+2,0_{1},1_{2} \rangle
+V_{n+4}(t)|n+4,0_{1}, 0_{2} \rangle]\biggr)
\]
\[
\times~(a-1)^{2}\biggl(e^{-it\Lambda_{n}}e^{-it\Lambda_{n-2}}[W_{n}(t)
|n,1_{1},0_{2}\rangle
+X_{n-2}(t)|n-2,1_{1},1_{2}\rangle]~~~
\]
$$
+ e^{-2it\Lambda_{n}}[Y_{n+2}(t) |n+2,0_{1},0_{2}\rangle
+Z_{n}(t)|n,0_{1},1_{2}\rangle]\biggr)\biggr\},~~\eqno(19)
$$
where the amplitudes $H_{n}(t)$, $T_{n+1}(t)$, $J_{n+1}(t)$, $V_{n+2}(t)$, $%
W_{n}(t)$, $X_{n-2}(t)$, $Y_{n+2}(t)$, and $Z_{n}(t)$ are given by:
$$
H_{n}(t)=[K_{n}(t)]^{2},\eqno(20)
$$
$$
T_{n+2}(t)=K_{n}(t)R_{n+2}(t),\eqno(21)
$$
$$
J_{n+2}(t)=K_{n+2}(t)R_{n+2}(t),\eqno(22)
$$
$$
V_{n+4}(t)=R_{n+2}(t)R_{n+4}(t).\eqno(23)
$$
$$
W(n,t)=K_{n}(t)K^{\ast}_{n-2}(t),\eqno(24)
$$
$$
X_{n-2}(t)=K_{n}(t)R_{n}(t),\eqno(25)
$$
$$
Y_{n+2}(t)=K^{\ast}_{n}(t) R_{n+2}(t),\eqno(26)
$$
and
$$
Z_{n}(t)=[R_{n+2}(t)]^{2},\eqno(27)
$$
respectively.
\section{Entanglement measure}
\label{sec:2}
For bipartite pure states, the partial (von Neumann) entropy of the reduced density matrices can provide a good measure of entanglement. However, for mixed states von Neumann entropy fails, because it can not distinguish classical and quantum mechanical correlations. For mixed states, the entanglement can be measured as the average entanglement of its pure-state decompositions $%
E_{f}(\rho)$:
$$
E_{f}(\rho)=min \sum_{i} p_{i} E(\psi_{i}),\eqno(28)
$$
with $E(\psi_{i})$ being the entanglement measure for the pure state $\psi_{i}$
corresponding to all the possible decompositions $\rho=\sum_{i}
p_{i}|\psi_{i}\rangle\langle\psi_{i}|$. The existence of an infinite number
of decompositions makes their minimization over this set difficult.
Wooters~\cite{W98} succeeded in deriving an analytical solution to this difficult
minimization procedure in terms of the eigenvalues of the non-Hermitian
operator
$$
R=\rho \tilde{\rho},\eqno(29)
$$
where the tilde denotes the spin-flip of the quantum state, which is defined
as:
$$
\tilde{\rho}=(\sigma_{y}\otimes\sigma_{y})\rho^{\ast}(\sigma_{y}\otimes%
\sigma_{y}),\eqno(30)
$$
where $\sigma_{y}$ is the Pauli matrix, and $\rho^{\ast}$ is the complex
conjugate of $\rho$ where both are expressed in a fixed basis such as $%
\{|e\rangle, |g\rangle\}$.
\newline
In terms of the eigenvalues of $R=\rho \tilde{\rho}$, $E_{f}(\rho)$ (known
as the entanglement of formation) takes the form
$$
E_{f}(\rho)=H\biggl[\frac{1}{2}+\frac{1}{2}\sqrt{1-C^{2}(\rho)}\biggr],\eqno(31)
$$
where $C(\rho)$ is called the concurrence and is defined as:
$$
C(\rho)=max\biggl(0, \sqrt{\lambda_{1}}-\sqrt{\lambda_{2}}-\sqrt{\lambda_{3}}-%
\sqrt{\lambda_{4}}\biggr),\eqno(32)
$$
with the $\lambda$'s representing the eigenvalues of $R=\rho \tilde{\rho}$ in
descending order, and,
$$
H(z)=-z \log z-(1-z) \log (1-z)\eqno(33)
$$
is the binary entropy. The concurrence is associated with the entanglement of
formation $E_{f}(\rho)$, Eq.(31), but it is by itself a good measure for
entanglement. The range of concurrence is from 0 to 1. For unentangled atoms
$C(\rho)=0$ whereas $C(\rho)=1$ for maximally entangled atoms.\newline
We consider special cases of the initial conditions, namely\newline
(i) only one atom excited; and (ii) initially both atoms excited.\\
We
apply two different excitations of the initial field, namely Fock state excitation and thermal field excitation.
\section{Special cases}
\label{sec:3}
\subsection{case 1. Excition in a Fock state}
\label{sec:3}
If the field is excited in a Fock state, the amplitudes $F_{n}$ in Eq. (6) obey the relation:
$$
F_{n}=\delta_{m,n},\eqno(34)
$$
where $m$ is the photon number of Fock state.
\subsubsection{Only one excited atom}
By setting $a=0$ in Eq. (17) and $F_{n}=\delta_{m,n}$ in Eq. (19), we obtain the wave
function of the system with field excited in a Fock state and with initially excited atom followed by an atom in the ground state.\newline
Having obtained the wave function of the full system, the corresponding
density operator of the total system can be easily obtained using Eq.(10). The
atom-atom system can be described in the basis of product states of the
individual atoms
$$
|1_{1},0_{2}\rangle=|1\rangle,\eqno(35a)
$$
$$
|1_{1},1_{2} \rangle=|2\rangle,\eqno(35b)
$$
$$
|0_{1},0_{2}\rangle=|3\rangle,\eqno(35c)
$$
$$
|0_{1},1_{2}\rangle=|4\rangle.\eqno(35d)
$$
Applying Eq. (11) to obtain the reduced density operator of the two
atoms, which can be written in this basis as:
$$
\rho_{a-a}(t)=\rho_{11}(t)|1\rangle\langle
1|+\rho_{14}(t)|1\rangle\langle 4| +\rho_{22}(t)|2\rangle\langle 2|
$$
$$
+\rho_{33}(t)|3\rangle\langle 3|+\rho_{41}(t)|4\rangle\langle
1|+\rho_{44}(t)|4\rangle\langle 4|,\eqno(36)
$$
with
$$
\rho_{11}(t)=|W_{n}(t)|^{2},\eqno(37)
$$
$$
\rho_{14}(t)=e^{-i\kappa\bigl[2\frac{\chi}{\kappa} (2n-1)+\frac{r^{2}+1}{2r}\bigr]t}~W_{n}(t)
Z^{\ast}_{n}(t)=\rho^{\ast}_{41}(t),\eqno(38)
$$
$$
\rho_{22}(t)=|X_{n-2}(t)|^{2},\eqno(39)
$$
$$
\rho_{33}(t)=|Y_{n+2}(t)|^{2},\eqno(40)
$$
$$
\rho_{44}(t)=|Z_{n}(t)|^{2}.\eqno(41)
$$
Expressing the reduced density state (36) in matrix form as:
$$
\mathbf{\rho_{a-a}(t)} = \left(
\begin{array}{cccc}
\rho_{11} & 0 & 0 & \rho_{14} \\
0 & \rho_{22} & 0 & 0 \\
0 & 0 & \rho_{33} & 0 \\
\rho_{41} & 0 & 0 & \rho_{44}%
\end{array}
\right),\eqno(42)
$$
one may write the spin-flip reduced density state $\tilde{\rho}$ of $%
\rho$ by applying Eq.(30) in the form:
$$
\mathbf{\tilde{\rho}_{a-a}(t)} = \left(
\begin{array}{cccc}
\rho_{44} & 0 & 0 & \rho_{41} \\
0 & \rho_{33} & 0 & 0 \\
0 & 0 & \rho_{22} & 0 \\
\rho_{14} & 0 & 0 & \rho_{11}%
\end{array}
\right).\eqno(43)
$$
From an easy procedure one can obtain the square roots of the
eigenvalues of the matrix $R$, given by Eq.(29), which are expressed by
the set:
$$
\{\sqrt{\lambda_{i}}\}=\biggl\{\sqrt{\rho_{22}\rho_{33}},~\sqrt{%
\rho_{22}\rho_{33}},
~Re(\rho_{14}) +\sqrt{\rho_{11}\rho_{44}-[Im(\rho_{14})]^{2}},
$$
$$
~Re(\rho_{14})-\sqrt{\rho_{11}\rho_{44}-[Im(\rho_{14})]^{2}}\biggr\}.\eqno%
(44)
$$
As found above, one may use the largest eigenvalue
using Eq.(32) to obtain the concurrence $C(\rho)$ as:
$$
C(\rho)=2\biggl(\sqrt{\rho_{11}\rho_{44}-[Im(\rho_{14})]^{2}}-\sqrt{%
\rho_{22}\rho_{33}}\biggr).\eqno(45)
$$
One of the interesting phenomenon described by two-level system is the
dynamical behavior. In the following, we examine the creation of entanglement in a
system that consisting of a pair of two-level atoms mediated by quantum field
contained in a cavity through which the two atoms pass successively. For the case when the initially excited atom is followed by the atom in the ground stste, the
resulting entanglement, measured by the concurrence $C$, as well as the sum of the 
populations $\rho_{22}+\rho_{33}$ are depicted in figures 1, 2, 3
and 4. We examine the effects of the level shifts as
well as of the Kerr-like medium on the creation of entanglement between the two
atoms mediated by the cavity field initially prepared in a Fock state. The
case of the effective vacuum is quite interesting since in this case $C$
oscillates between zeros and its maximum value (see Fig. 1a). It shows a 
two-peak periodical behavior with maxima $\approx 0.8$ that are reached at the maxima of $\rho_{22}+\rho_{33}$ which also shows periodical
behavior with maxima $\approx 0.25$. In this case , the concurrence $C$
reduces to $C=2\sqrt{\rho_{11}\rho_{44}}$ where $\rho_{22}=0$ and $\rho_{14}
$ is always real. In fact, $C$ attains the value zero (corresponding to disentangled
atoms) when $\rho_{22}+\rho_{33}=0$ (corresponding to atoms in pure state) while strong
entanglement occurs at $\rho_{22}+\rho_{33}=0.25$ (corresponding to atoms in coherent atomic
state). It is worth mentioning that a similar behavior was shown in
Ref.~\cite{Ateto07}, for one-quantum process, but with more
oscillations in the same interval of time. 
\begin{figure}
\begin{center}
\includegraphics[width=5.5cm,height=4.5cm]{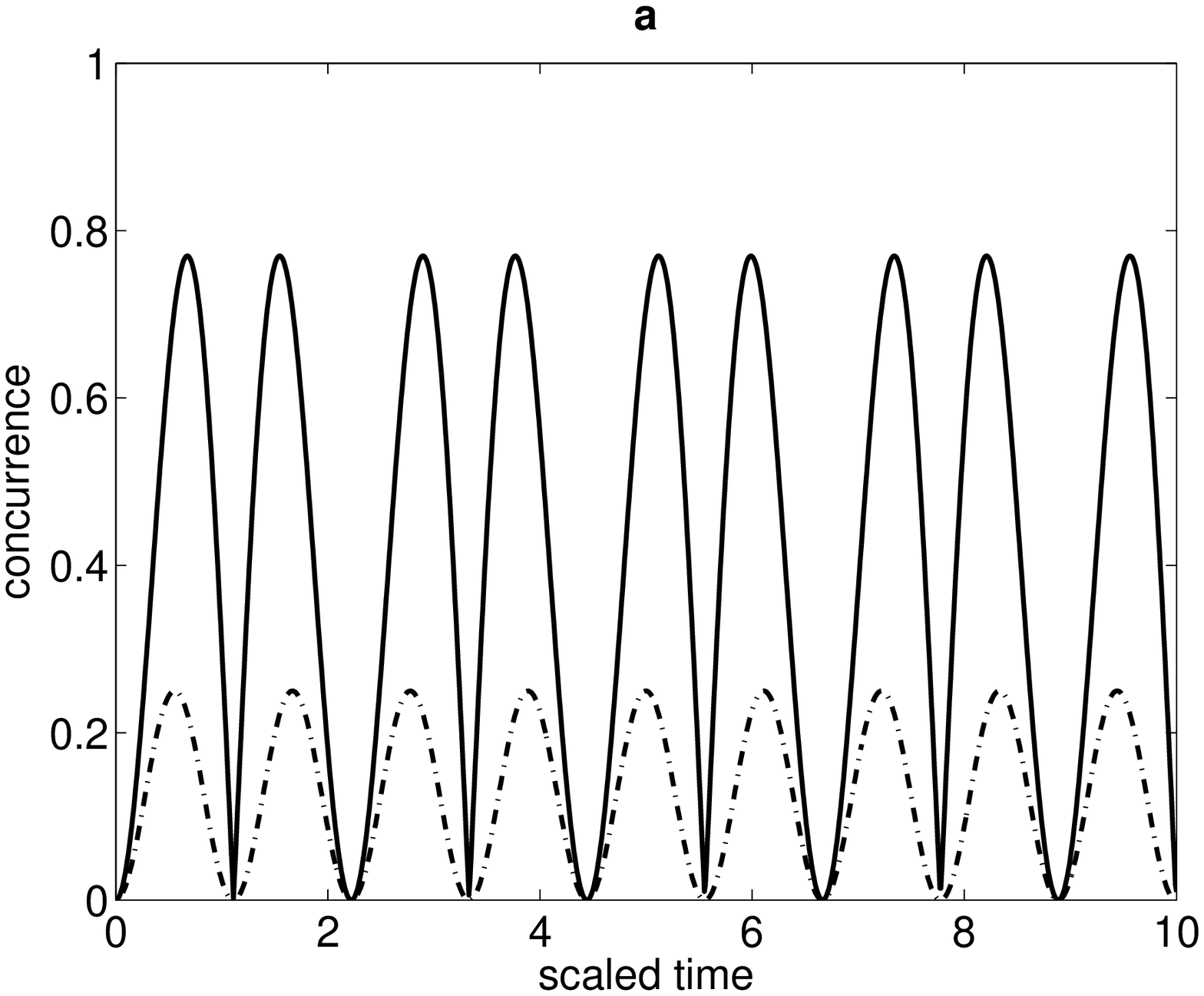}
\includegraphics[width=5.5cm,height=4.5cm]{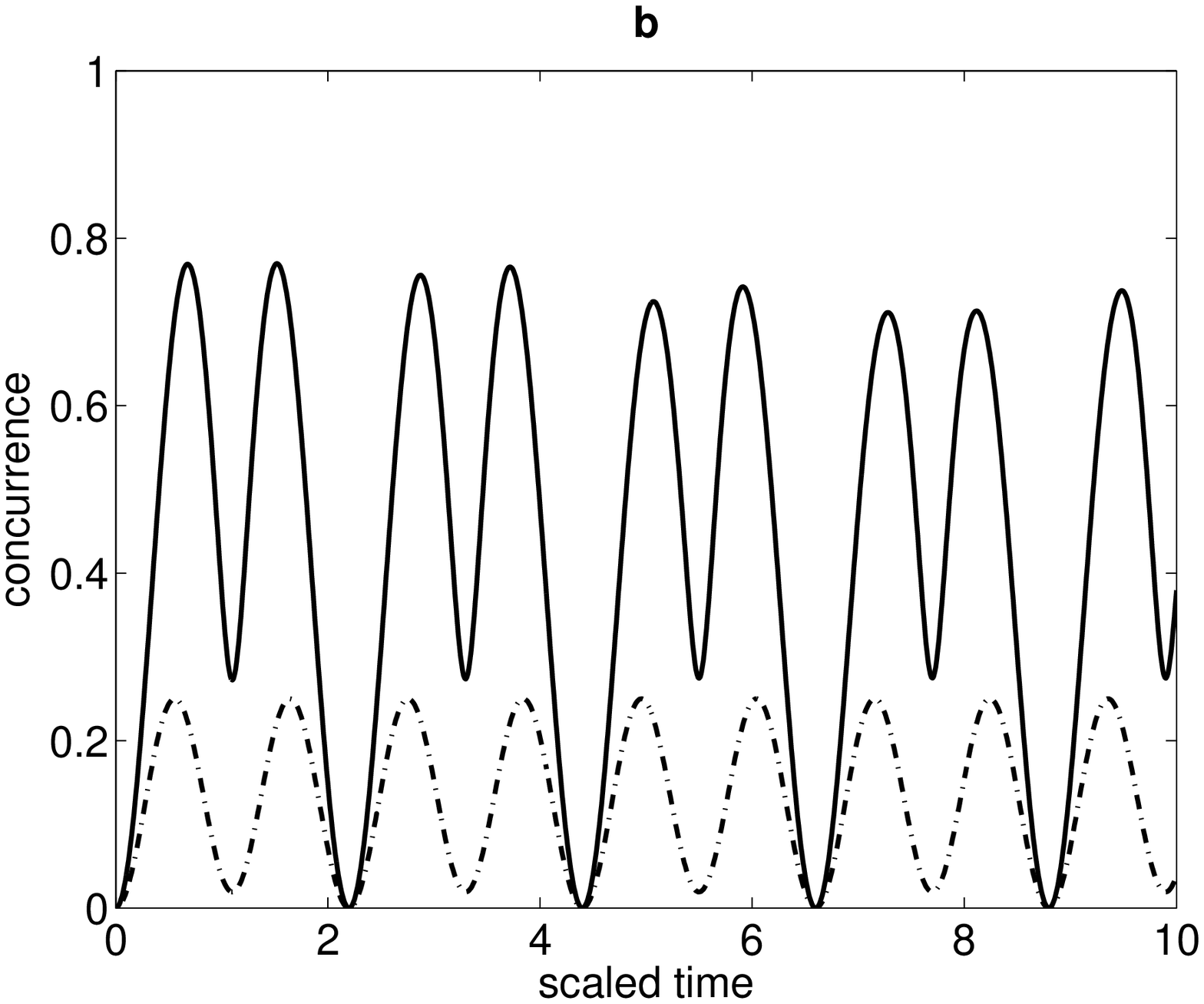}
\includegraphics[width=5.5cm,height=4.5cm]{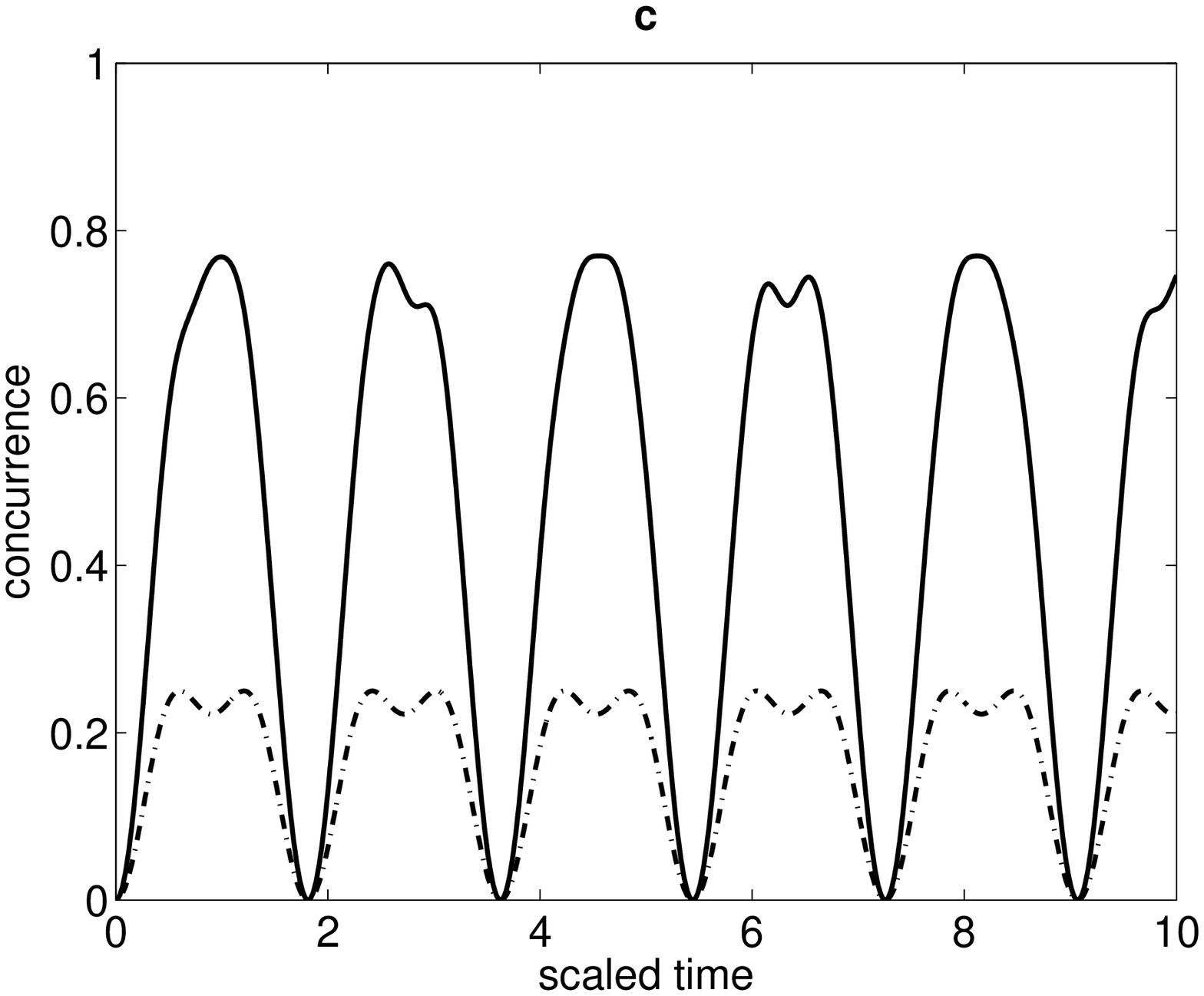}
\includegraphics[width=5.5cm,height=4.5cm]{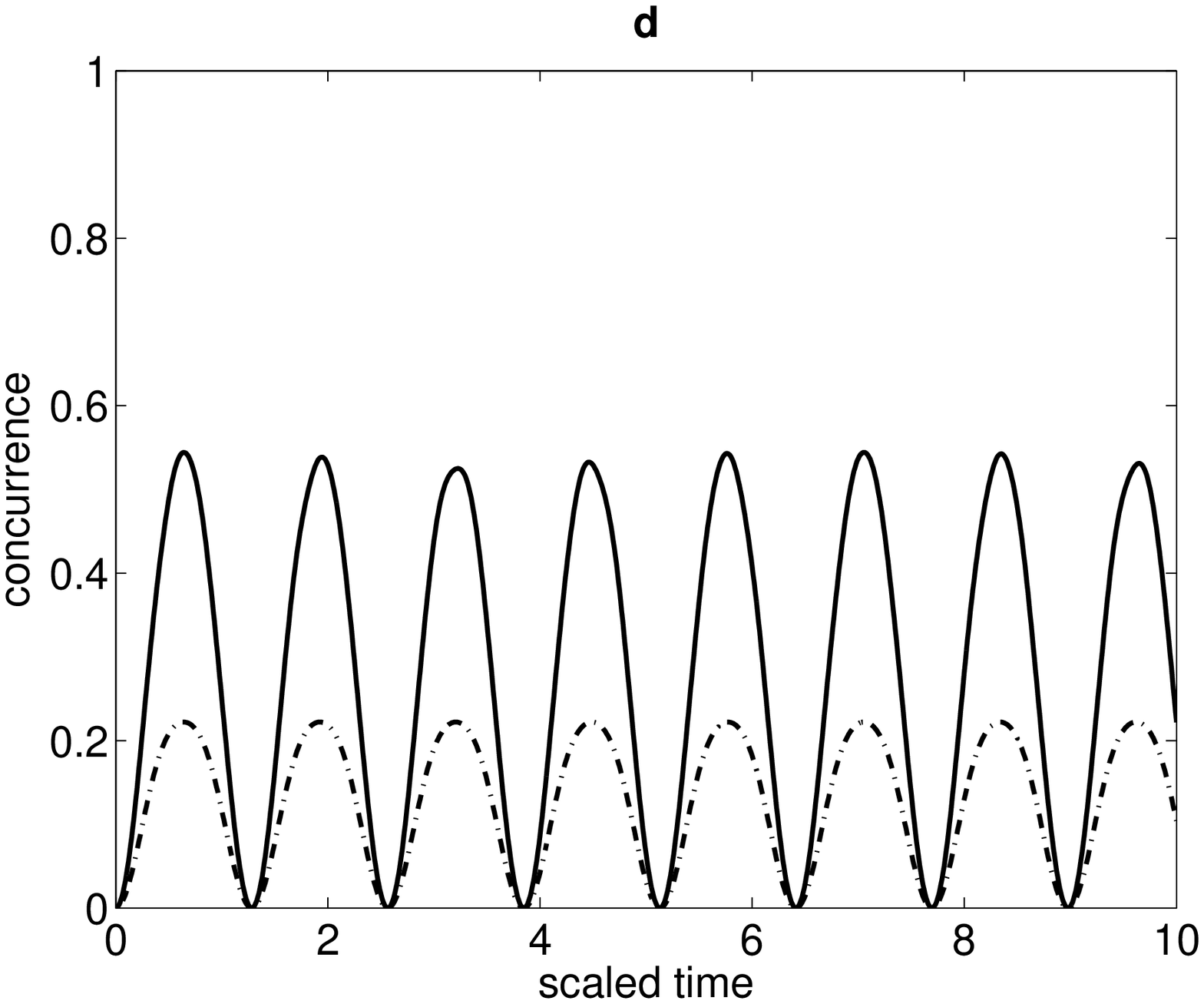}
\caption{Concurrence $C$ (solid curve) and $\rho_{22}+\rho_{33}$ (dotted curve) as functions of the scaled time $\kappa t$. The cavity field start from a Fock state with $n=0.0$ where $r=0.0$. (a) $\chi/\kappa=0.0$. (b) $\chi/\kappa=0.2$. (c) $%
\chi/\kappa=1.0$ (d)$\chi/\kappa=2.0$}
\label{fig:1}       
\end{center}
\end{figure}
%
%
%
As soon as we apply the nonlinear medium, an interesting result can be
observed. The splitting disentanglement point of the two peaks begins to
disappear gradually by increasing $\chi/\kappa$ till $\chi/\kappa$ reaches unity,
which shows one peak periodical behavior in approximately same time
intervals, Figs. 1b and 1c. Moreover, the same behavior is also observed for  $\rho_{22}+\rho_{33}$. In this case, the atomic system exhibits long
time intervals of entanglement with strong Kerr medium. This is because the fact that the higher values of the Kerr parameter allow a complete transmission of the
interaction field incident on the atomic system. A very strong Kerr medium
decreases the entanglement maxima, while the periodical behavior is
preserved (see Fig. 1d). The above results show that the two atoms
exhibit long time intervals of entanglement the application of Kerr medium
with matching Kerr parameter.
\\
An interesting case is the one when we assume that $%
\kappa_{2}\gg\kappa_{1}$ so that the coupling parameter ratio, $r=\frac{%
\kappa_{1}}{\kappa_{2}}<10^{-2}$, where the effect of one of the coupling
parameters is very weak (see Fig. 2a). \newline
We notice that $C$ as well as $\rho_{22}+\rho_{33}$ evolve identically with
fixed-amplitude periodical evolution. Moreover, $C$ shows two-peak
oscillatory behavior with maxima ($\simeq 0.8$) that are reached at the
maxima of $\rho_{22}+\rho_{33}$, Fig. 2a. Also, $C$ shows very small rapid
oscillations around its maxima before it collapses to its minimum (see Fig.
2a). This implies a longer time of strong entanglement between the two atoms.
When the Stark shift parameter $r$ increases, a considerable
decrease of the maxima of $C$ is found comparing with that of $%
\rho_{22}+\rho_{33}$, especially when $r=0.2$, similar to the effect of very
high Kerr parameter, Fig. 1d. \newline
Further interesting results are found when we take into account the effects of
both Kerr and Stark parameters. One peak periodical behavior with clearly
remarkable interval ( $\approx$ twice the case when no Kerr and stark) of
time with very small rapid oscillations around $C$ maxima (Fig. 1a, 2b). In
this case the two atoms are in the entangled state for longer periods of time than in the previous cases before collapse to disentangled state. 
\begin{figure}
\begin{center}
\includegraphics[width=5.5cm,height=4.5cm]{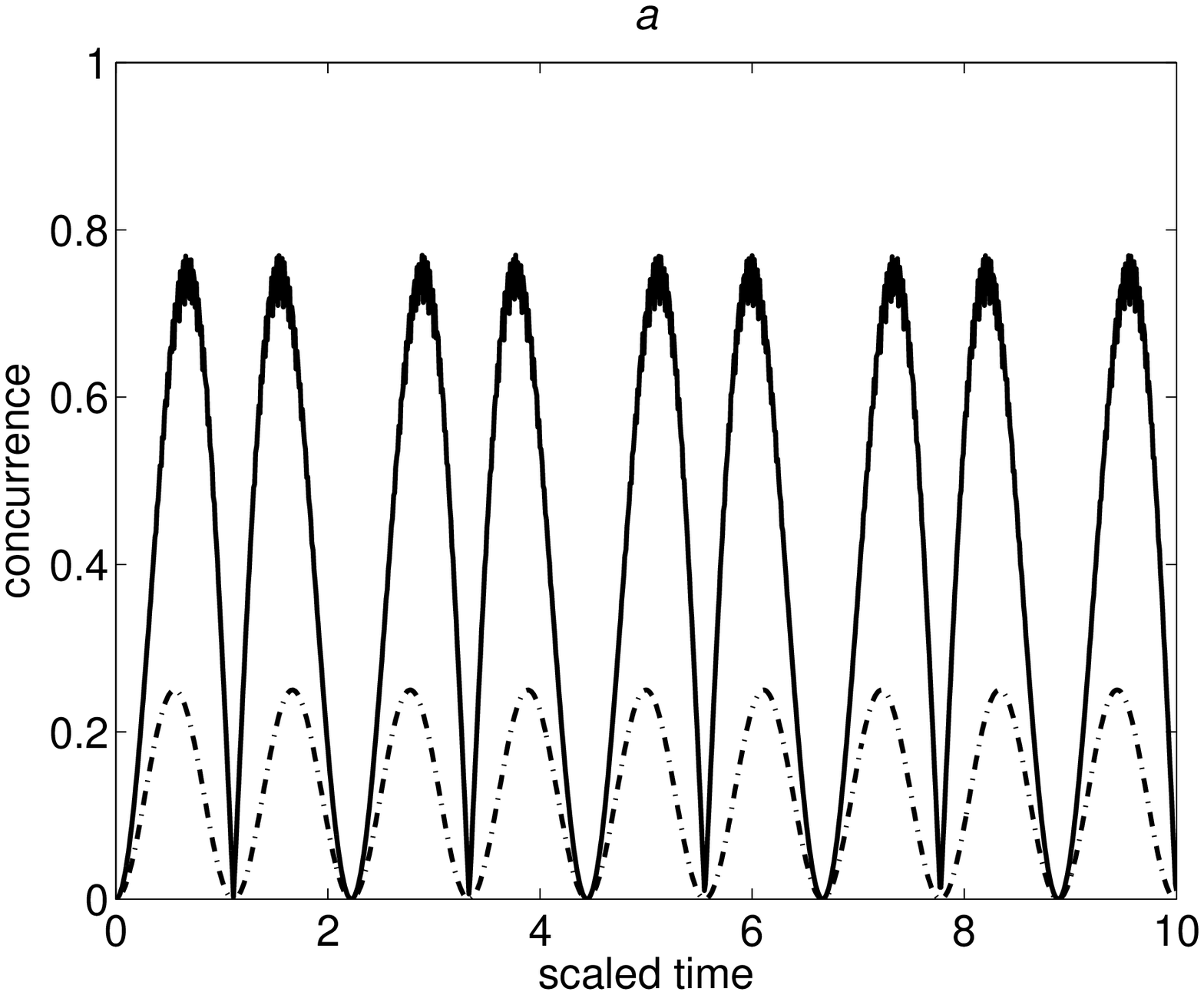}
\includegraphics[width=5.5cm,height=4.5cm]{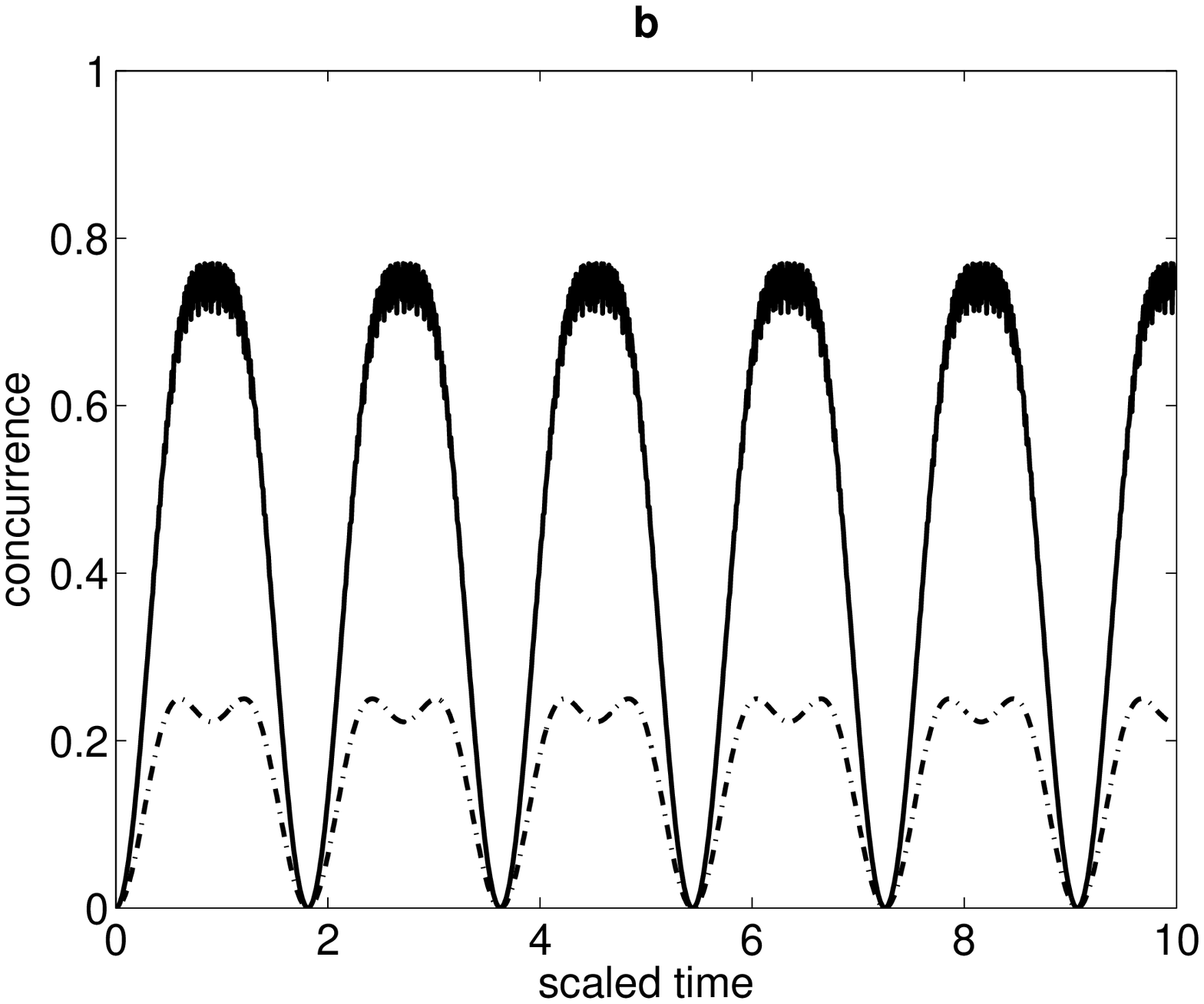}
\caption{The same as Fig. 1 but for (a) $\protect\chi/\kappa=0.0$, $r=0.001$. (b) $%
\protect\chi/\kappa=1.0$, $r=0.001$}
\label{fig:1}       
\end{center}
\end{figure}
When $n=2.0$, $C$ falls off rapidly except for some
revivals in irregular intervals with the minima of $C$ being reached at
the maxima of $\rho_{22}+\rho_{33}$ (Fig. 3a). Moreover, a weak stark
constant, $r=0.5$, reduces the revival intervals with a remarkable reduction of the $C
$ maxima, Fig. 3b. \newline
A surprising result is found when the excitation number $n=2.0$, while $%
\frac{\kappa_{1}}{\kappa_{2}}<10^{-2}$. The two atoms remain in a pure state
forever although the populations sum oscillates periodically with time
with maxima equal to unity. In this case, due to the strong tendency of the
ground atom to become excited, its interaction with the field dominates, while
neither interaction between field and the excited atom (since no role of its
level shifts) nor the two atoms themselves. In this case, the atomic states oscillates between $|1_{1},0_{2}\rangle$ and $|1_{1},1_{2}\rangle$,
i. e., both probabilities $\rho_{11}$ and $\rho_{22}$ contribute to $C$,
while $\rho_{33}$, $\rho_{44}$ and $Im[\rho_{14}]$ are always zero, which implies $%
C=0$ in all times (Fig. 4a).
\begin{figure}
\begin{center}
\includegraphics[width=5.5cm,height=4.5cm]{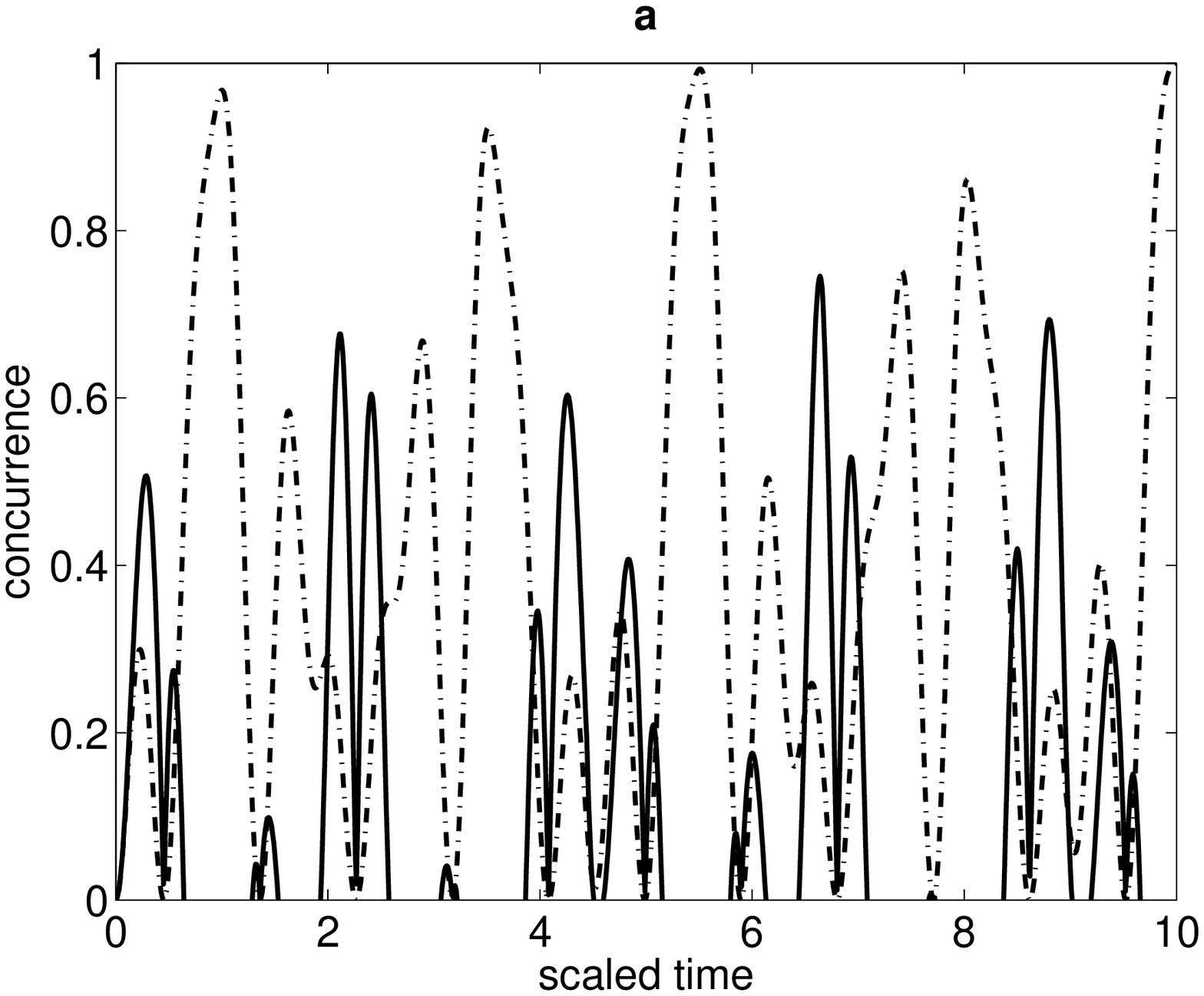}
\includegraphics[width=5.5cm,height=4.5cm]{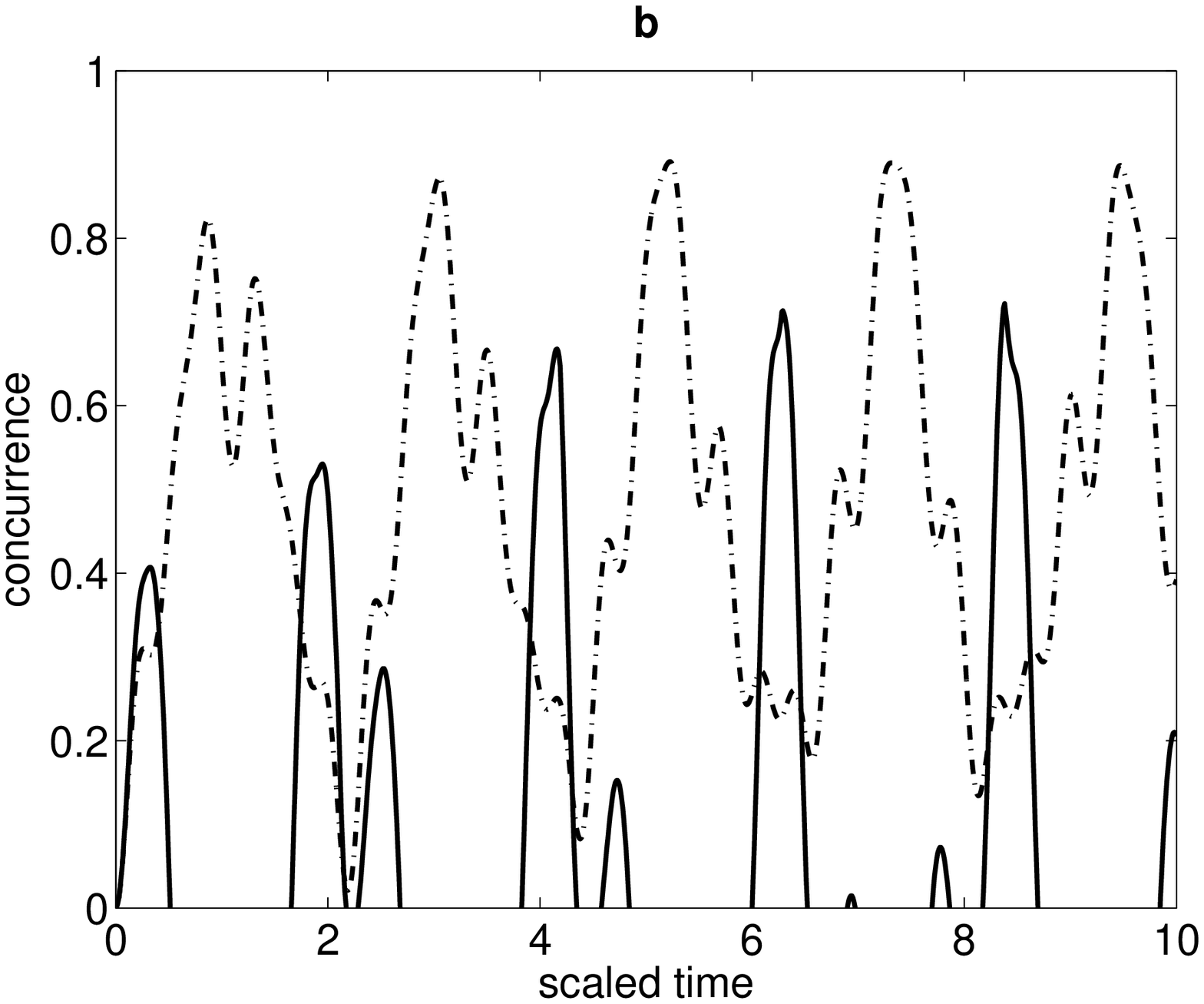}
\caption{The same as Fig. 1 but when $n=2.0$, (a) $\protect\chi/\kappa=0.0$. (b) $%
\protect\chi/\kappa=0.5$.}
\label{fig:1}       
\end{center}
\end{figure}
\begin{figure}
\begin{center}
\includegraphics[width=5.5cm,height=4.5cm]{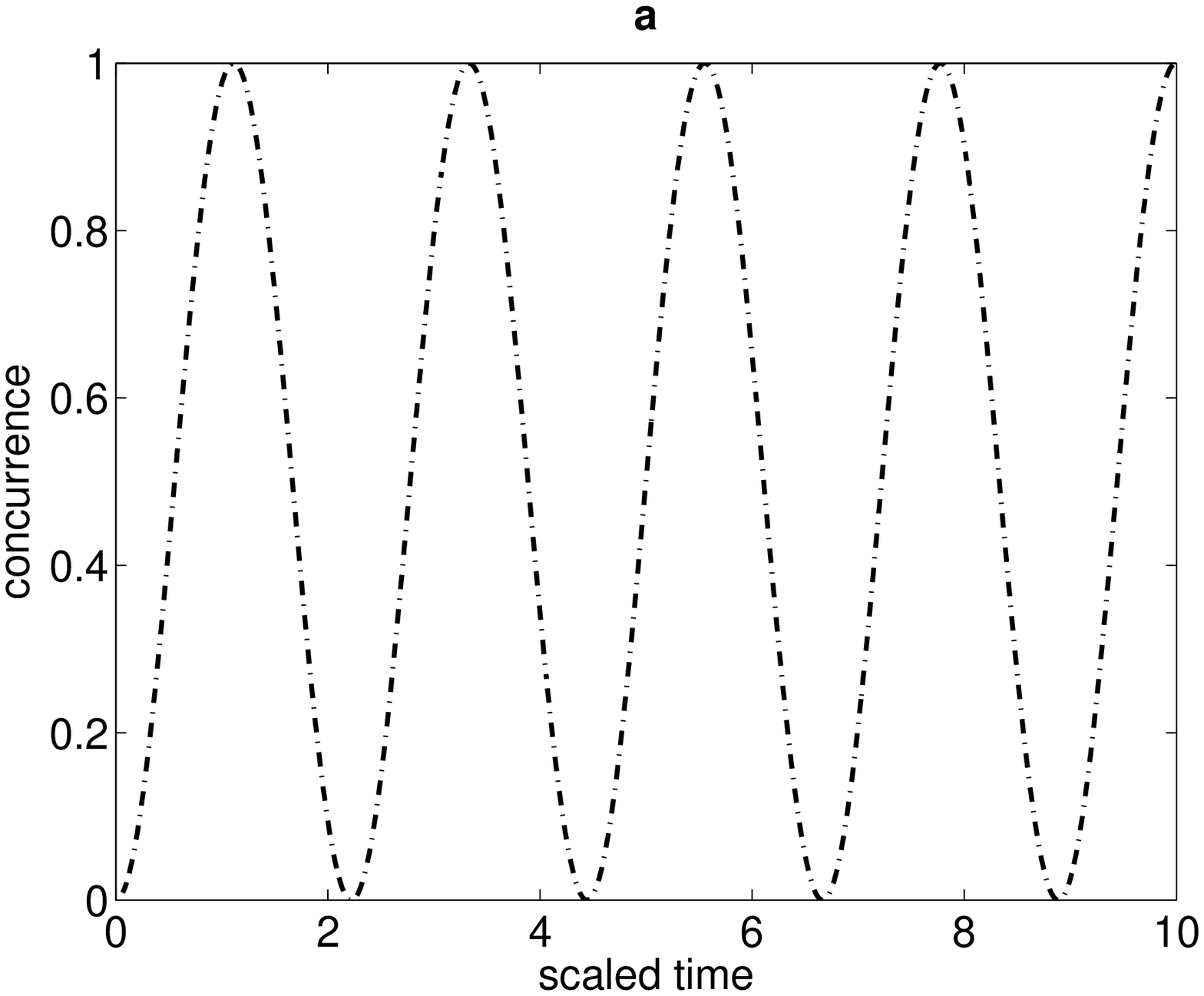}
\includegraphics[width=5.5cm,height=4.5cm]{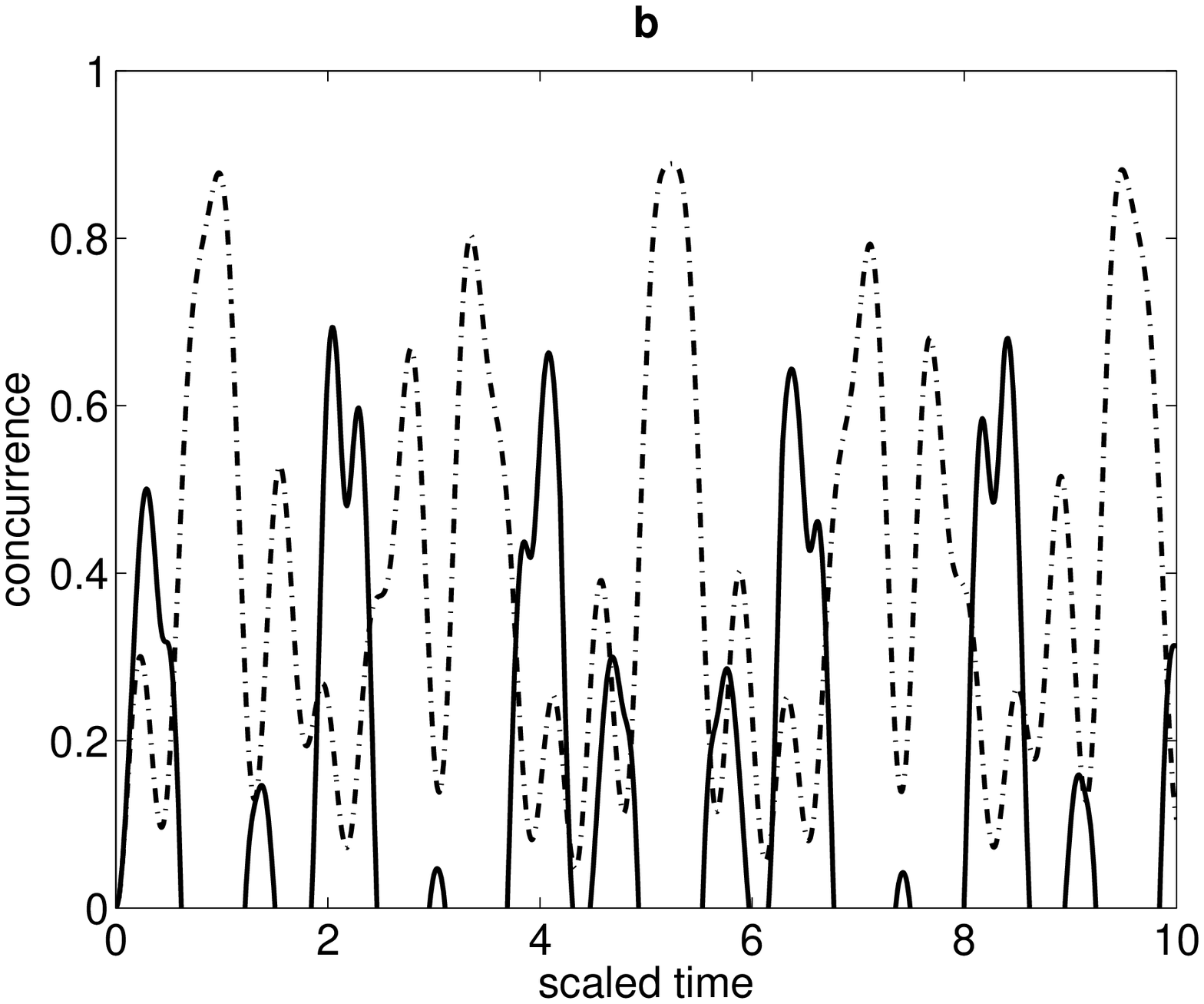}
\includegraphics[width=5.5cm,height=4.5cm]{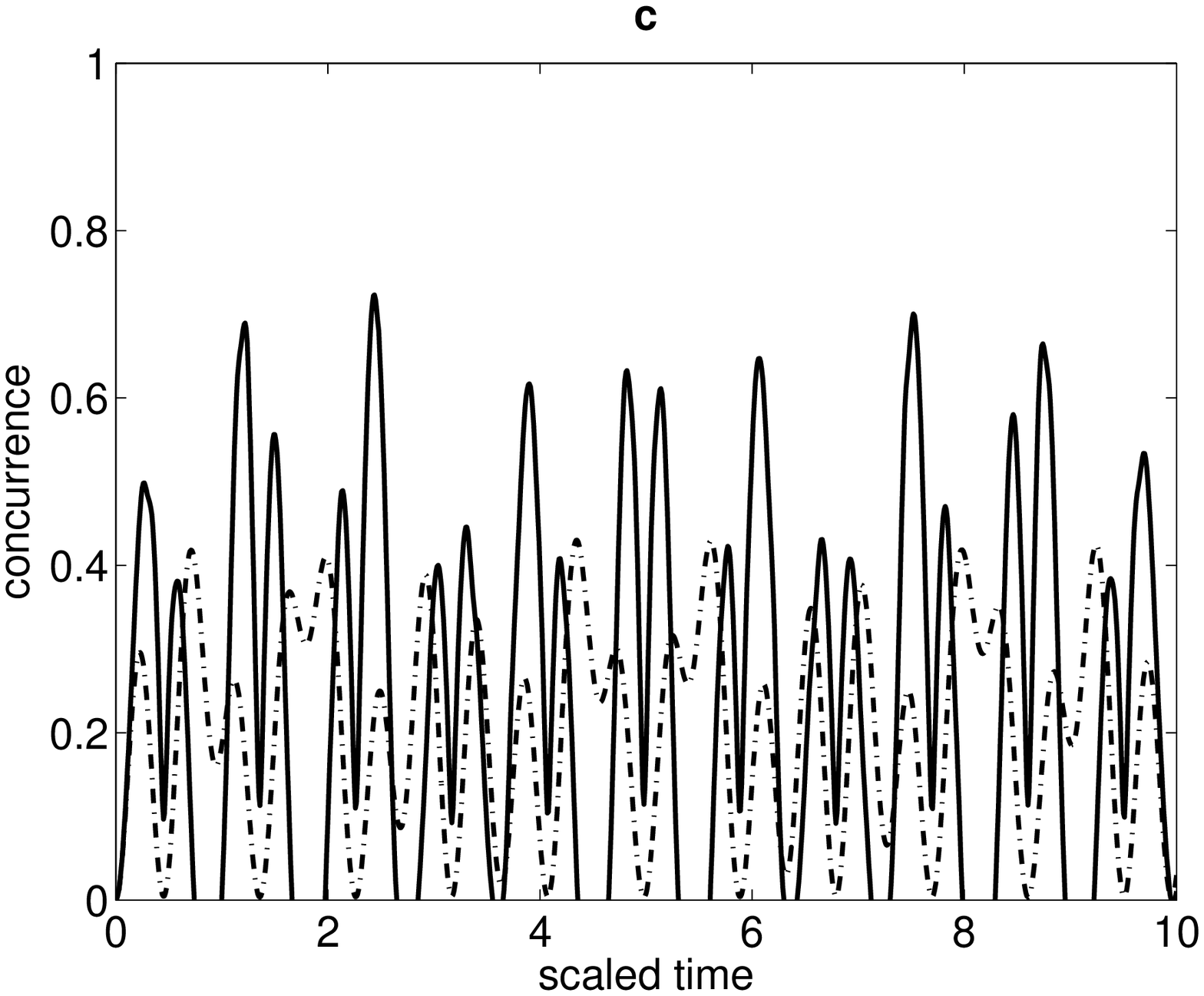}
\caption{Concurrence $C$ (solid curve) and $\rho_{22}+\rho_{33}$ (dotted curve) as functions of the scaled time $\kappa t$. The cavity field start from a Fock state with  $n=2.0$ where $\protect\chi/\kappa=0.0$. (a) $r=0.001$. (b) $r=0.5$. (c) $\protect%
\chi=2.0$, $r=0.1$}
\label{fig:1}       
\end{center}
\end{figure}
So we do not need a full population to get strong entanglement between the
two atoms as illustrated in figure 1a which shows the opposite behaviors of  $\rho_{22}+\rho_{33}$ and $C$, i. e., the smaller the
the populations sum, the higher the concurrence and stronger entanglement
between the atoms. With increasing $r$, a kind of entanglement between
the two atoms is created due to the role of the level shifts of both atoms.
It is evident that the strongest degree of entanglement occurs when $%
\rho_{22}+\rho_{33}$ reaches its minimum. Moreover, the amplitudes of both $C
$ and $\rho_{22}+\rho_{33}$ decrease considerably as $|r|> 1$. A case with
high Kerr parameter, $\chi/\kappa=2.0$, with values of $r=0.1$ is considered in figure 4c, which shows the tendency of both atoms to be entangled
more quickly with the maxima ($\simeq 0.75$) being reached when $%
\rho_{22}+\rho_{33}=0.25$ and any value of $\rho_{22}+\rho_{33}$ less or
greater than this value means low degree of entanglement between the two
atoms.
\subsubsection{Two excited atoms}
\label{sec:3}
By setting $F_{n}=\delta_{m,n}$ and $a=1$, we obtain the wave
function of the system with field excited initially in a Fock state and with initially the two atoms are excited.%
\newline
In this case, the atom-atom system can be described in the basis %
$$
|1_{1},1_{2}\rangle=|1\rangle,\eqno(46a)
$$
$$
|1_{1},0_{2} \rangle=|2\rangle,\eqno(46b)
$$
$$
|0_{1},1_{2}\rangle=|3\rangle,\eqno(46c)
$$
$$
|0_{1},0_{2}\rangle=|4\rangle,\eqno(46d)
$$
where the reduced density operator of the two atoms can be written in this
basis as
$$
\rho_{a-a}(t)=\rho_{11}(t)|1\rangle\langle
1|+\rho_{22}(t)|2\rangle\langle 2| +\rho_{33}(t)|3\rangle\langle 3|
$$
$$
+\rho_{23}(t)|2\rangle\langle 3|+\rho_{32}(t)|3\rangle\langle
2|+\rho_{44}(t)|4\rangle\langle 4|,\eqno(47)
$$
$$
\rho_{11}(t)=|H_{n}(t)|^{2},\eqno(48)
$$
$$
\rho_{22}(t)=|T_{n+2}(t)|^{2},\eqno(49)
$$
$$
\rho_{23}(t)=e^{i\kappa[2\frac{\chi}{\kappa} (2n+3)+\frac{r^{2}+1}{r}]t}~T_{n+2}(t)
J^{\ast}_{n+2}(t)=\rho^{\ast}_{32}(t),\eqno(50)
$$
$$
\rho_{33}(t)=|J_{n+2}(t)|^{2},\eqno(51)
$$
$$
\rho_{44}(t)=|V_{n+4}(t)|^{2}.\eqno(52)
$$
One may write $\rho_{a-a}(t)$ in the form:
$$
\mathbf{\rho_{a-a}(t)} = \left(
\begin{array}{cccc}
\rho_{11} & 0 & 0 & 0 \\
0 & \rho_{22} & \rho_{23} & 0 \\
0 & \rho_{32} & \rho_{33} & 0 \\
0 & 0 & 0 & \rho_{44}%
\end{array}
\right),\eqno(53)
$$
while the spin-flip reduced density operator $\tilde{\rho}$ can be obtained by applying Eq.(30)
$$
\mathbf{\tilde{\rho}_{a-a}(t)} = \left(
\begin{array}{cccc}
\rho_{44} & 0 & 0 & 0 \\
0 & \rho_{33} & \rho_{32} & 0 \\
0 & \rho_{23} & \rho_{22} & 0 \\
0 & 0 & 0 & \rho_{11}%
\end{array}
\right),\eqno(54)
$$
and the square roots of the eigenvalues of the matrix $R$, given by
Eq.(29), are the following
$$
\{\sqrt{\lambda_{i}}\}=\biggl\{Re(\rho_{23})+\sqrt{\rho_{22}\rho_{33}-
[Im(\rho_{23})]^{2}},
Re(\rho_{23})-\sqrt{\rho_{22}\rho_{33}-[Im(\rho_{23})]^{2}},
$$
$$
~\sqrt{\rho_{11}\rho_{44}},~\sqrt{\rho_{11}\rho_{44}}\biggr\}.\eqno(55)
$$
By using of Eq.(32) to obtain the largest eigenvalue, the concurrence $%
C(\rho)$ takes the from
\[
C(\rho)=max(0, \sqrt{\lambda_{1}}-\sqrt{\lambda_{2}}-\sqrt{\lambda_{3}}-%
\sqrt{\lambda_{4}})
\]
$$
=2\biggl(\sqrt{\rho_{22}\rho_{33}-[Im(\rho_{23})]^{2}} -\sqrt{%
\rho_{11}\rho_{44}}\biggr),\eqno(56)
$$
\begin{figure}
\begin{center}
\includegraphics[width=5.5cm,height=4.5cm]{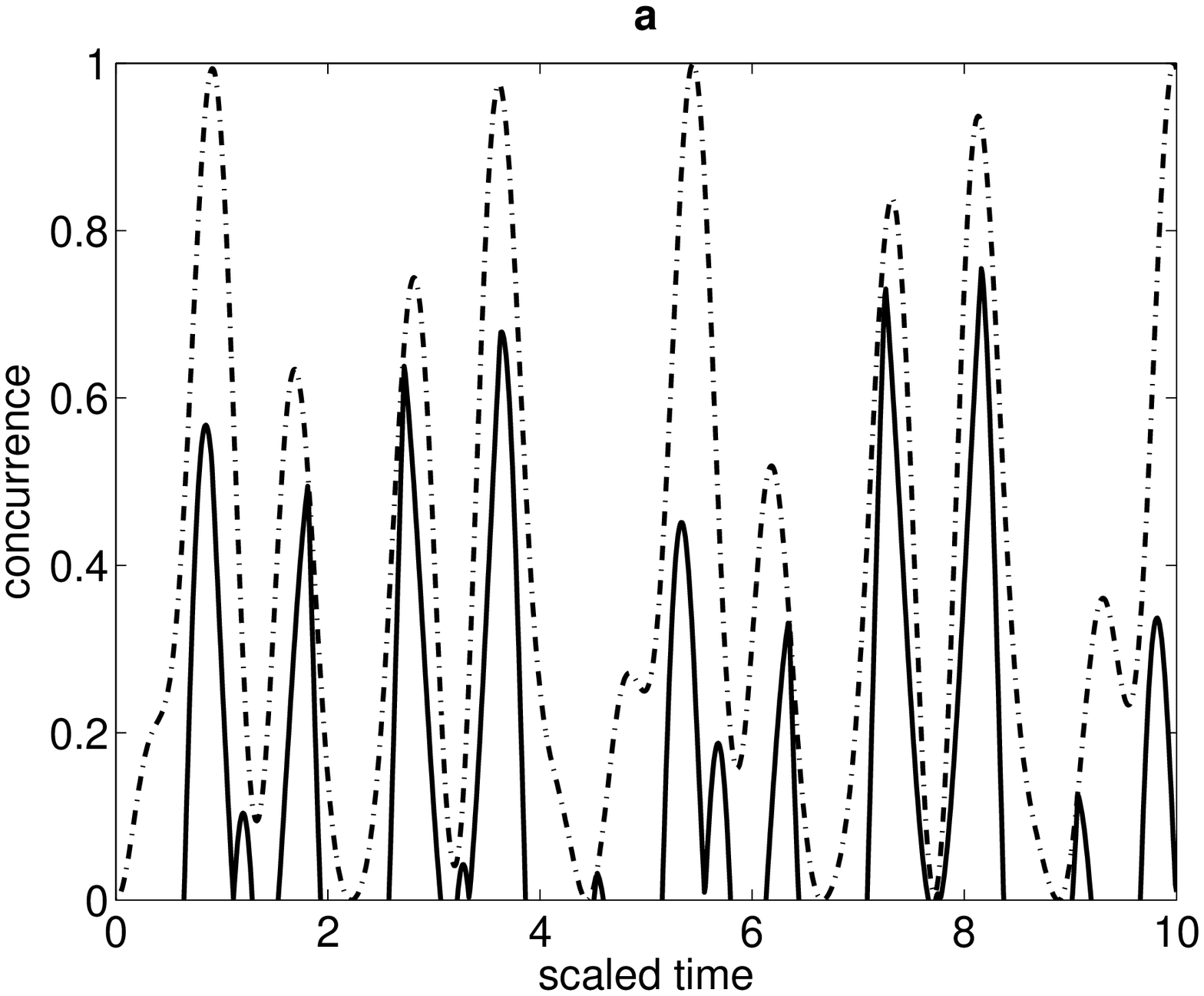}
\includegraphics[width=5.5cm,height=4.5cm]{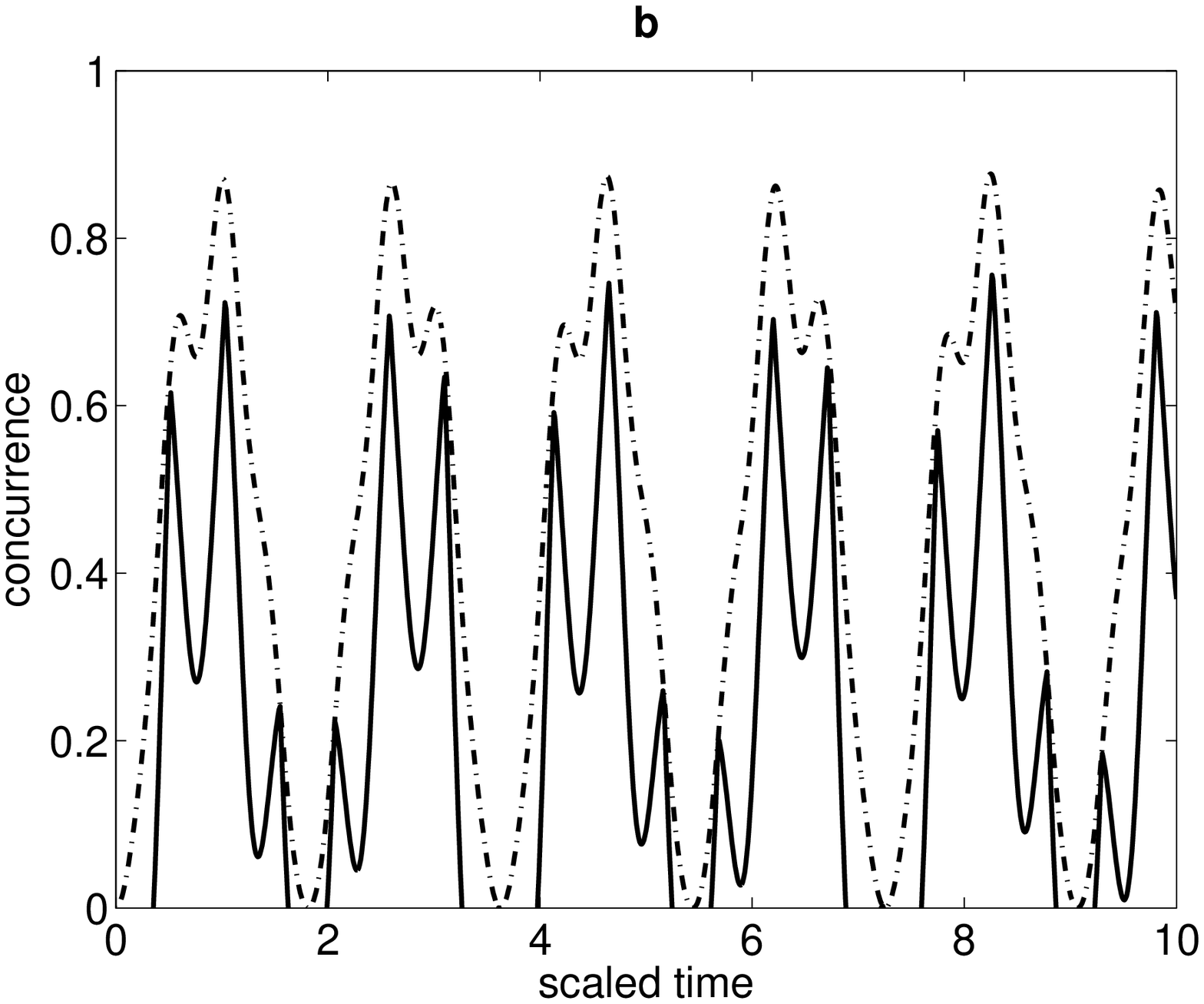}
\caption{Concurrence $C$ (solid curve) and $\rho_{22}+\rho_{33}$ (dotted curve) as functions of the scaled time $\kappa t$. The cavity field start from a Fock state with $n=0.0$ where $r=0.0$. (a) $\chi/\kappa=0.0$. (b) $\chi/\kappa=1.0$.}
\label{fig:1}       
\end{center}
\end{figure}
To obtain a clear understanding of the situation we examine the concurrence and
population dynamics when both successive atoms enter the cavity initially in
an excited state. In vacuum, one can clearly notice
the strong positive effect of the nonlinear medium on the degree of
entanglement of the atomic system. When the Kerr parameter is absolutely
zero, the maximum degree of entanglement ($\simeq 0.75$ ) is reached
near the end of the scaled time (at $t\simeq 8/\kappa$) where the two atoms in some kind of opposite states. Moreover, by increasing the value of the Kerr parameter, $\chi/\kappa=0.5$, the concurrence begins with maximum $C\simeq 0.72$ is reached in the near the begin of time scale with wider intervals of entanglement of the atomic system (Figs. 5a,b).
\newline
On taking into consideration the Stark shift, the obtained results are illustrated in Fig. 6. We can notice clearly the similar behavior as in section (4.1.1) except for that $C$ reaches its minimum where $\rho_{22}+\rho_{33}=1.0$ , and when $r=0.2$, little small
shift parameter, a quasi-periodical behavior with wider temporal intervals of
atomic entanglement are showed due to the Stark shift in opposite to its
effect in corresponding case. 
\begin{figure}
\begin{center}
\includegraphics[width=5.5cm,height=4.5cm]{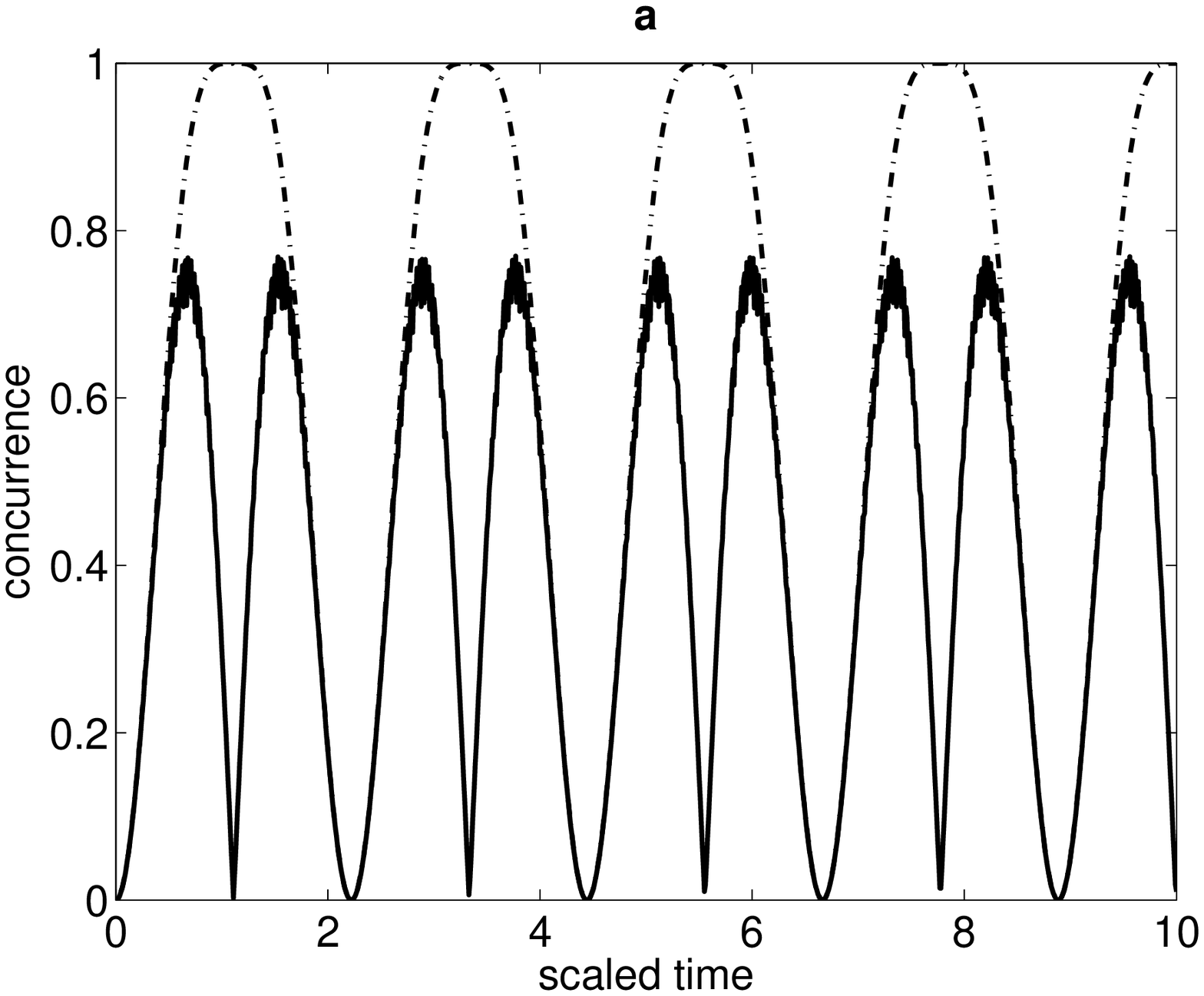} %
\includegraphics[width=5.5cm,height=4.5cm]{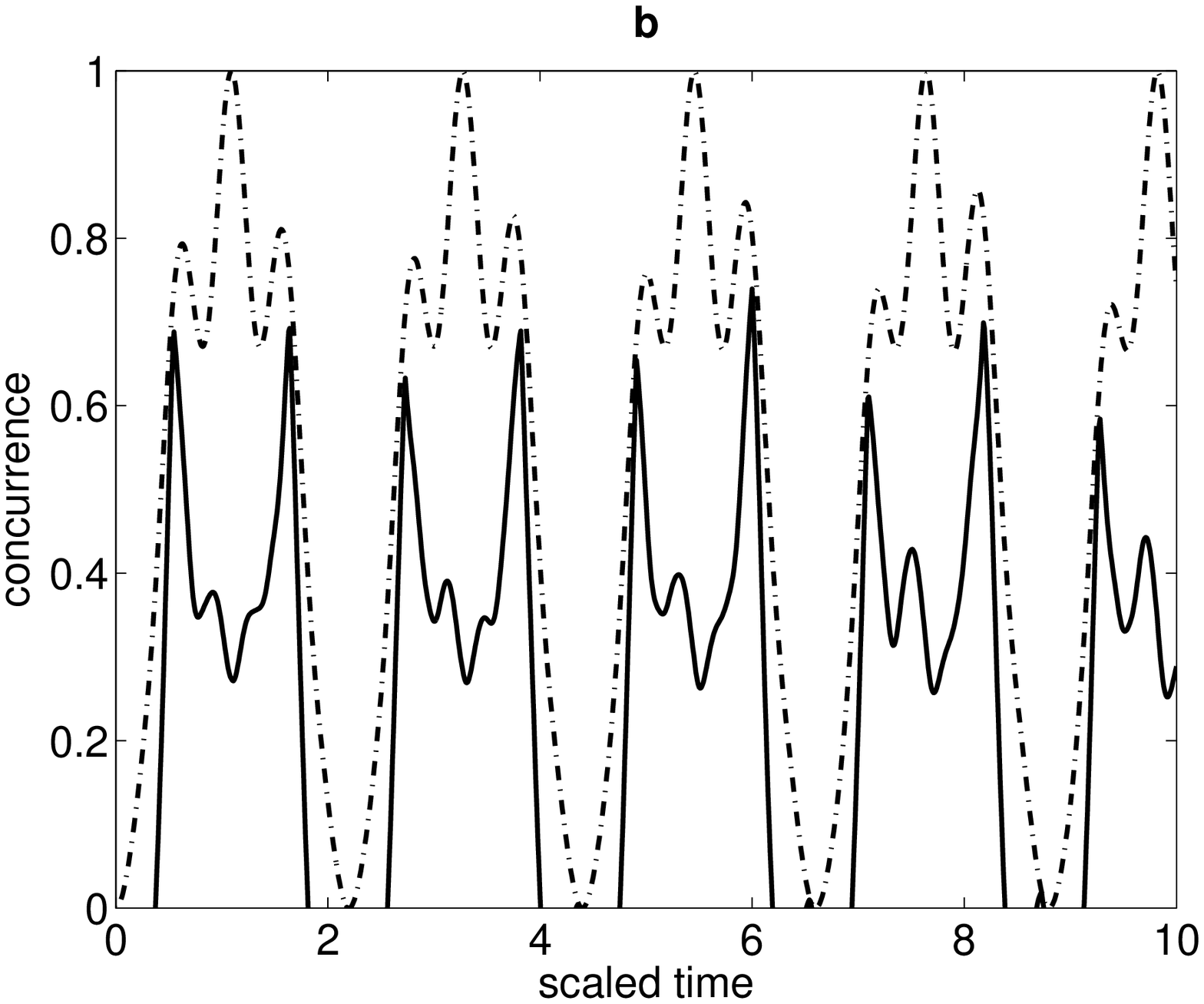} %
\includegraphics[width=5.5cm,height=4.5cm]{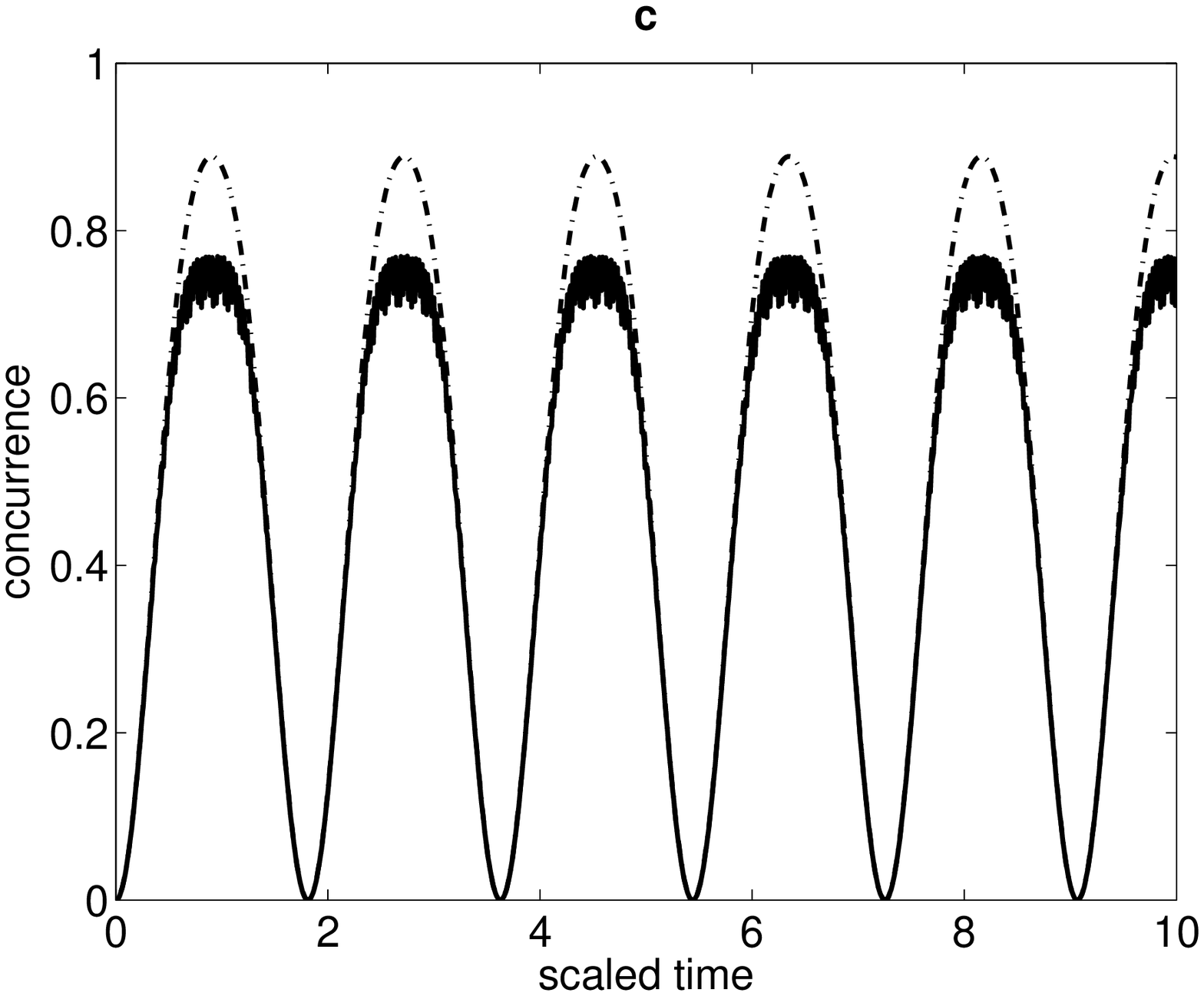}

\caption{The same as Fig. 5, but for $\protect\chi/\kappa=0.0$. (a) $r=0.001$. (b) $%
r=0.2$. (c) $\protect\chi/\kappa=1.0$, $r=0.001$}
\label{fig:1}       
\end{center}
\end{figure}
\begin{figure}
\begin{center}
\includegraphics[width=5.5cm,height=4.5cm]{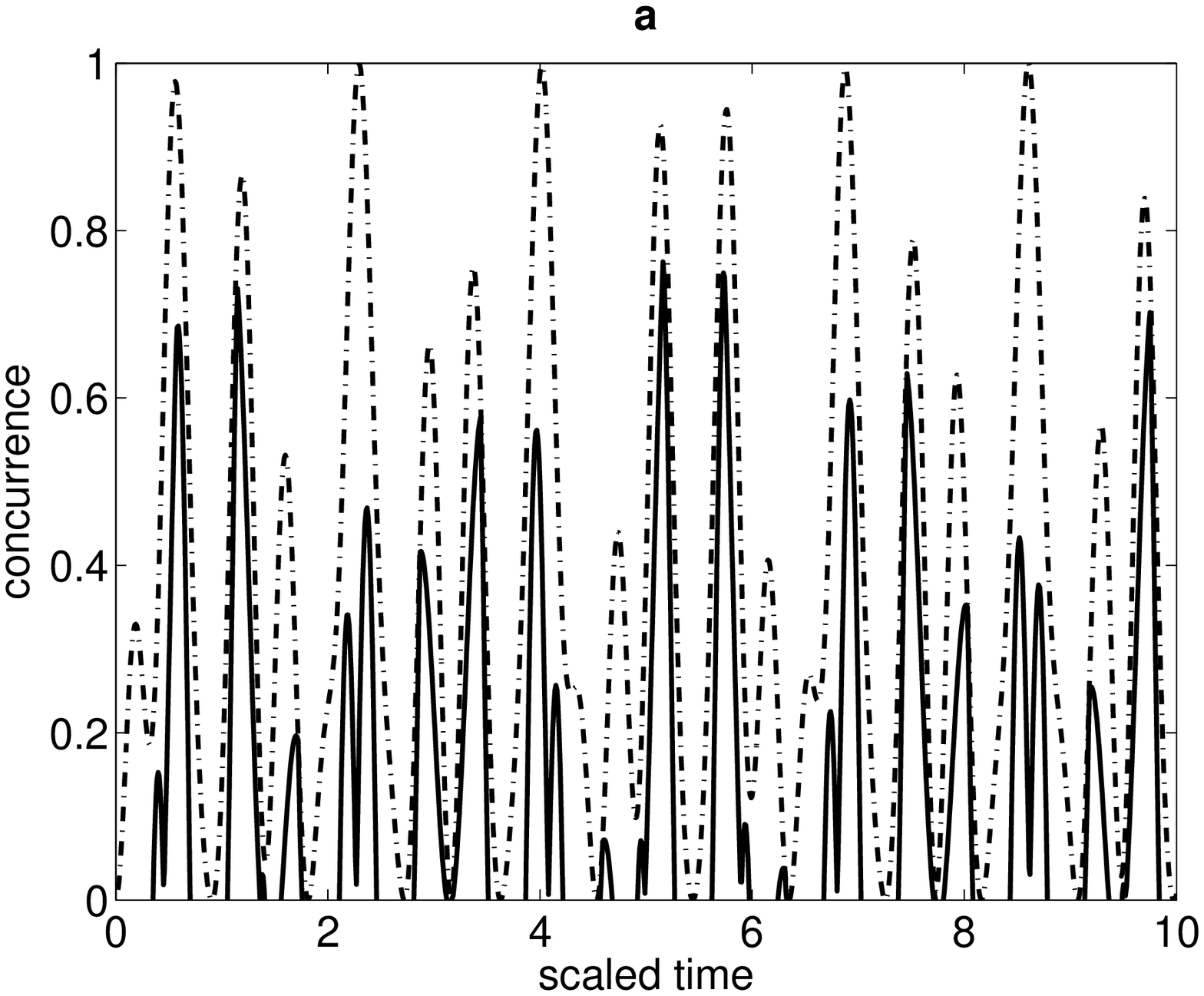} %
\includegraphics[width=5.5cm,height=4.5cm]{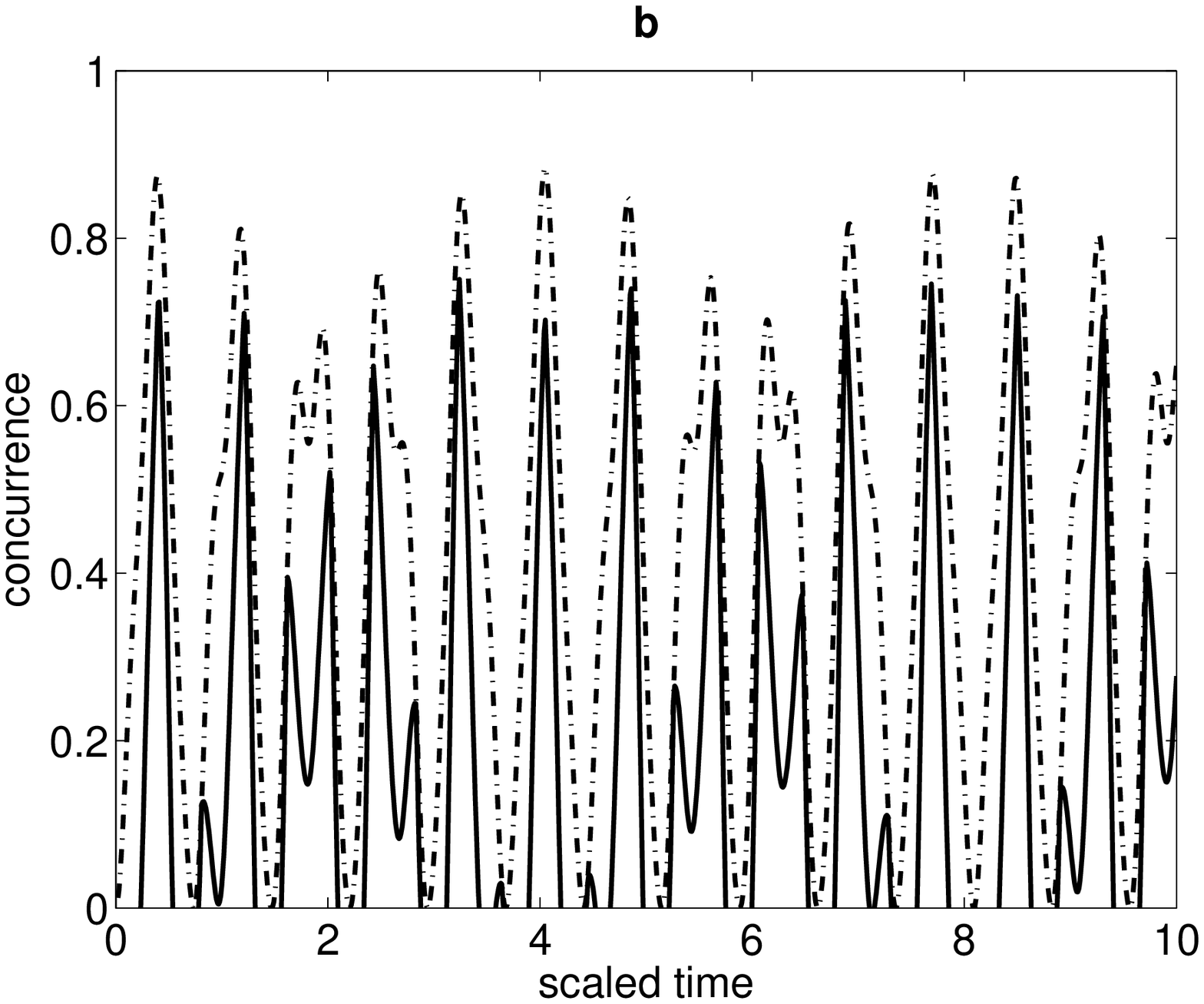} %
\caption{The same as Fig. 5, but for $n=2.0$ and $\protect\chi/\kappa=0.5$ for (b).}
\label{fig:1}       
\end{center}
\end{figure}
\begin{figure}
\begin{center}
\includegraphics[width=5.5cm,height=4.5cm]{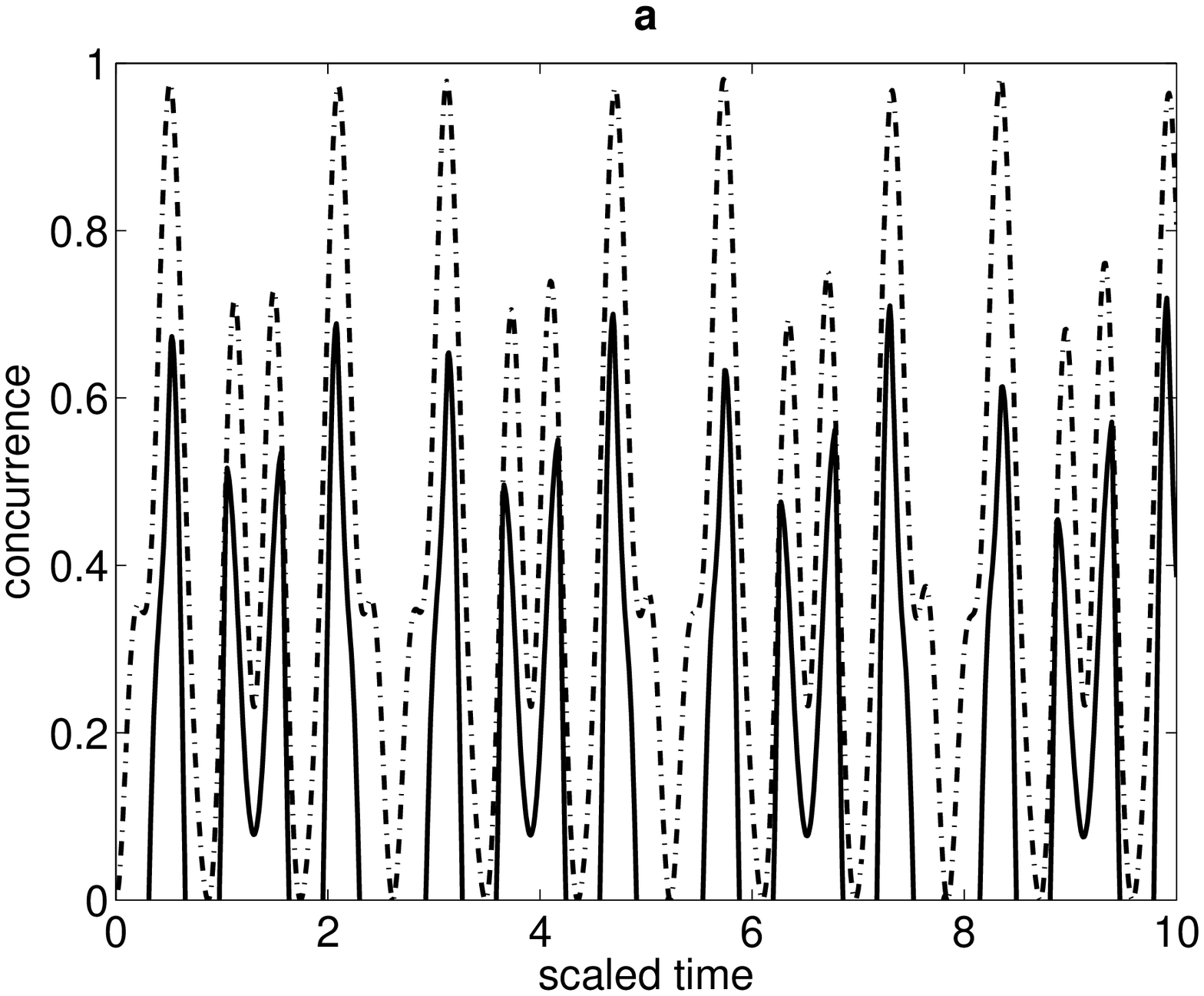} %
\includegraphics[width=5.5cm,height=4.5cm]{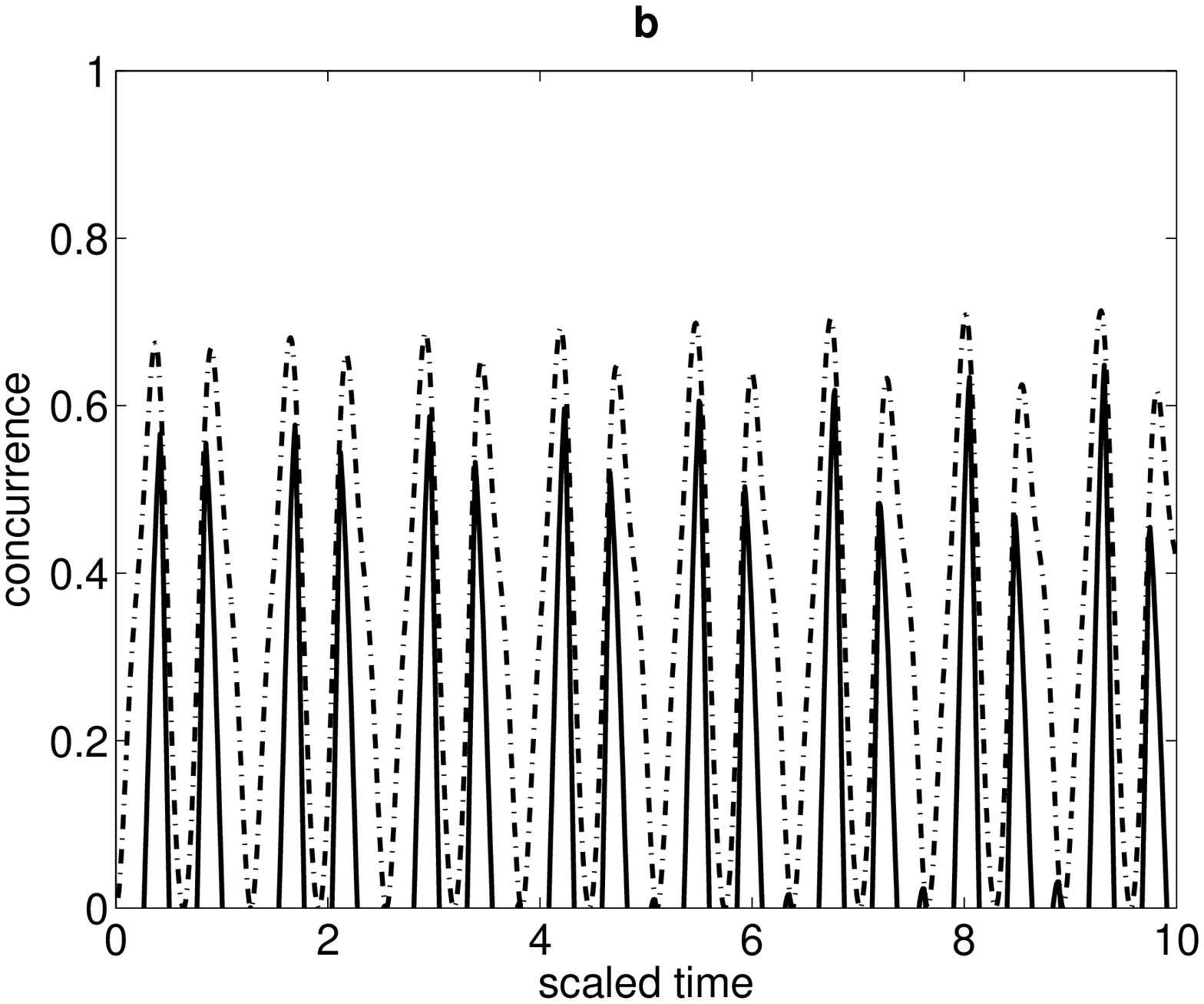} %
\includegraphics[width=5.5cm,height=4.5cm]{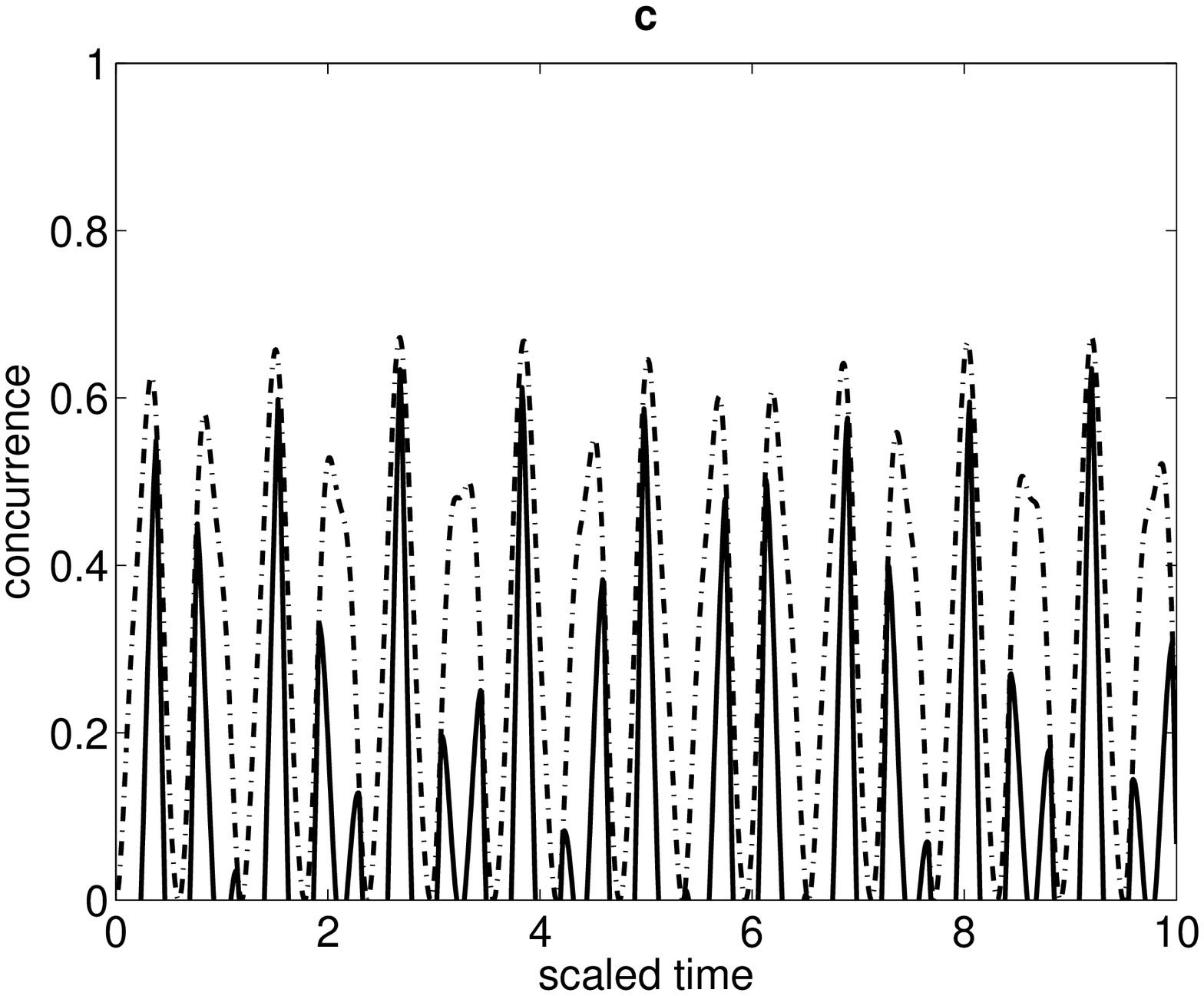}
\caption{Concurrence $C$ (solid curve) and $\rho_{22}+\rho_{33}$ (dotted curve) as functions of the scaled time $\kappa t$. The cavity field start from a Fock state with  $n=2.0$ where $\chi/\kappa=0.0$. (a) $%
r=0.5$. (b) $r=2.0$ (c) $\chi/\kappa=0.1$, $r=2.0$.}
\label{fig:1}       
\end{center}
\end{figure}
When the cavity is excited with number state $n=2$, and $\chi/\kappa=0.0$, $\chi/\kappa=0.5$, we notice the strong positive effect of the excitation number on the entanglement of the atomic system. We can clearly notice more oscillations of $C$ in the same time intervals, see Fig. 7. This implies that the Kerr medium acts as factor of enhancement of the entanglement between the two atoms in opposite to the same situation of one cavity Fock state~\cite{Ateto07}. Moreover, the maxima of $C$ depend crucially on the maxima of $\rho_{22}+\rho_{33}$. \newline
More surprising is the case when $\frac{\kappa_{1}}{\kappa_{2}}<10^{-2}$, in
this case, $\rho_{11}\simeq 1.0$, while $\rho_{22}$, $\rho_{33}$, $\rho_{44}$
and $Im[\rho_{23}]$ are always zero, which implies that the two atoms remain in
their initial excited states and the cavity field plays no role and entanglement of the two atoms is not observed. As the atomic system has a level
shift, the atomic system shows entanglement whose maxima are reached at
the maxima of $\rho_{22}$+$\rho_{33}$. The entanglement amplitue decreases as the Stark shift parameter increases and as possible as the Kerr parameter is still small, see Fig. 8.\newline
Opppsite to the case of only one excited atom, when $n>0$, the number of
photons in the cavity destroys the entanglement between the two atoms in case
of small Stark shift. As the level shift between the two atomic levels
increases, more intervals of entanglement between the atoms are created
associated with increasing of the maxima of $C$ which occur at the maxima of
$\rho_{22}+\rho_{33}$ till $r=2.0$ by which a periodical evolution of $C$
appears.
\subsection{case 2. Excition in a Thermal state}
\label{sec:4}
The thermal field is the most easily available radiation field. At thermal
equilibrium, the field has an average photon number given by:
$$
\bar{n}=(e^{\hbar \omega/kT}-1)^{-1},\eqno(57)
$$
with Boltzmann constant $k$ and absolute temperature $T$. The photon distribution $p(n)$ is given by
$$
p(n)=\frac{\bar{n}^{n}}{(1+\bar{n})^{n+1}},\eqno(58)
$$
which has a peak at zero, i.e., $n_{peak}=0$.
\subsubsection{Only one excited atom}
\label{sec:4}
Setting $F_{n}=F_{n}\delta_{n,n}$ and $a=0$ the wave
function that governs the system in a thermal state, with initially excited atom followed by the one in the ground stste, can be obtained.
With the condition that $p(n)=|F_{n}|^{2}$ is the photon distribution
function of the thermal cavity given by Eq. (58), the reduced density operator of the  atom-field system after taking the trace over the field variables has the
form of Eq. (36) with the coefficients given by:
$$
\rho_{11}(t)=\sum_{n} p(n)|W_{n}(t)|^{2},\eqno(59)
$$
$$
\rho_{14}(t)= \sum_{n} p(n) e^{-i\kappa[2\frac{\chi}{\kappa} (2n-1)+\frac{r^{2}+1}{2r}]t}~W_{n}(t)
Z^{\ast}_{n}(t)=\rho^{\ast}_{41}(t),\eqno(60)
$$
$$
\rho_{22}(t)=\sum_{n} p(n) |X_{n-2}(t)|^{2},\eqno(61)
$$
$$
\rho_{33}(t)=\sum_{n} p(n) |Y_{n+2}(t)|^{2},\eqno(62)
$$
$$
\rho_{44}(t)=\sum_{n} p(n) |Z_{n}(t)|^{2}.\eqno(63)
$$
With these elements, following the same procedure, one can easily compute the concurrence $C(\rho)$ given by Eq. (32).\newline
In the following we compare the results obtained when the cavity field is excited in the 
thermal field with various mean photon numbers. The results are
depicted in Figs. 9, 10, 11 and 12.\newline
A small average photon number, $\bar n=0.5$ creates a high degree of
entanglement of chaotic behavior of the atomic system with many maxima
of the highest value ($\simeq 0.88$) is reached when $\rho_{22}+%
\rho_{33}=0.0$. Moreover, the atomic system remains entangled forever, Fig.
9a. A similar effect can be noticed by increasing the average photon number,
$\bar n=2.0$, with higher degree of entanglement $C\simeq 0.93$ accompanied
by increase of its minima, Fig. 11a. %
\begin{figure}
\begin{center}
\includegraphics[width=5.5cm,height=4.5cm]{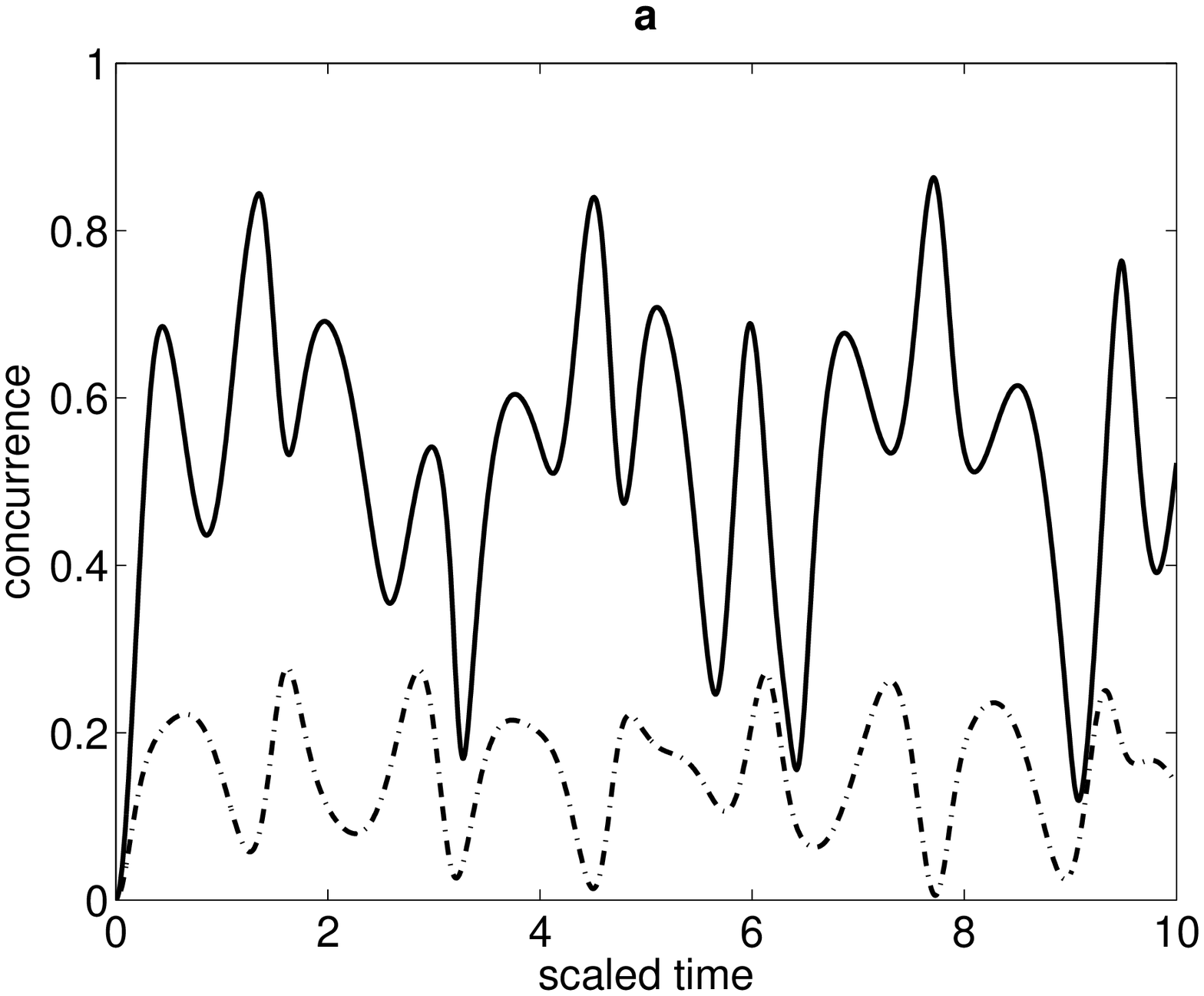} %
\includegraphics[width=5.5cm,height=4.5cm]{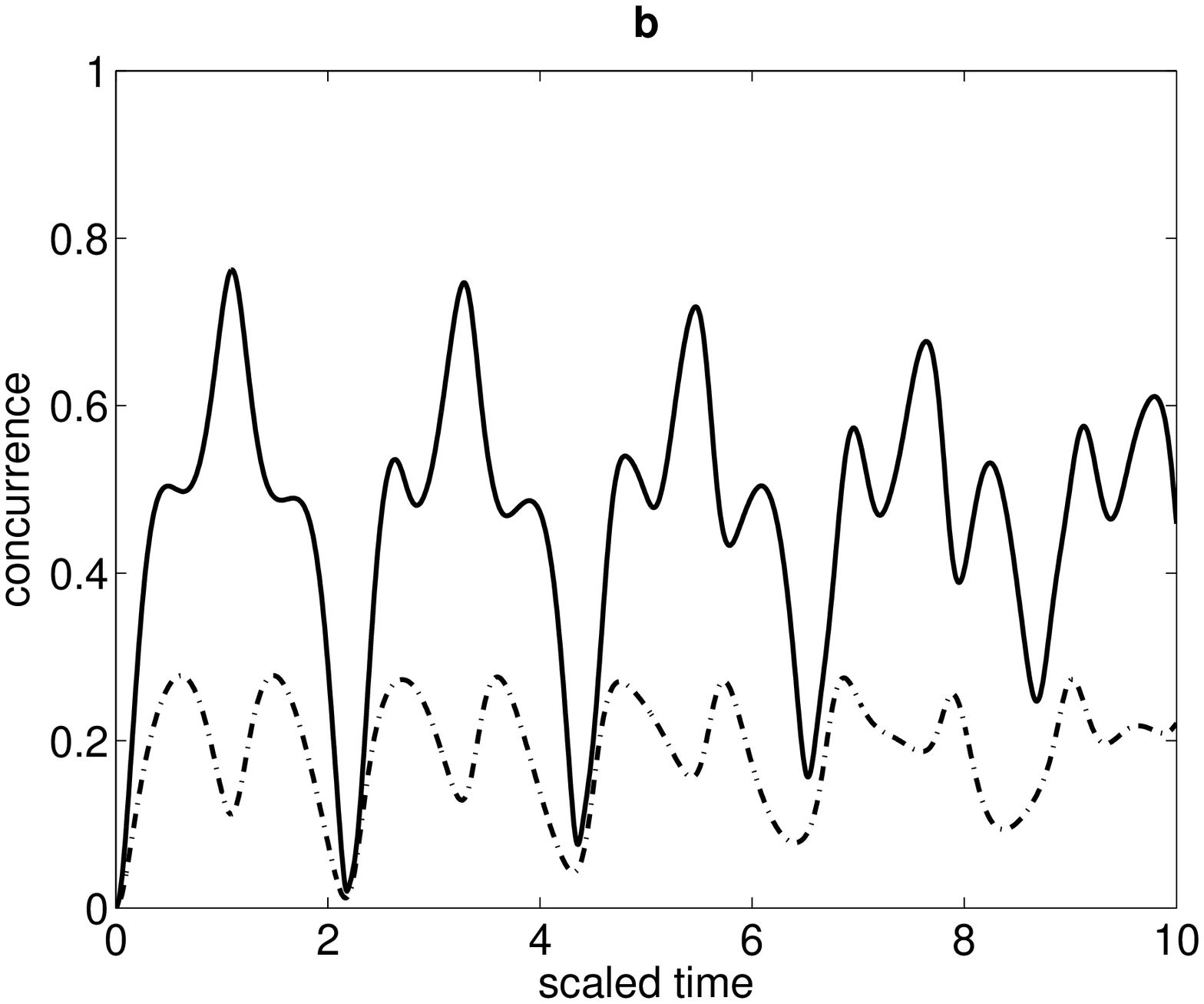}
\caption{Concurrence $C$ (solid curve) and $\rho_{22}+\rho_{33}$ (dotted curve) as functions of the scaled time $\kappa t$. The cavity field start from a thermal state with average photon number $\bar{n}=0.5$ where $r=0.0$. (a) $\chi/\kappa%
=0.0$. (b) $\chi/\kappa=0.5$.}
\label{fig:1}       
\end{center}
\end{figure}
Choosing a Kerr parameter of value $\chi/\kappa=0.5$, affects the general behavior of $C$ negatively, where $C$ goes to zero after one period of $\rho_{22}+\rho_{33}$%
, while its maxima are remarkably reduced. However, increasing $\bar n$
decreases the maxima of $C$ remarkably while the general behavior is preserved, see
Fig. 11b.\newline
A Stark shift with parameter $r<10^{-2}$ creates periodical entanglement with maxima ($%
\simeq 0.82$) with period $t=0.7 n \pi/\kappa$, $n=0,1,2,...$
collapse to minima slower than when $\bar n=2.0$, see Figs.
10a,12a.\newline
Increasing the shift parameter, $r=0.1$, reduces the maxima of $C$, while a quasi-periodical behavior can be noticed while similarly to the case of $\bar n=2.0$, the state of the two atoms is not a pure stste, except for the case when the maxima of $C$ reduce remarkably,
Figs. 10b, 12b.\\
\begin{figure}
\begin{center}
\includegraphics[width=5.5cm,height=4.5cm]{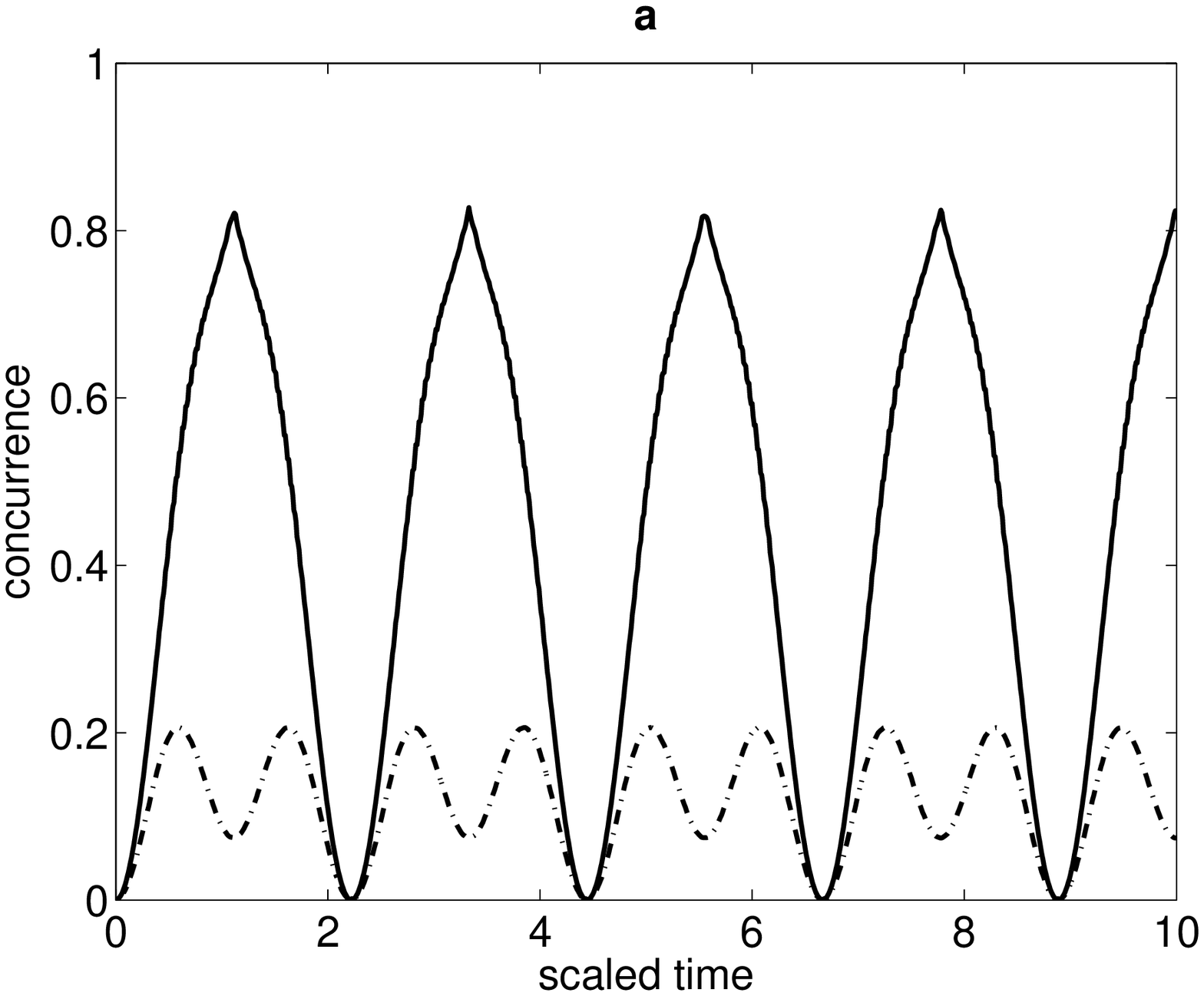} %
\includegraphics[width=5.5cm,height=4.5cm]{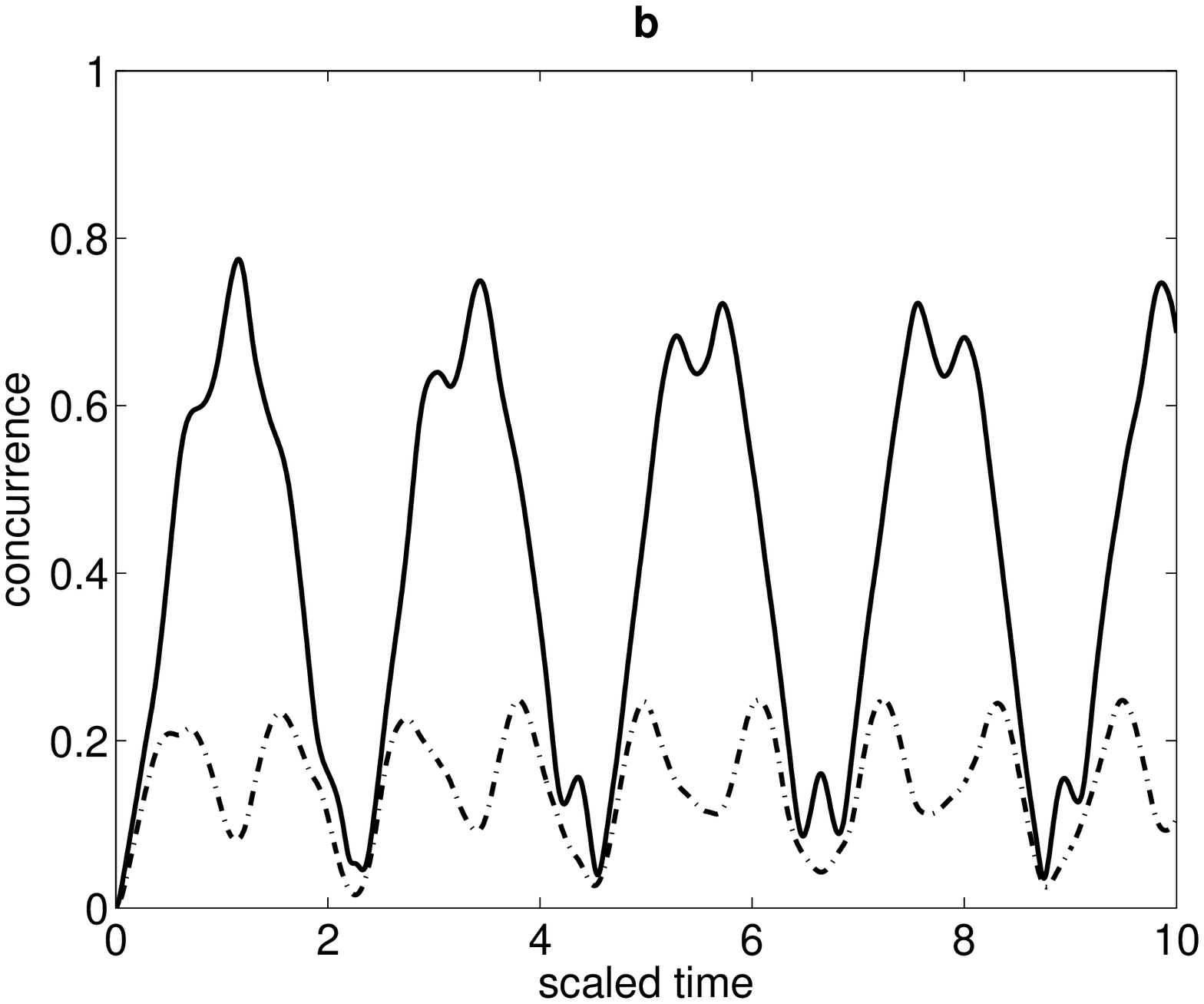}
\includegraphics[width=5.5cm,height=4.5cm]{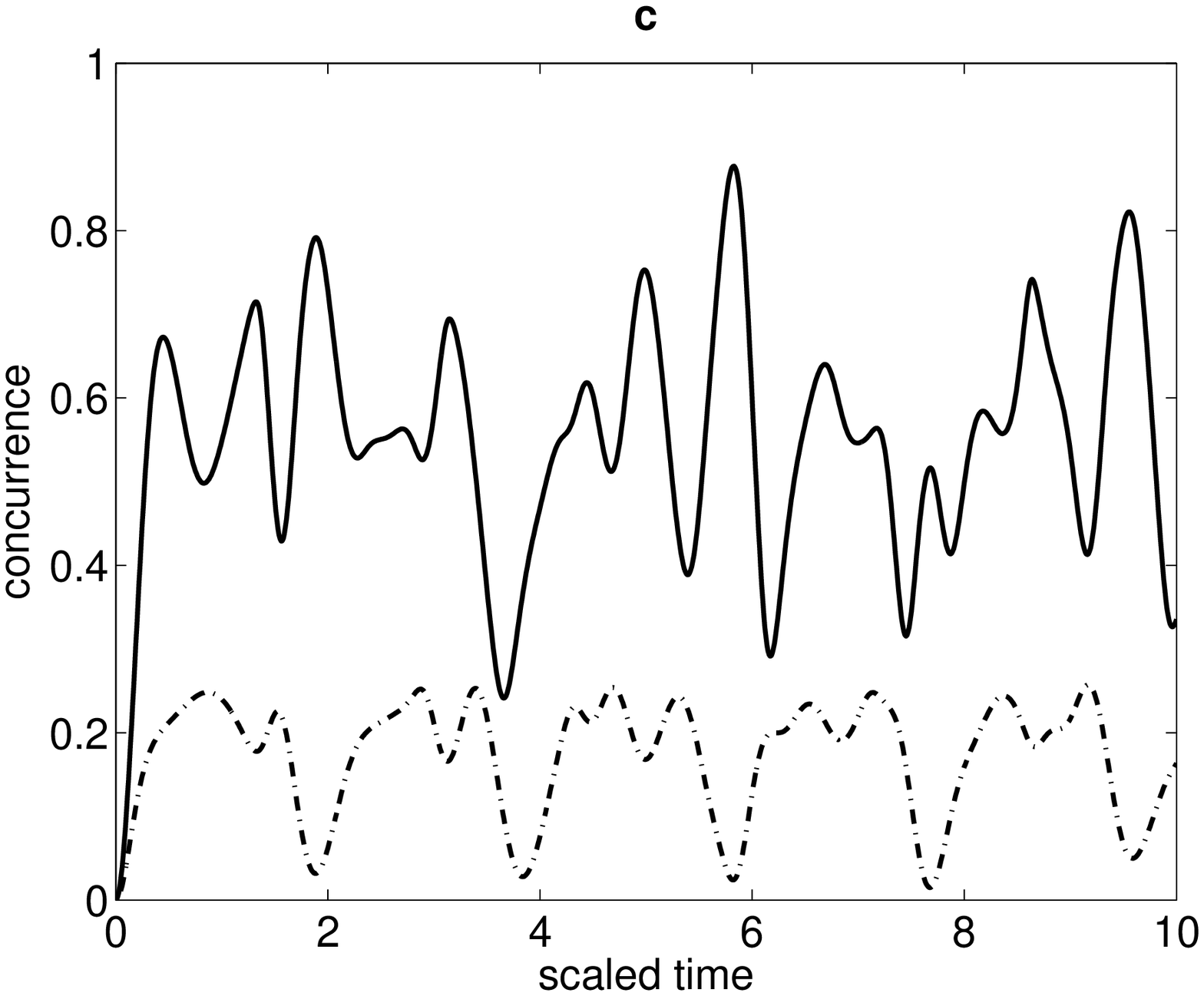}
\caption{The same as Fig. 9 but for $\protect\chi/\kappa=0.0$. (a) $r=0.01$. (b) $%
r=0.1$. (c) $\protect\chi/\kappa=0.5$, $r=0.3$}
\label{fig:1}       
\end{center}
\end{figure}
\begin{figure}
\begin{center}
\includegraphics[width=5.5cm,height=4.5cm]{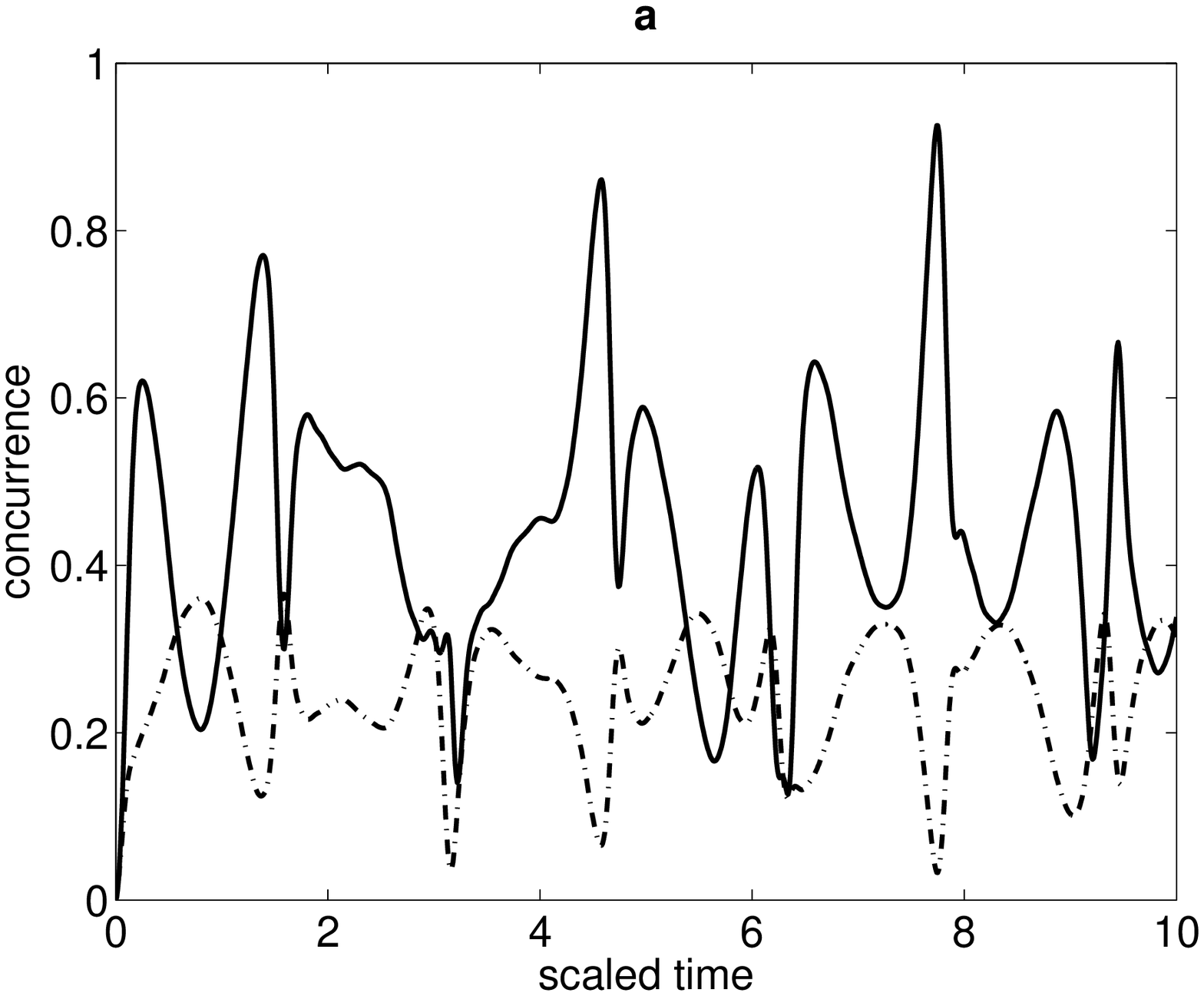} %
\includegraphics[width=5.5cm,height=4.5cm]{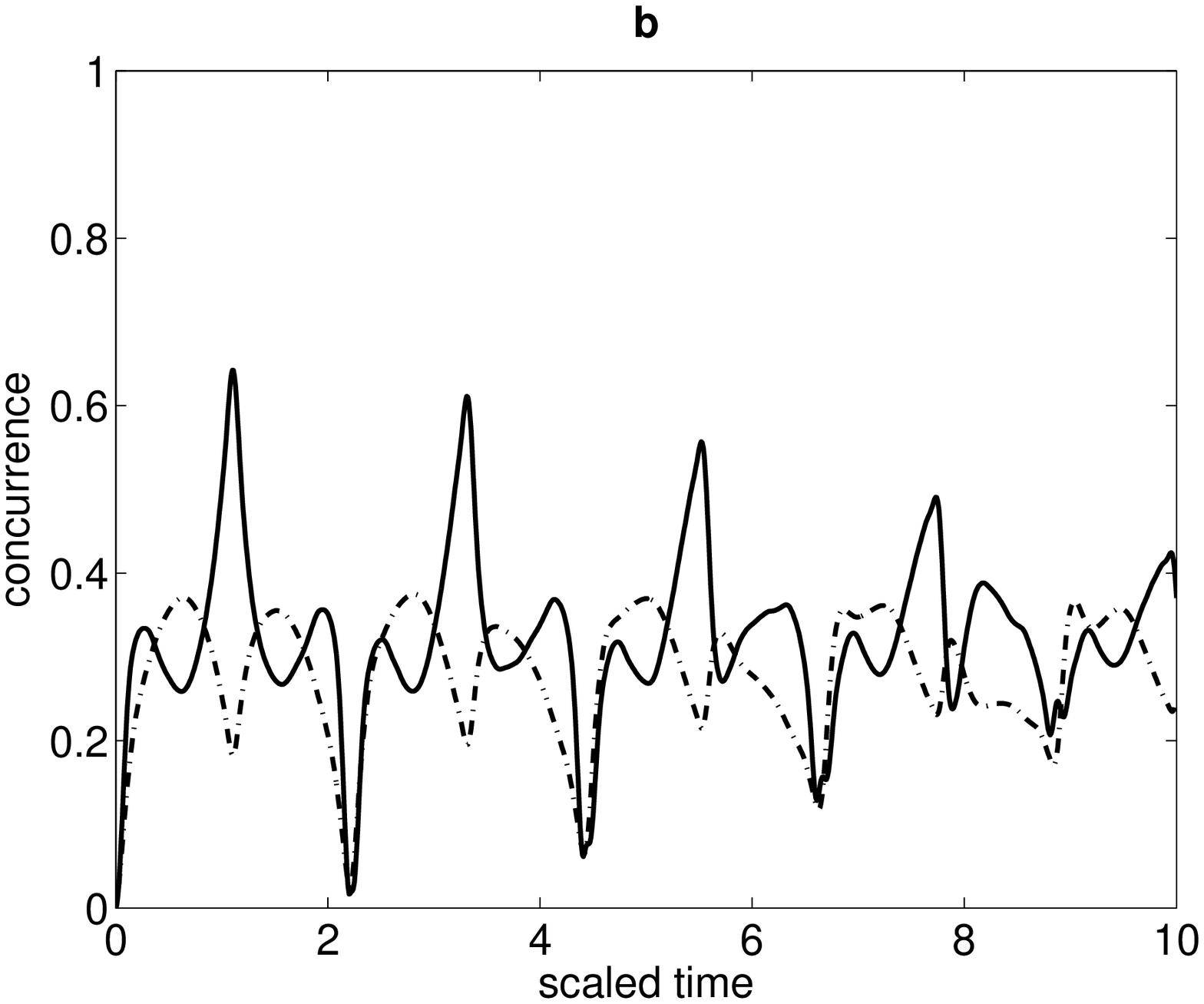}
\caption{The same as Fig. 9 but for $\bar n=2.0$.}
\label{fig:1}       
\end{center}
\end{figure}
\begin{figure}
\begin{center}
\includegraphics[width=5.5cm,height=4.5cm]{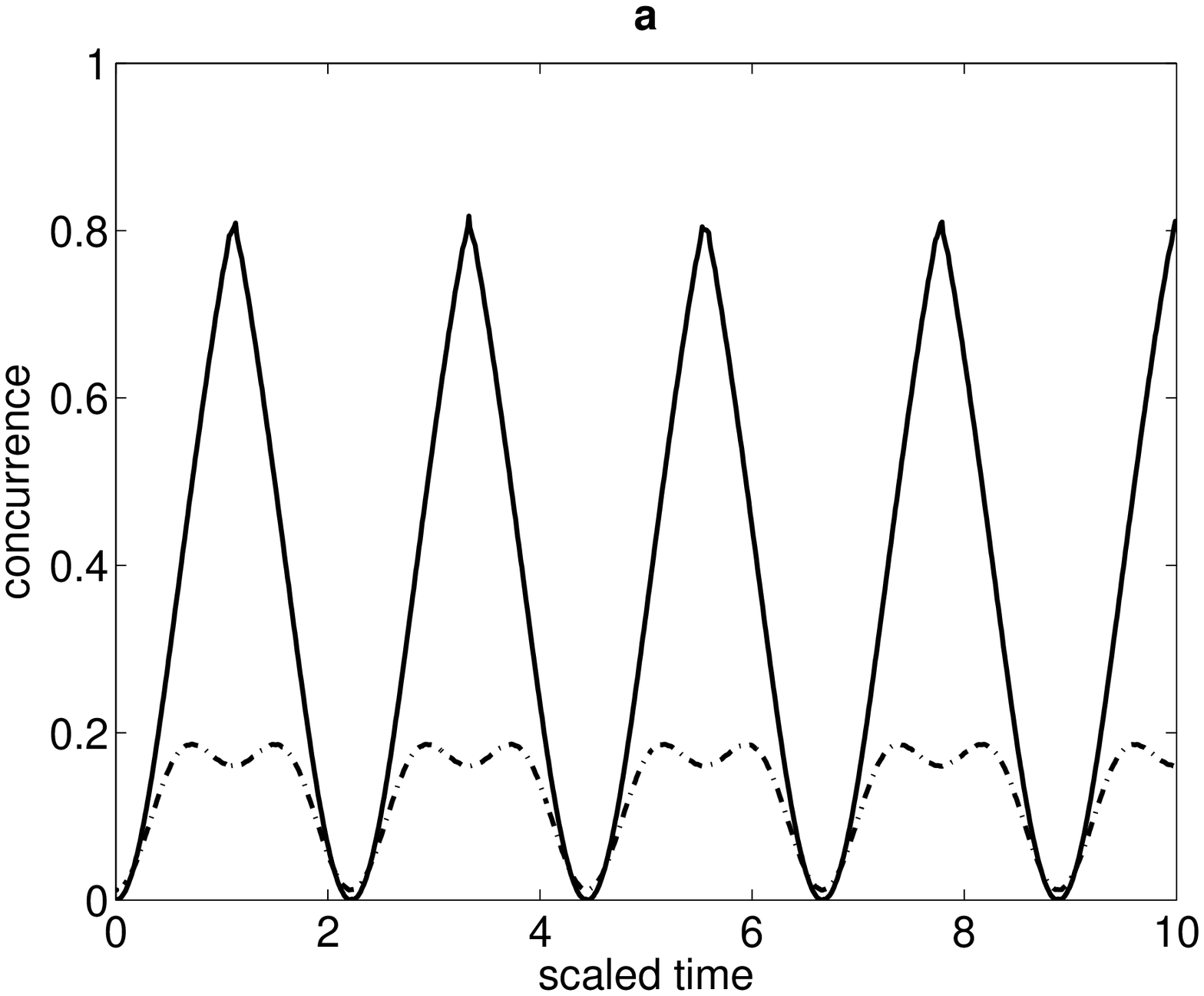} %
\includegraphics[width=5.5cm,height=4.5cm]{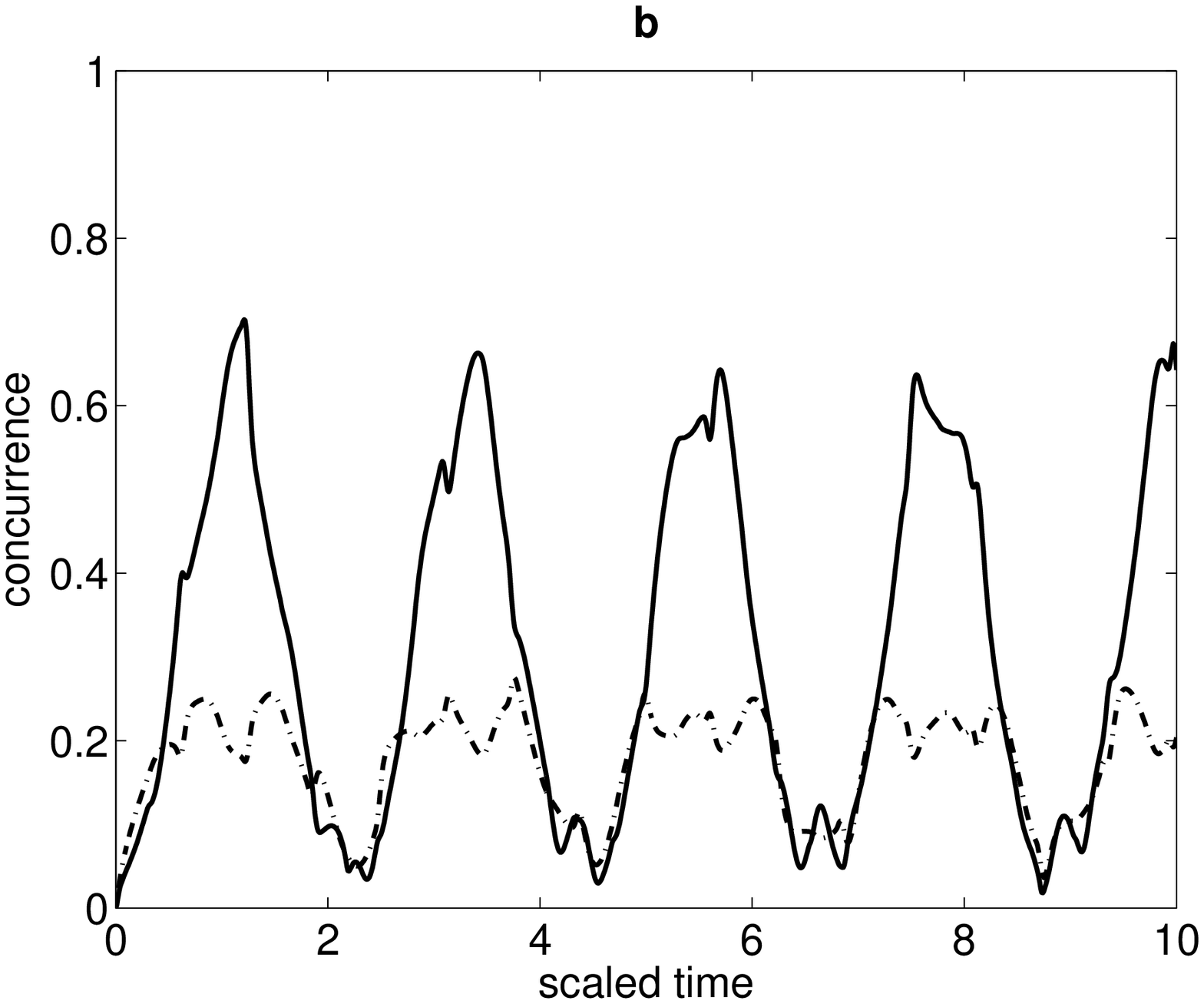}
\includegraphics[width=5.5cm,height=4.5cm]{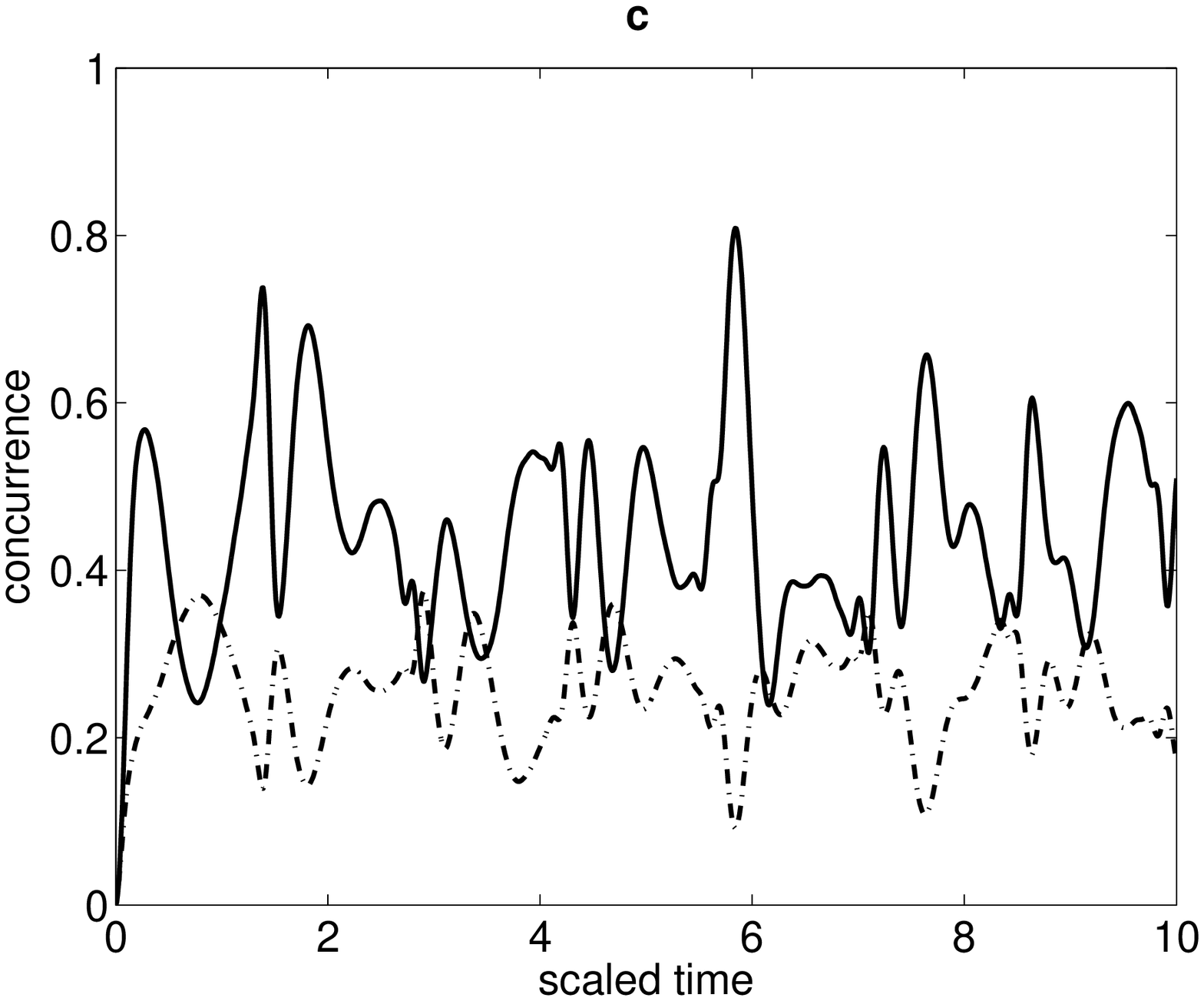}
\caption{The same as Fig. 10 but for $\bar n=2.0$.}
\label{fig:1}       
\end{center}
\end{figure}
When the effects of both Kerr-like medium and Stark shift are taken into
account, the only noticeable effect is the reduction of the minima of $C$ while
the general behavior as the case when no stark is present, Fig. 9c, 12d.
\subsubsection{Two excited atoms}
\label{sec:4}
To obtain the wave function of this case, we set $F_{n}=F_{n}\delta_{n,n}$ and $a=1$ in Eq. (19).\newline
With the same condition $p(n)=|F_{n}|^{2}$, the reduced density state of the
atom-system after taking the trace over the field variables has the form of
Eq. (28) with the coefficients given by:
$$
\rho_{11}(t)=\sum_{n} p(n)|H_{n}(t)|^{2},\eqno(64)
$$
$$
\rho_{22}(t)=\sum_{n} p(n) |T_{n+2}(t)|^{2},\eqno(65)
$$
$$
\rho_{23}(t)= \sum_{n} p(n) e^{i\kappa[2\frac{\chi}{\kappa} (2n+3)+\frac{r^{2}+1}{r}]t}~T_{n+2}(t)
J^{\ast}_{n+2}(t)=\rho^{\ast}_{32}(t),\eqno(66)
$$
$$
\rho_{33}(t)=\sum_{n} p(n) |J_{n+2}(t)|^{2},\eqno(67)
$$
$$
\rho_{44}(t)=\sum_{n} p(n) |V_{n+4}(t)|^{2}.\eqno(68)
$$
With these elements the concurrence $C(\rho)$ can be easily computed. 
\\
Remarkably interesting results are found when the injected thermal field
interacts with two atoms passing through it in excited states. The
results are shown in figures 13, 14, 15, and 16. We
notice clearly the average photon number reducing the general
behavior of the concurrence $C$, while similar behaviors to the
corresponding cases of Fock state field are noticed. Moreover, the
effect of nonlinear medium on increasing the maxima of $C$, and
creating a periodical entanglement with small oscillations around
its maximum with wider periods by taking into account the effect of
level shifts are preserved. Also, the behavior of $C$, where reaches
its maxima at the maxima of $\rho_{22}+\rho_{33}$ is also preserved,
see Figs.13-16 and 5-8. %
\begin{figure}
\begin{center}
\includegraphics[width=5.5cm,height=4.5cm]{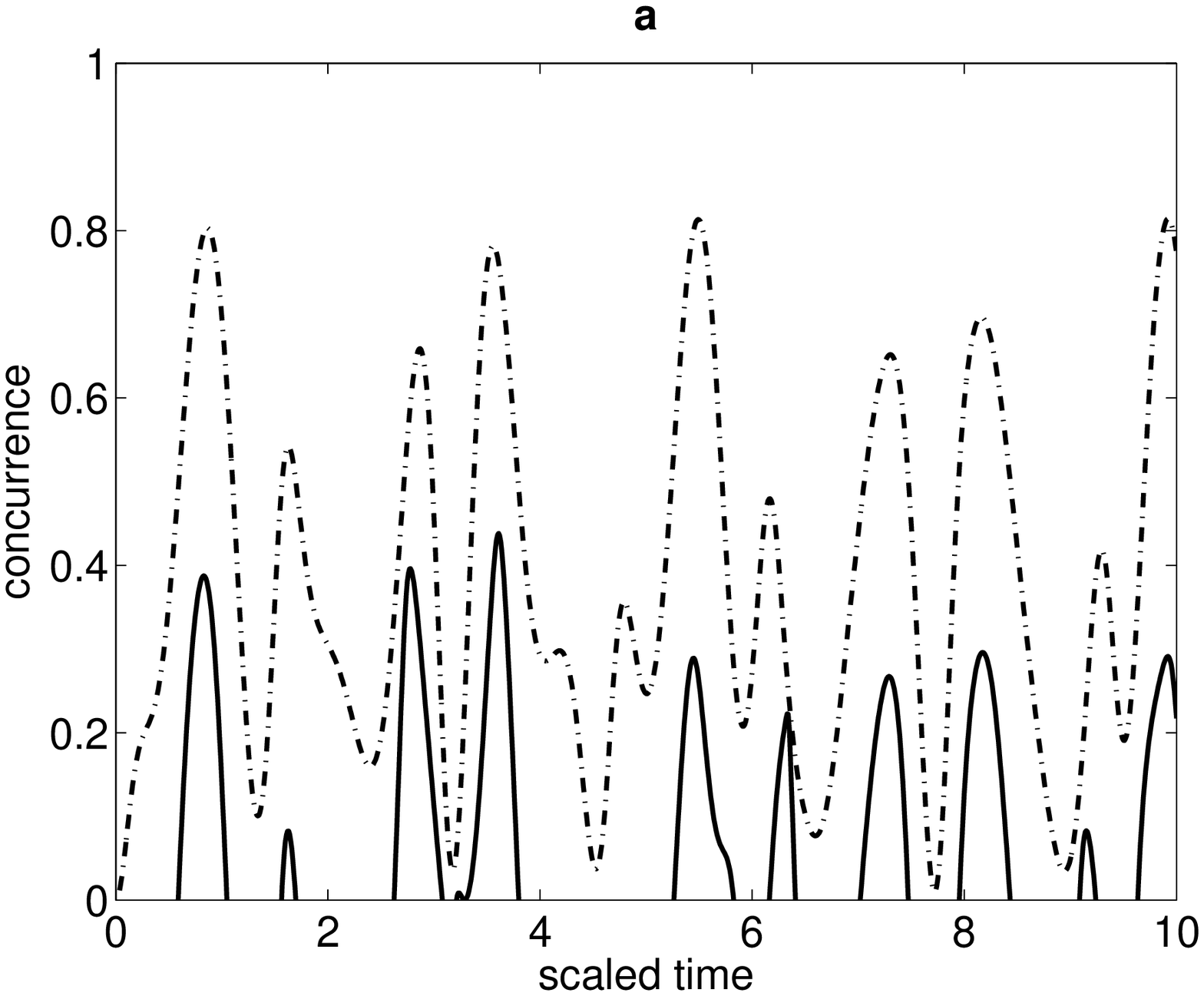} %
\includegraphics[width=5.5cm,height=4.5cm]{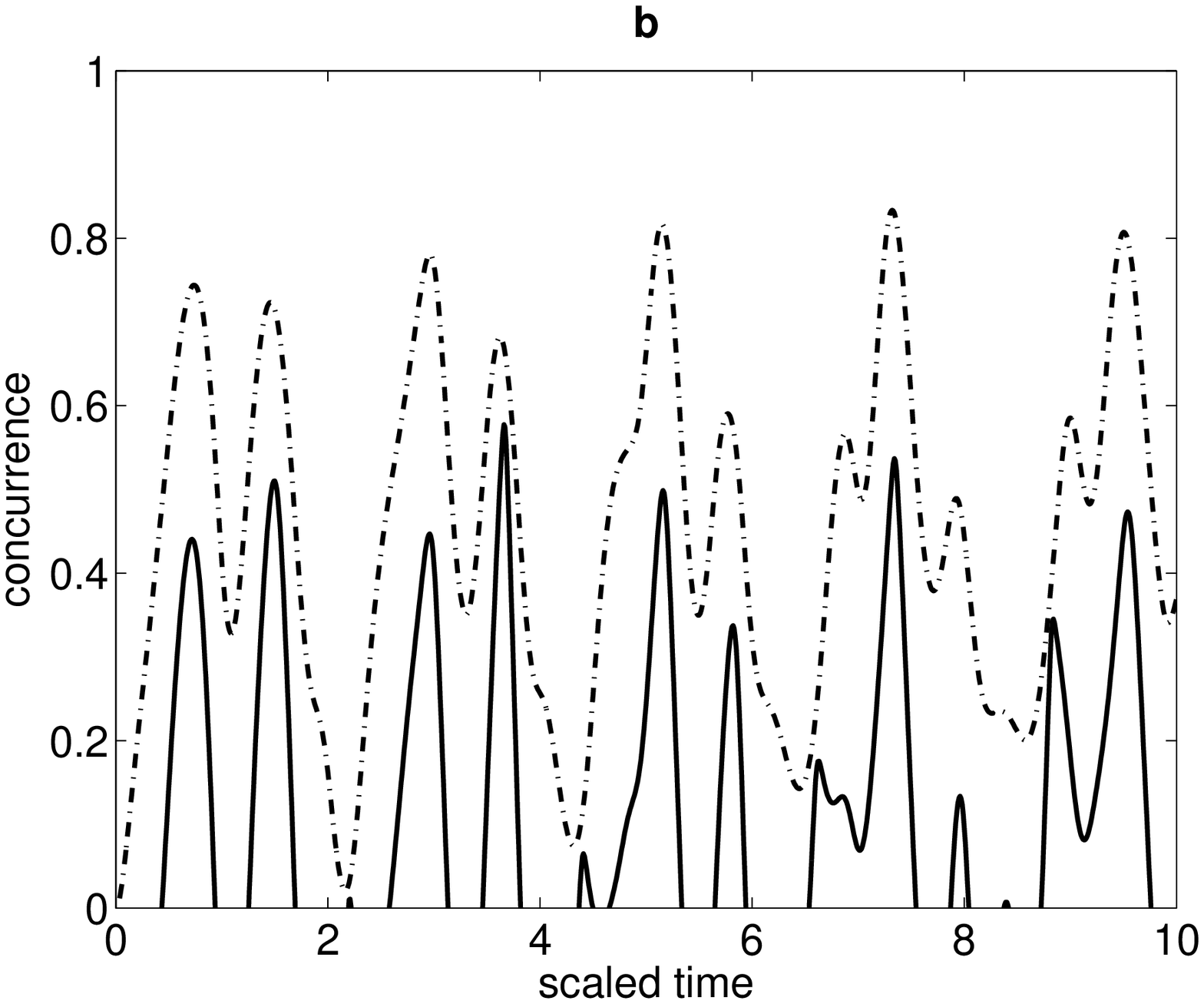}
\caption{Concurrence $C$ (solid curve) and $\rho_{22}+\rho_{33}$ (dotted curve) as functions of the scaled time $\kappa t$. The cavity field start from a thermal state with average photon number $\bar{n}=0.5$ where $r=0.0$. (a) $\protect\chi/\kappa%
=0.0$. (b) $\protect\chi/\kappa=0.5$.}
\label{fig:1}       
\end{center}
\end{figure}
\begin{figure}
\begin{center}
\includegraphics[width=5.5cm,height=4.5cm]{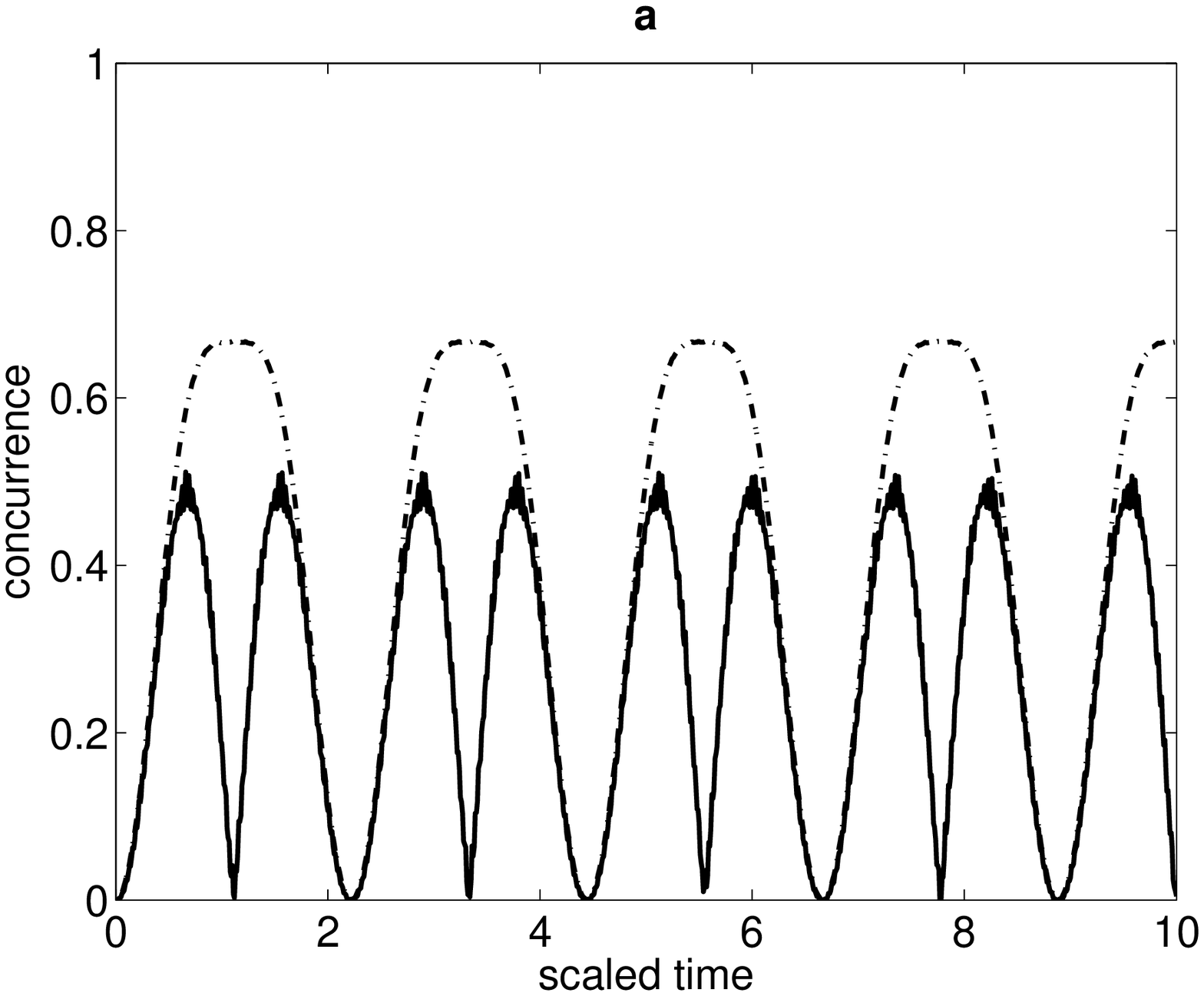} %
\includegraphics[width=5.5cm,height=4.5cm]{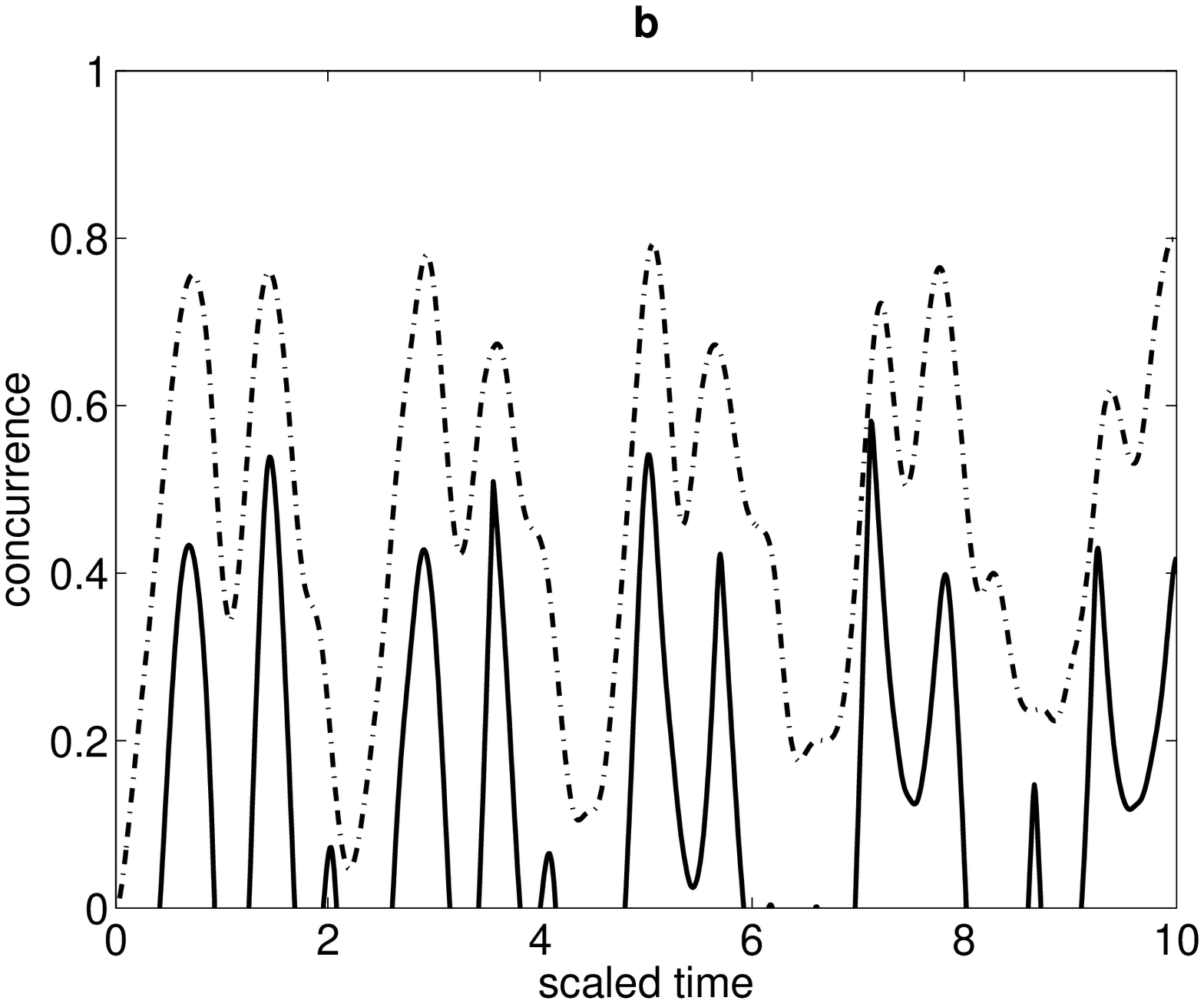}
\includegraphics[width=5.5cm,height=4.5cm]{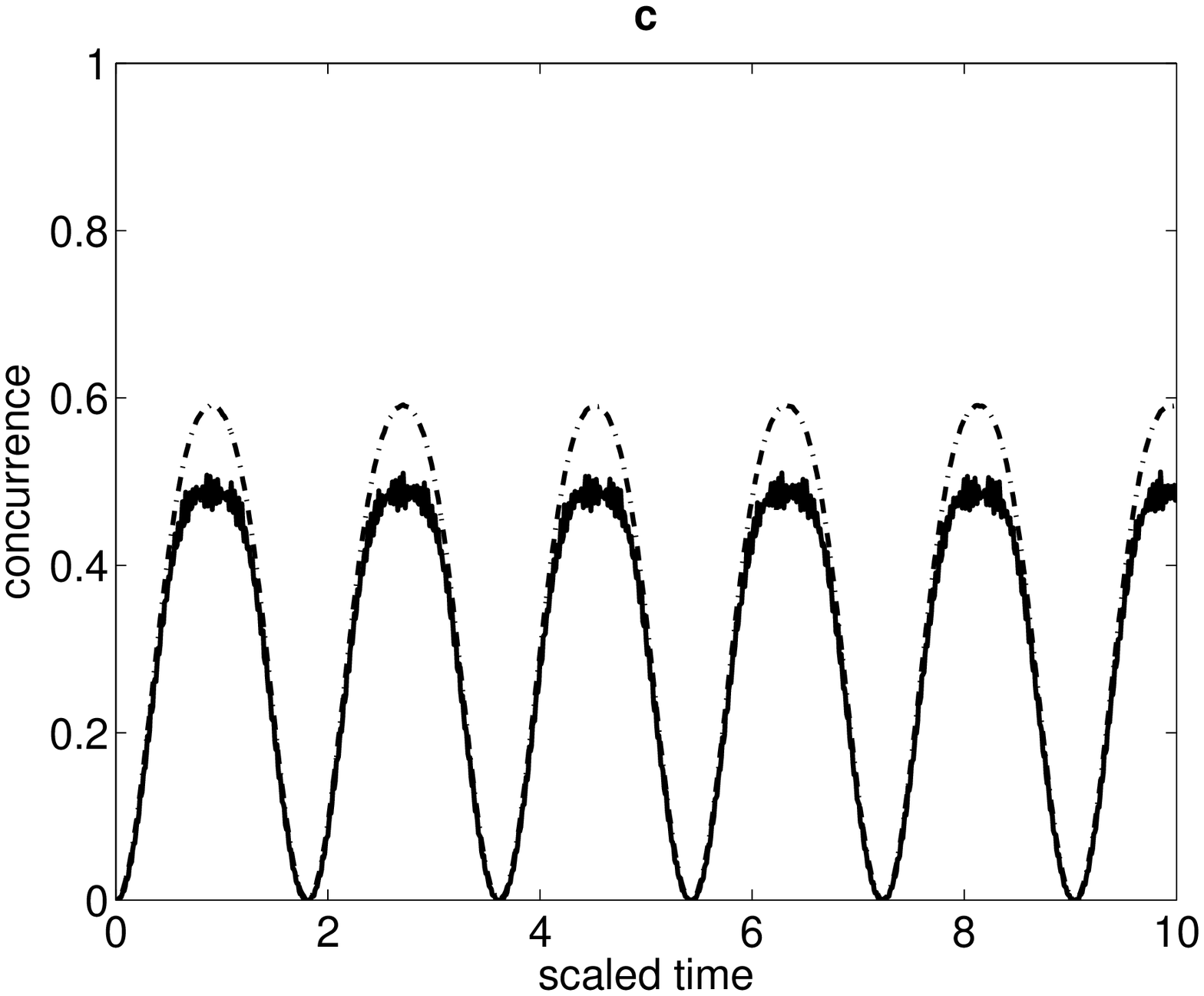} %
\includegraphics[width=5.5cm,height=4.5cm]{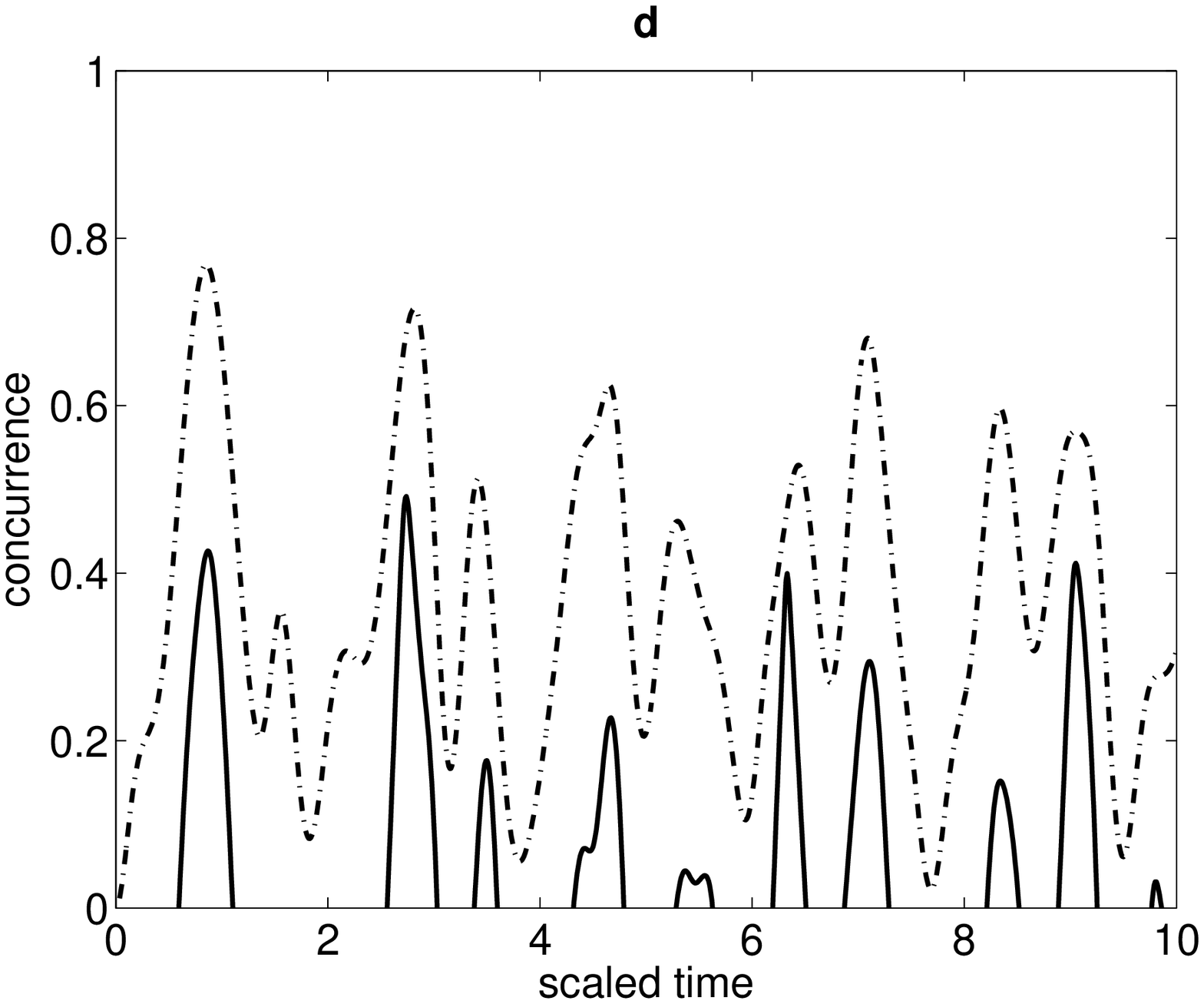}
\caption{The same as Fig. 13 but for $\protect\chi/\kappa=0.0$. (a) $r=0.01$. (b) $%
r=0.3$ (c) $\protect\chi/\kappa=1.0$, $r=0.01$ (d) $\protect\chi/\kappa=0.5$, $r=0.3$.}
\label{fig:1}       
\end{center}
\end{figure}
\begin{figure}
\begin{center}
\includegraphics[width=5.5cm,height=4.5cm]{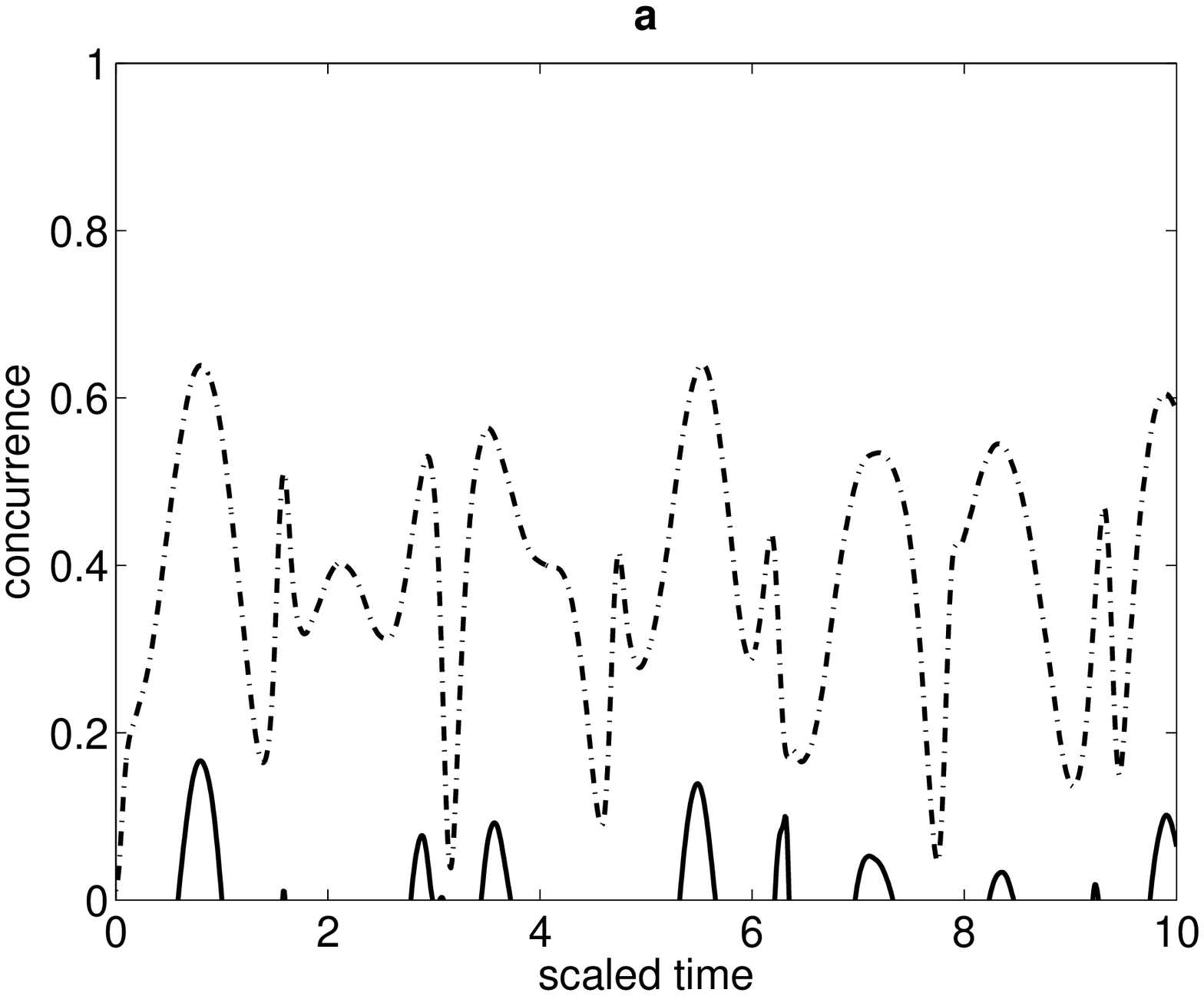} %
\includegraphics[width=5.5cm,height=4.5cm]{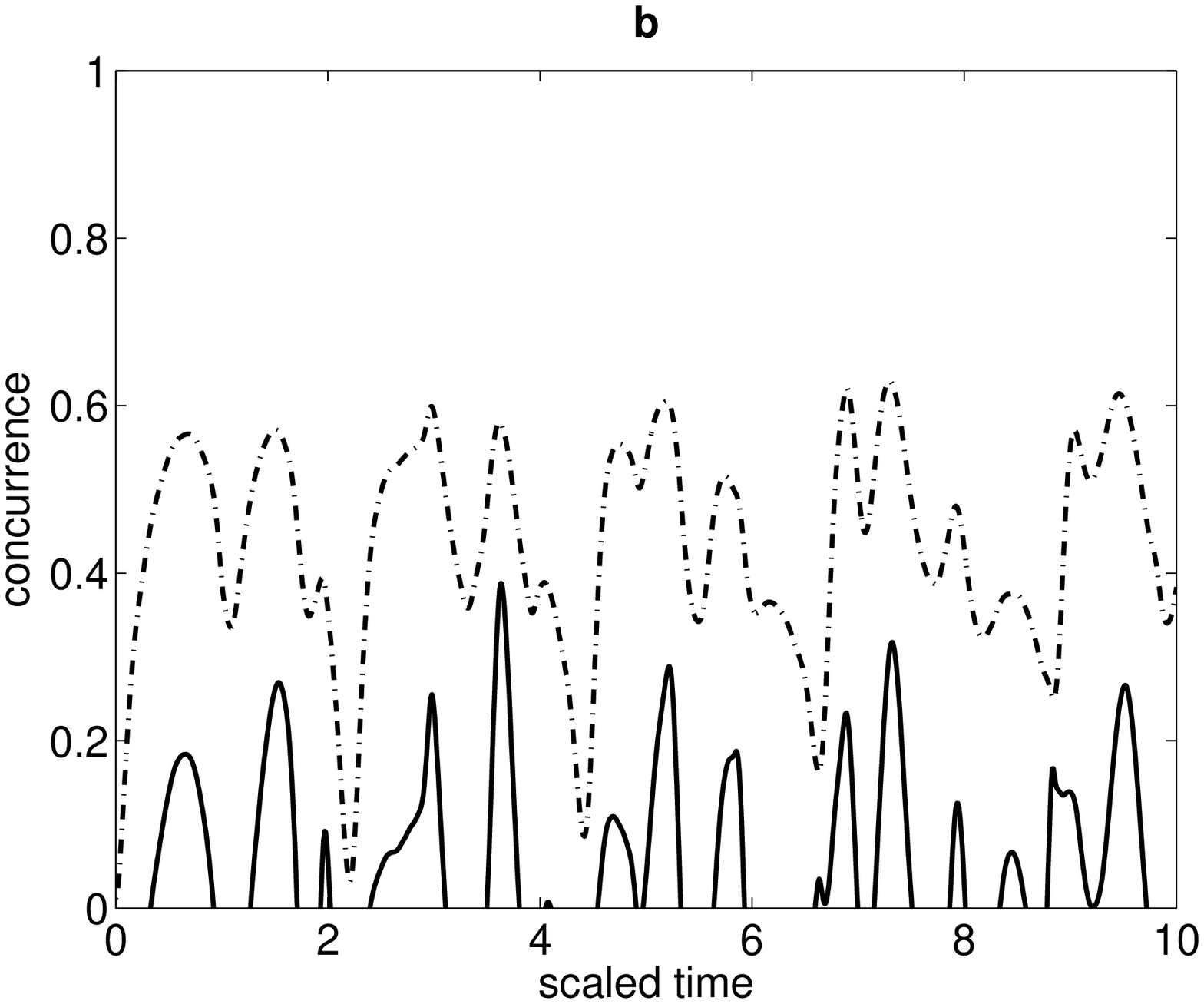}
\caption{The same as Fig. 13 but for $\bar n=2.0$.}
\label{fig:1}       
\end{center}
\end{figure}
\begin{figure}
\begin{center}
\includegraphics[width=5.5cm,height=4.5cm]{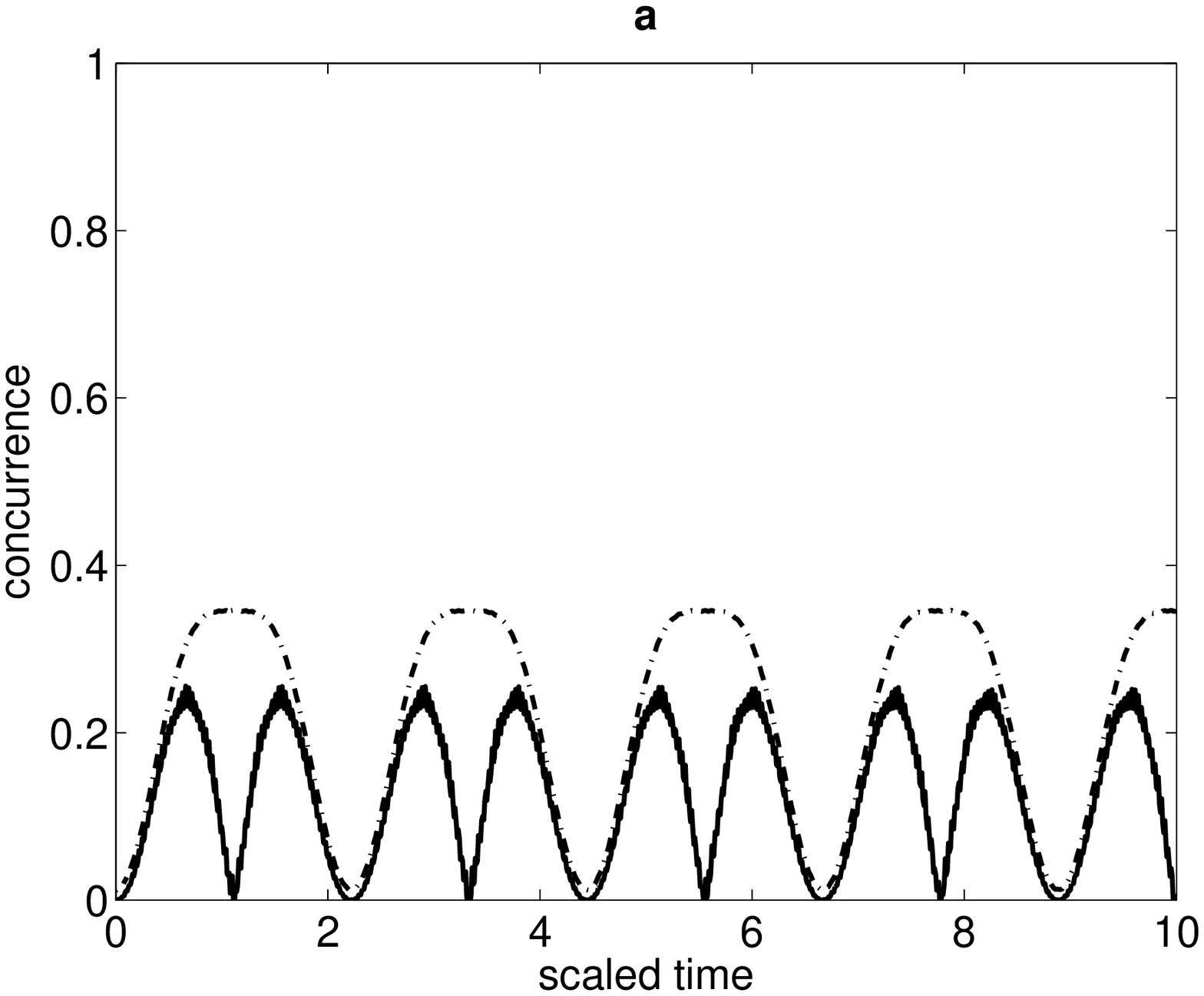} %
\includegraphics[width=5.5cm,height=4.5cm]{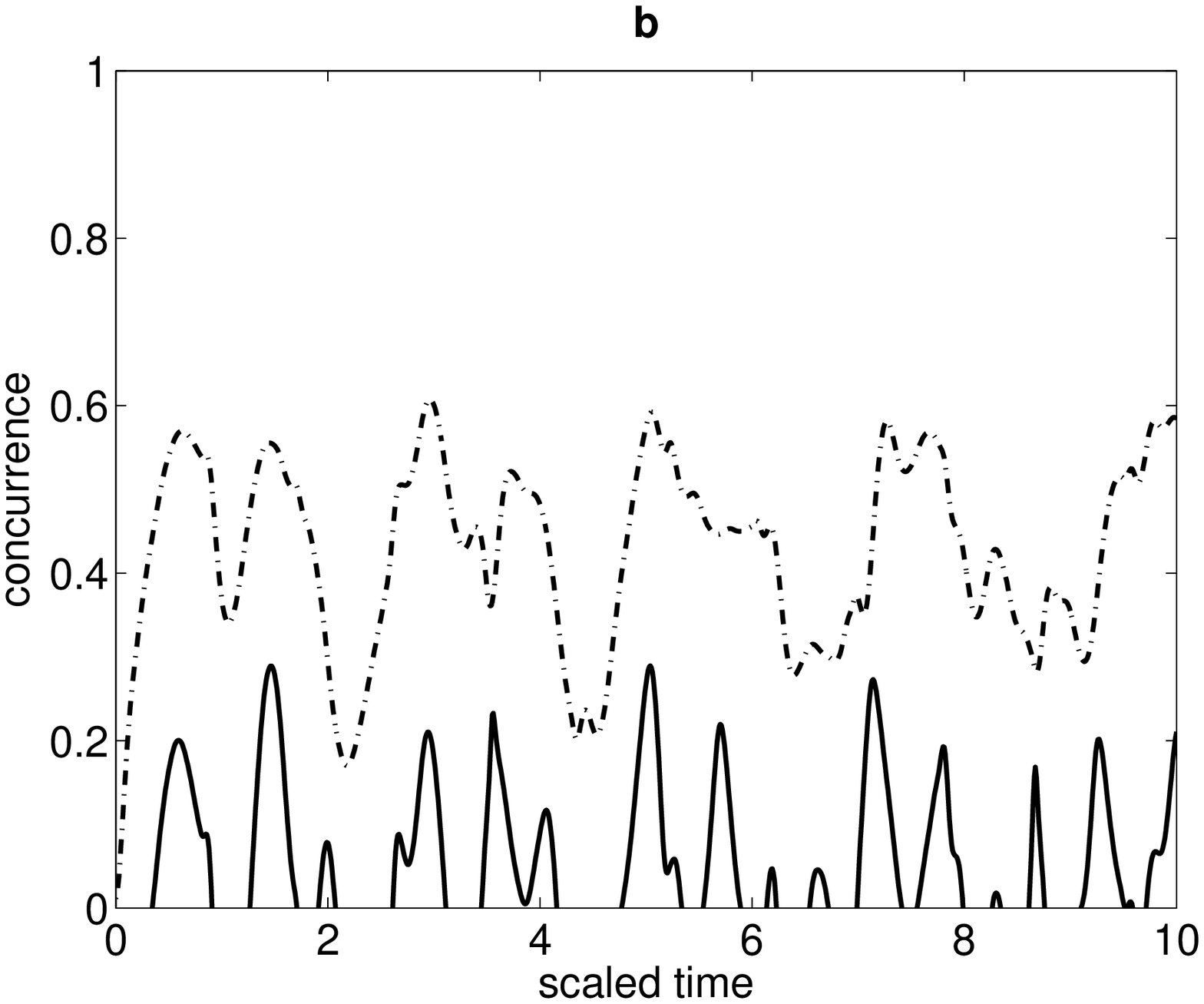}
\includegraphics[width=5.5cm,height=4.5cm]{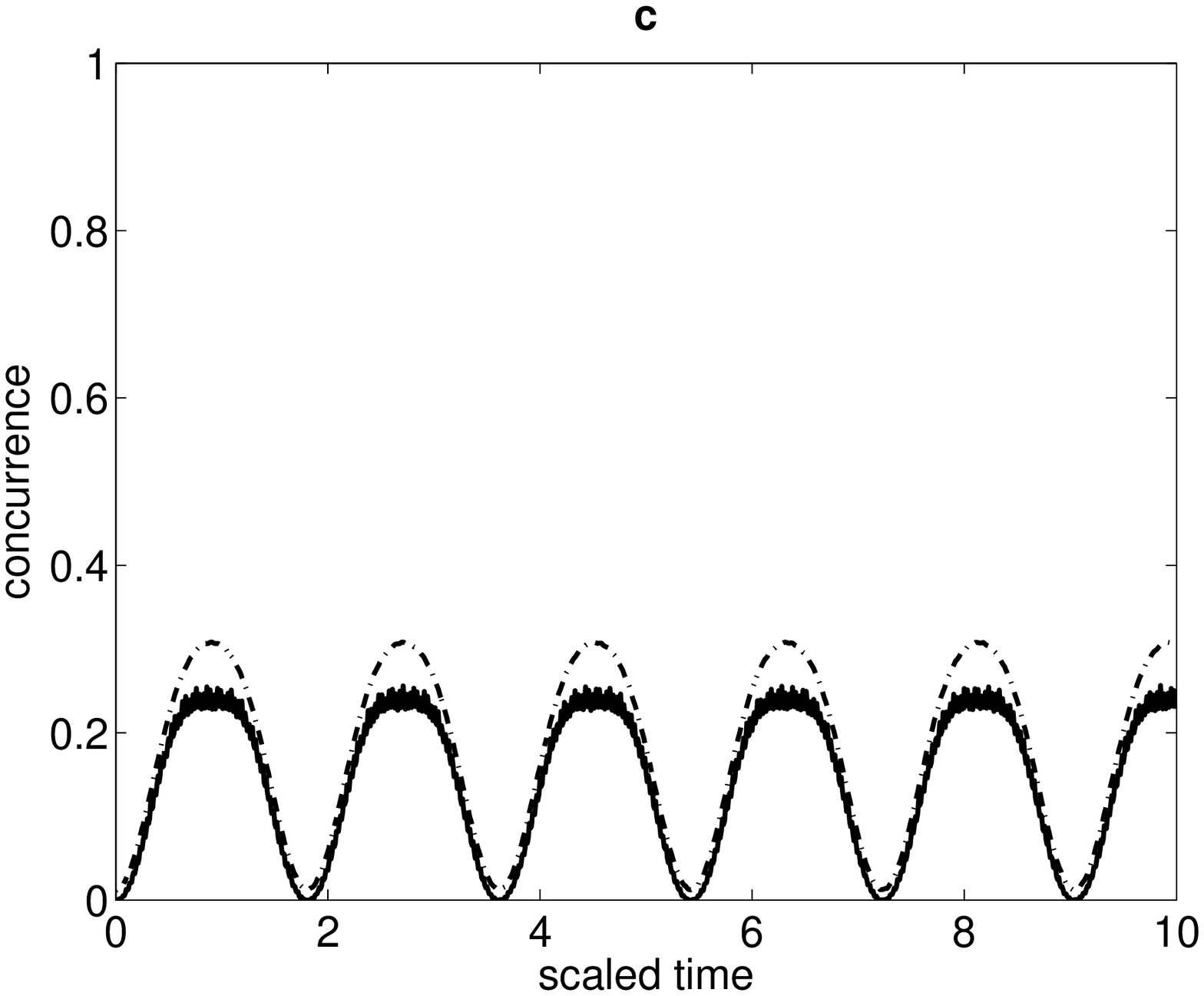} %
\includegraphics[width=5.5cm,height=4.5cm]{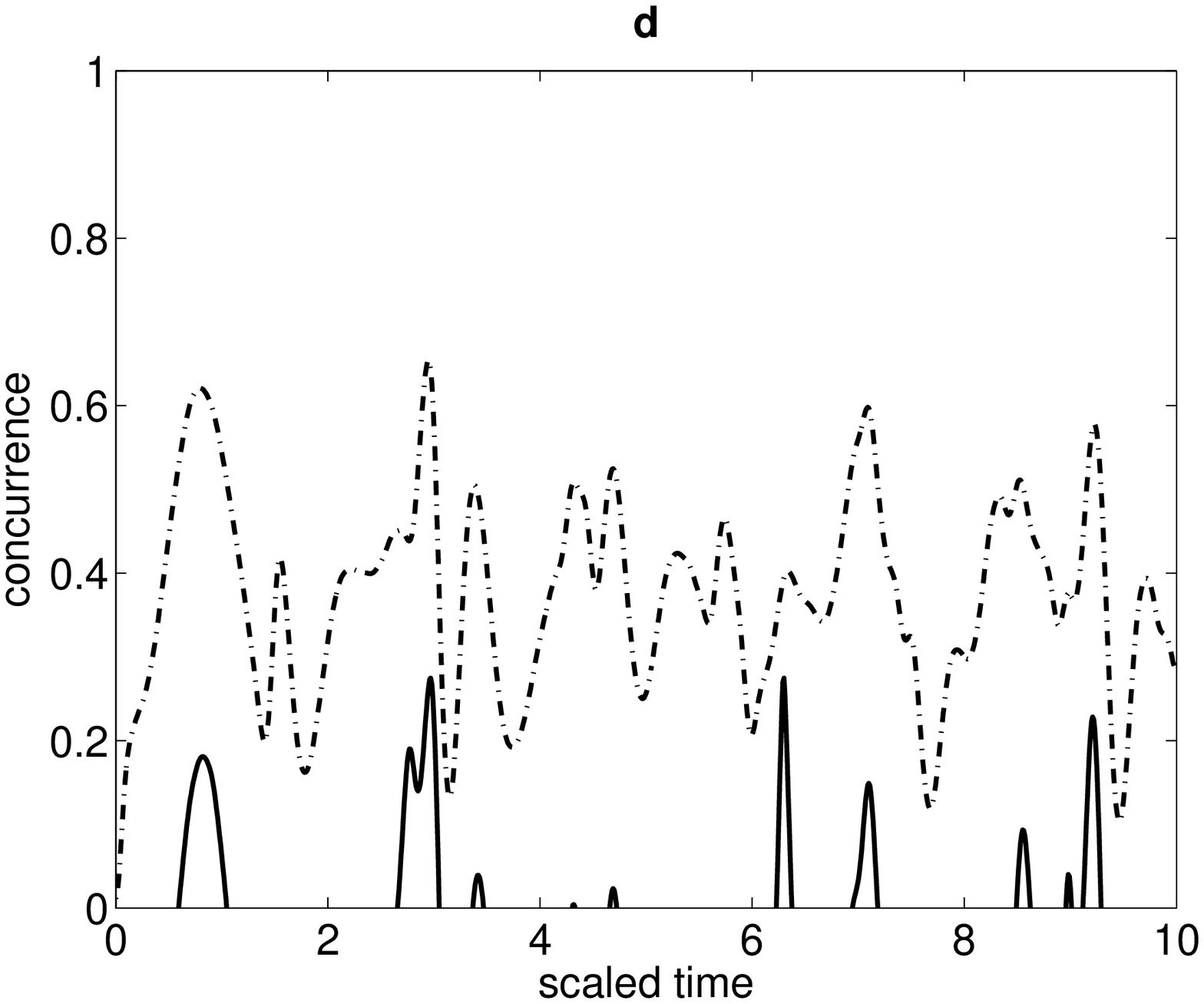}
\caption{The same as Fig. 14 but for $\bar n=2.0$.}
\label{fig:1}       
\end{center}
\end{figure}
\newpage
\section{Conclusion}
In conclusion, from the results illustrated in the previous
sections, we can conclude that two atoms (two-qubits) entanglement via
two-photon process is more sensitive to the initial conditions than
one photon process. Long-time intervals of two-qubit entanglement
can be achieved by filling the vacuum cavity with a Kerr-like medium with parameter close to unity and taking into account a slight level shift regardless of the initial states of the two atoms. Moreover, when the cavity contains only one
atom excited, long-time periods of two-qubits entanglement with no decay to zero can
be achieved by applying a weak Kerr medium with small level shift when the 
cavity is excited in the thermal state and contains only one excited atom. Furthermore, the two atoms become less entangled in excited cavity, while they become stronger entangled as well as the
effects of both Kerr-like medium and Stark shift taken into consideration.
\begin{acknowledgements}
The author would like to thank the referees for their constrictive
comments that helped to improve the text in many ways.
\end{acknowledgements}


\begin{thebibliography}{99}
%
%
%
%
%
\bibitem{Nielsen00} M.~Nielsen and I.~Chang,\emph{Quantum Computation
and Quantum Communication} (Cambridge University Press, Cambridge,
England, 2000).

\bibitem {ZZHE93} M. \.{Z}ukowski, A. Zeilinger, M. A. Horne, and A. K. Ekert, Phys.
Rev. Lett. 71, 4287 (1993); S. Bose, V. Vedral, and P. L. Knight, Phys. Rev.
A 57, 822 (1998); 60, 194 (1999); B.-S. Shi, Y.-K. Jiang, G.-C. Guo, Phys.
Rev. A 62, 054301 (2000); L. Hardy and D. D. Song, Phys. Rev. A 62, 052315
(2000).

\bibitem{DBCZ99} W. D\"{u}r, H.- J. Briegel, J. I. Cirac and P. Zoller, Phys. Rev. A
59, 169 (1999).

\bibitem {D99} D. P. DiVincenzo et al, "The entanglement of assistance", in Lecture
Notes in Computer Science 1509 (Springer- Verlag, Berlin, 1999), pp. 247-257.%


\bibitem {A04} M. Abdel-Aty, J. Opt. B 6,  201 (2004); O. Cohen, Phys. Rev. Lett. 80, 2493 (1998).

\bibitem {VPC04} F. Verstraete, M. Popp, and J. I. Cirac, Phys. Rev. Lett. 92, 027901
(2004); M. Popp, F. Verstraete, M. A.Martin- Delgado and J. I. Cirac Phys.
Rev. A 71, 042306 (2005).

\bibitem{GS04} G. Gour and B. C. Senders, Phys. Rev. Lett. 93, 260501(2004).

\bibitem {OA07} A.-S. F. Obada, and  M. Abdel-Aty, Phys. Rev. B, 75, 195310
(2007); G. Gour, Phys. Rev. A , 71, 012318 (2005).

\bibitem{PWK04} N. A. Peters, T.-C. Wei, and P. G. Kwiat, Phys. Rev. A
70, 052309 (2004).
%
\bibitem {BBPS96} C. H. Bennett, H. J. Bernstein, S. Popescu, and B. Schumacher, Phys.
Rev. A 53, 2046 (1996).

\bibitem {BDSW96} C. H. Bennett, D. P. DiVincenzo, J. A. Smolin, and W. K. Wootters,
Phys. Rev. A 54, 3824 (1996).

\bibitem {VP96} V. Vedral and M. B. Plenio, Phys. Rev. A 57, 1619 (1996).

\bibitem {V02} V. Vedral, Rev. Mod. Phys. 74, 197 (2002), M. Abdel-Aty, Prog. Quant. Elect. 31, 1 ( 2007)

\bibitem {VT99} G. Vidal and R. Tarrach, Phys. Rev. A 59, 141 (1999).

\bibitem {VPRK95} V. Vedral, M. B. Pienio, M. A. Rippin and P. L. Knight, Phys. Rev.
Lett. 78, 2275 (1995).

\bibitem {W98} W. K. Wootters, Phys. Rev. Lett. 80 2245 (1998).

\bibitem {GMN06} B. Ghosh, A. S. Majumdar and N. Nayak, Phys. Rev. A
74, 052315 (2006)
\bibitem {GMN07} B. Ghosh, A. S. Majumdar, and N. Nayak, Int. J. Quant.
Inf. 5, 169 (2007)
%
%
\bibitem{JCPZ94} J. I. Cirac and P. Zoller, Phys. Rev. A 50, R2799 1994; J.
I. Cirac, P. Zoller, H. J. Kimble, and H. Mabuchi, Phys.
Rev. Lett. 78, 3221 1997; L.-M. Duan, M. D. Lukin, J.
I. Cirac, and P. Zoller, Nature London 414, 413 2001; E.
Solano, G. S. Agarwal, and H. Walther, Phys. Rev. Lett.
90, 027903 2003; S. G. Clark and A. S. Parkins, ibid. 90,
047905 2003; L.-M. Duan, B. Wang, and H. J. Kimble,
Phys. Rev. A 72, 032333 2005; G. S. Agarwal and K. T.
Kapale, ibid. 73, 022315 2006.
\bibitem{SPPK91} S. J. D. Phoenix and P. L. Knight, Phys. Rev. A 44, 6023
(1991)
\bibitem{JPLK91} S. J. D. Phoenix and P. L. Knight, Phys. Rev. Lett. 66,
2833 (1991).
\bibitem{KLMCK93} I. K. Kudryavtsev, A. Lambrecht, H. Moya-Cessa and P. L. knight, J. Mod. Opt., 40, 1605 (1993).
\bibitem {RBH01} J. M. Raimond, M. Brune, and S. Haroche, Rev. Mod. Phys. 73, 565
(2001).

\bibitem {MCJLKK06} D. N. Matsukevich, T. Chanelire, S. D. Jenkins, S.-Y. Lan, T. A. B.
Kennedy, and A. Kuzmich, Phys. Rev. Lett., 96, 030405 (2006).
%
\bibitem {DGMN04} A. Datta, B. Ghosh, A. S. Majumdar and N. Nayak, Europhys. Lett. 67,
934 (2004).
%
\bibitem {TDD03} T. E. Tessier,. I. H. Deutsch, and A. Delgado, quant-ph/0306015 v4,
2003.

\bibitem {M02} P. Masiak, Phys. Rev. A 66, 023804 (2002).
%
\bibitem {TDFD03} T. Tessier, A. Delgado, I. Fuentes-Guridi, and I. H. Deutsch, Phys.
Rev.A 68, 062316 (2003).
%
\bibitem{HMNWBRH97} E. Hagley, X. Maitre, G. Nogues, C. Wunderlich, M.
Brune, J. M. Raimond, and S. Haroche, Phys. Rev. Lett.
79, 1 1997; A. Rauschenbeutel, P. Bertet, S. Osnaghi,
G. Nogues, M. Brune, J. M. Raimond, and S. Haroche,
Phys. Rev. A 64, 050301 R 2001; A. Auffeves, P. Maioli,
T. Meunier, S. Gleyzes, G. Nogues, M. Brune, J. M.
Raimond, and S. Haroche, Phys. Rev. Lett. 91, 230405
2003; S. Nussmann, M. Hijlkema, B. Weber, F. Rohde,
G. Rempe, and A. Kuhn, ibid. 95, 173602 2005; D. N.
Matsukevich, T. Chaneliere, S. D. Jenkins, S.-Y. Lan, T.
A. B. Kennedy, and A. Kuzmich, ibid. 96, 030405 2006.
%
\bibitem {KLAK02} M. S. Kim, Jinhyoung Lee, D. Ahn and P. L. Knight, Phys. Rev. A 65,
040101(R) (2002); L. Zhou, H. S. Song and C. Li, J. Opt. B: Quantum
Semiclass. Opt. 4, 425 (2002).

\bibitem{Braun02}  D. Braun, Phys. Rev. Lett. 89, 277901 2002.
%
\bibitem{LEW96} M. L$\ddot{o}$ffler, B.-G. Englert and H Walther, Appl. Phys. B
63, 511 (1996).
\bibitem{AMNN01} A. S. Majumdar and N. Nayak, Phys. Rev. A 64, 013821
(2001)
\bibitem{RSKW90} G. Rempe, F. Schmidt-Kaler, and H.Walther, Phys. Rev.
Lett. 64, 2783 1990; G. Rempe, and H. Waklther, Phys.
Rev. A 42, 1650 1990; H. Paul, and Th. Richter, Optics
Commun. 85, 508 1991; J. D. Cresser, Phys. Rev. A, 46,
5913 1992.
%

\bibitem{PB93} S. J. D. Phoenix and S. M. Barentt, J. Mod. Opt., 40(6), 979 (1993).
%
\bibitem{FIJAME86} P. Filipowicz, J. Javanainen, and P. Meystre, Phys. Rev. A 34, 3077 (1986).
%
\bibitem{PK88} S. J. D. Phoenix and P. L. Knight, Ann. Phys. 186, 381-407 (1988).
%
\bibitem{DMHW85} D. Meschede, H.Walther, and G. Muller, Phys. Rev Lett.
54, 551 1985 .
\bibitem {FJM86} P. Filipowicz, J. Javanainen, and P. Meystre, J. Opt. Soc. Am. B 3, 906 (1986).
\bibitem{LNT65} M. Lipeles, R. Novick, and N. Tolk, Phys. Rev. Lett. 15,
690 (1965); K. J. Mcneil and D. F. Walls, J. Phys. A 7,
617 (1974); N. Nayak and B. K. Mohanty, Phys. Rev. A
19, 1204 (1979).
\bibitem{MCKRD94} H. Moya-Cessa, P. L. Knight and A. Rosenhouse-Dantsker, Phys. Rev. A 50, 1814 (1994).
\bibitem{Yuen76} H. P. Yuen, Phys. Rev. A 13, 2226 (1976); C. M. Caves,
Phys. Rev. D 23, 1693 (1981).
\bibitem{FALI95} M.-F. Fang and H.-E. Liu, Phys. Lett. A 200, 250 (1995)
\bibitem{AFAT02} M. Abdel-Aty, S. Furuichi and M. Ateto, Jpn. J. Appl. Phys. 41, 111 (2002)
%
\bibitem {WLG06} C.-Z. Wang, C.-X. Li and G.-C. Guo, Eur. Phys. J. 37, 267 (2006); (M. Abdel-Aty, S. Abdel-Khalek and A.-S. F. Obada, Chaos, Solitons and Fractals 12, 2015 (2001) and papers therein). 
%
%
\bibitem{JOPU92} A. Joshi and R. R. Puri, Phys. Rev. A 45, 5056 (1992).
\bibitem{Hillaer91} M. Hillery, Phys. Rev. A 44, 4578 (1991).
\bibitem{CHCOWO92} A.N. Chaba, M.J. Collett, D.F. Walls, Phys. Rev. A 46,
1499 (1992).
\bibitem{ZHGCLM00} R. Zambrini, M. Hoyuelos, A. Gatti, P. Colet, L. Lugiato,
M.S. Miguel, Phys. Rev. A 62, 063801 (2000).
\bibitem{Gerry99} C.C. Gerry, Phys. Rev. A 59, 4095 (1999).
\bibitem{PKH03} M. Paternostro, M.S. Kim, B.S. Ham, Phys. Rev. A 67,
023811 (2003).
\bibitem{PACH00} J. Pachos, M. Chountasis, Phys. Rev. A 62, 052318
(2000).
\bibitem{VFT00} D. Vitali, M. Fortunato, P. Tombesi, Phys. Rev. Lett. 85,
445 (2000).
\bibitem{Ateto07} M. S. Ateto, Int. J. Quant. Inf. 5(4), 535 (2007).
\bibitem{EGSWH96} B.-G. Englert T. Gantsog, A. Schenzle, C. Wagner and
H Walther, Phys. Rev. A 53, 4386 (1996).
\bibitem{MHNWBRH97} X. Maˆitre, E. Hagley, G. Nogues, C. Wunderlich, P. Goy,
M. Brune, J. M. Raimond and S. Haroche, Phys. Rev.
Lett. 79, 769 (1997).
\bibitem{ELSW002} B.-G. Englert, P. Lougovski, E. Solano and H. Walther,
e-print: quant-ph/0209128v1 24Sep 2002.

\bibitem{HHH95} R. Horodecki, P. Horodecki and M. Horodecki, Phys.
Lett. A 200, 340(1995).
\bibitem{FNLU96} M.-F. Fang and H.-E. Liu, Phys. Lett. A 210, 11 (1996)
%
%
%
\end{thebibliography}


\end{document}